\documentclass{article} 
\usepackage[preprint]{colm2026_conference}

\usepackage{microtype}
\usepackage{hyperref}
\usepackage{url}
\usepackage{booktabs}
\usepackage{amsmath}
\usepackage{paralist}
\usepackage{multirow}
\usepackage{tabularx}
\usepackage{amssymb}
\usepackage{pifont}
\usepackage{xcolor} 
\usepackage{subcaption}
\usepackage{algorithm}
\usepackage{algpseudocode}
\usepackage[normalem]{ulem}
\usepackage{fontawesome}
\usepackage{graphicx}
\usepackage{colortbl}
\usepackage{wrapfig}
\usepackage{pifont}
\usepackage{float}
\usepackage{afterpage}
\usepackage{wrapfig}
\definecolor{rankfirst}{RGB}{198, 239, 206}
\definecolor{ranksecond}{RGB}{189, 215, 238}
\definecolor{modeldeepseek}{RGB}{218, 218, 255}
\definecolor{modelllama}{RGB}{255, 235, 205}
\newcommand{\cmark}{\textcolor{green!70!black}{\ding{51}}}
\newcommand{\xmark}{\textcolor{red}{\ding{55}}}

\usepackage{lineno}

\definecolor{darkblue}{rgb}{0, 0, 0.5}
\hypersetup{colorlinks=true, citecolor=darkblue, linkcolor=darkblue, urlcolor=darkblue}

\title{RAGRouter-Bench: \\A Dataset and Benchmark for Adaptive RAG Routing}

\author{Ziqi Wang\textsuperscript{1}\thanks{Equal contribution.}, Xi Zhu\textsuperscript{1}\footnotemark[1], Shuhang Lin\textsuperscript{1}, Haochen Xue\textsuperscript{2}, 
\textbf{Minghao Guo\textsuperscript{1}}, \textbf{{Yongfeng Zhang}\textsuperscript{1}\thanks{Corresponding author.}}\\
\textsuperscript{1}Rutgers University \space\space\space \textsuperscript{2}University of Liverpool \\
\texttt{\{ziqi.wang0908, xi.zhu, shuhang.lin, minghao.guo, yongfeng.zhang\}@rutgers.edu} \\
\texttt{haochen@liverpool.ac.uk} \\
}

\begin{document}

\ifcolmsubmission
\linenumbers
\fi

\maketitle

\begin{abstract}
Retrieval-augmented generation (RAG) has evolved into a family of paradigms with distinct performance profiles and resource demands, turning paradigm selection into a multi-criteria, context-dependent decision problem. Nevertheless, existing studies largely focus on isolated method improvements or query-only benchmarking, without systematically examining how RAG paradigms behave across diverse query–corpus contexts and effectiveness–efficiency trade-offs.
In this work, we introduce \text{RAGRouter-Bench}, the first dataset and benchmark for adaptive RAG routing. Grounded in query–corpus compatibility, the benchmark integrates three canonical query types, fine-grained corpus indicators capturing structural and semantic properties, and a unified protocol for evaluating both generation quality and resource consumption. Then, we implement standardized RAG paradigms with multiple backbone LLMs across all query–corpus combinations, constructing a comprehensive benchmark with quantitative metrics and LLM-as-a-Judge evaluations to inform context-aware and cost-effective RAG routing decisions.
We further formulate routing as context-dependent paradigm selection and benchmark a range of query–corpus routers on the constructed dataset. Extensive experiments demonstrate that no one-size-fits-all paradigm exists across query–corpus pairs, and that adaptive routing yields more favorable effectiveness–efficiency trade-offs than fixed paradigm selection. These findings establish query–corpus compatibility as a central principle for adaptive RAG routing and position RAGRouter-Bench as a systematic testbed for next-generation RAG systems.
\footnote{\faGithub~\textbf{Code:} \url{https://github.com/ziqiwang0908/RAGRouter-Bench}}
\footnote{\faDatabase~\textbf{Dataset:} \url{https://huggingface.co/datasets/Chaplain0908/RAGRouter}}
\footnote{\faTrophy~\textbf{Leaderboard:} \url{https://huggingface.co/spaces/Chaplain0908/RAGRouter-Leaderboard}}
\end{abstract}

\section{Introduction}
While LLMs are prone to hallucinations in specialized or evolving domains \citep{ji2023survey, mallen2023trust}, retrieval-augmented generation (RAG) has emerged to provide factual and faithful grounding through a two-stage ``retrieve-then-generate'' pipeline \citep{lewis2020rag, guu2020realm, gao2023survey}. 
Recent efforts have demonstrated that retrieval constitutes the primary bottleneck in RAG \citep{jin2024flashrag}, as it not only defines the information boundary \citep{cao2024retrieval} but also dominates the system's computational overhead \citep{jin2024flashrag}, making the strategic selection of RAG paradigms critical.


\begin{figure}[b]
  \centering
  \vspace{-3mm}
  \includegraphics[width=\columnwidth]{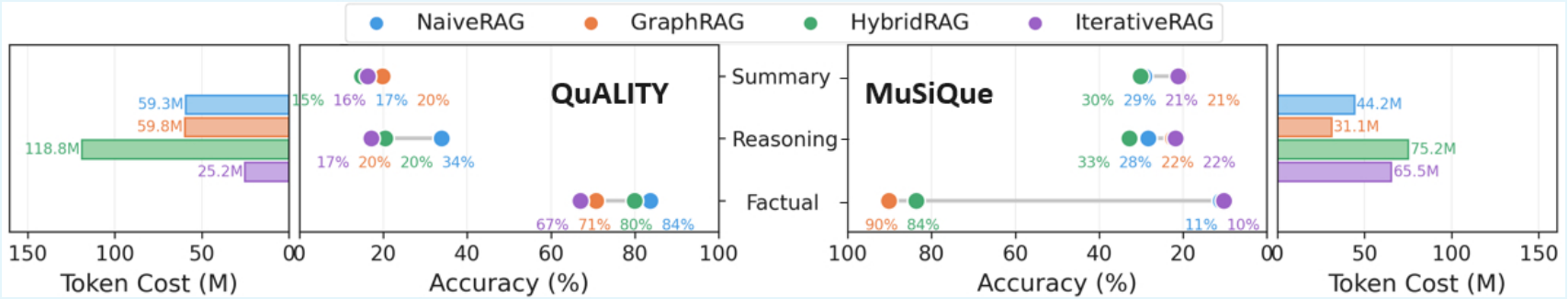}
  \caption{
   Paradigm conflict in accuracy and token consumption across four RAG paradigms.
  }
  \label{fig:motivation}
\end{figure}

Existing RAG paradigms represent an evolution of retrieval strategies \citep{gao2023survey}. 
NaiveRAG relies on similarity-based retrieval for efficient factoid QA and summarization \citep{karpukhin2020dense}, while GraphRAG adopts graph retrieval to enable multi-hop reasoning \citep{edge2024graphrag, gutierrez2024hipporag, he2024gretriever, sun2024think}. 
HybridRAG integrates complementary vector and graph retrievals \citep{sarmah2024hybridrag}, whereas IterativeRAG dynamically invokes retrieval modules to enhance reasoning at the cost of efficiency \citep{asai2024selfrag}. 
Together, their distinct strengths and resource demands turn retrieval into a multi-criteria decision problem, highlighting the necessity of adaptive RAG routing \citep{jeong2024adaptiverag, tang2025mbarag}.


To ground this discussion, we conduct preliminary experiments comparing representative RAG paradigms across different corpora (Figure~\ref{fig:motivation}), yielding three key insights.
(i) No single paradigm consistently dominates across settings, indicating the absence of a one-size-fits-all solution;
(ii) The optimal RAG depends not only on query characteristics but also critically on the underlying corpus;
(iii) Increased methodological sophistication does not necessarily translate into better performance, as simpler alternatives can achieve comparable results with substantially lower overhead.
These findings highlight that adaptive RAG routing hinges on query–corpus compatibility and effectiveness–efficiency trade-offs, calling for systematic benchmarking across queries, corpora, and retrieval strategies.

Nevertheless, existing datasets and benchmarks exhibits several limitations, as summarized in Table \ref{tab:benchmark_comparison}.
First, prior studies often attribute optimal RAG selection solely to query complexity, overlooking the semantic and structural properties of the corpus and, more fundamentally, the role of query–corpus compatibility that governs RAG selection \citep{jeong2024adaptiverag, tang2025mbarag}. Moreover, they fail to provide fine-grained, quantifiable indicators from both query and corpus perspectives, hindering accurate and interpretable decisions \citep{gao2023survey, peng2024graph}. In addition, real-world RAG instances are highly heterogeneous without standardized design protocols, making fair comparison across paradigms inherently difficult.
Overall, current benchmarks suffer from coarse analysis, incomparable designs, and inadequate effectiveness–efficiency trade-offs, which preclude systematic evaluation across query–corpus combinations and ultimately impede the development of adaptive RAG systems \citep{chen2023benchmarking, lyu2024crud, friel2024ragbench, jin2024flashrag}.

\begin{table*}[t]
    \centering
    \vspace{-5mm}
    \caption{
    \textbf{Comparison with existing RAG benchmarks.} Query and Corpus columns indicate whether the benchmark provides multi-type query coverage and corpus-level analysis. Routing indicates whether the benchmark supports adaptive paradigm selection.}
    \small
    \setlength{\tabcolsep}{4pt}
    \renewcommand{\arraystretch}{1.25}
    \resizebox{\textwidth}{!}{%
    \begin{tabular}{l ccc | ccc | cc | cc}
        \toprule
        \multirow{2}{*}{\textbf{Benchmark}} & \multicolumn{3}{c|}{\textbf{Design}} & \multicolumn{3}{c|}{\textbf{Query Type}} & \multicolumn{2}{c|}{\textbf{Corpus Analysis}} & \multicolumn{2}{c}{\textbf{Evaluation}} \\
        \cmidrule(lr){2-4} \cmidrule(lr){5-7} \cmidrule(lr){8-9} \cmidrule(lr){10-11}
         & Query & Corpus & Routing & Factual & Reasoning & Summary & Semantic & Structural & Effectiveness & Efficiency \\
        \midrule
        MultiHop-RAG \citep{tang2024multihop} & \cmark & \xmark & Fixed & \xmark & \cmark & \xmark & \xmark & \xmark & \cmark & \xmark \\
        WebQSP \citep{yih2016value} & \cmark & \xmark & Fixed & \cmark & \cmark & \xmark & \xmark & \xmark & \cmark & \cmark \\
        GraphRAG-Bench \citep{xiang2025whentouse} & \cmark & \xmark & Fixed & \cmark & \cmark & \cmark & \xmark & \cmark & \cmark & \xmark \\
        GraphRAG-Bench \citep{xiao2025graphragbench} & \cmark & \xmark & Fixed & \cmark & \cmark & \cmark & \xmark & \cmark & \cmark & \cmark \\
        MemoRAG \citep{qian2024memorag} & \xmark & \cmark & Fixed & \xmark & \cmark & \xmark & \xmark & \xmark & \cmark & \cmark \\
        \midrule
        \textbf{RAGRouter-Bench (Ours)} & \textbf{\cmark} & \textbf{\cmark} & \textbf{Adaptive} & \textbf{\cmark} & \textbf{\cmark} & \textbf{\cmark} & \textbf{\cmark} & \textbf{\cmark} & \textbf{\cmark} & \textbf{\cmark} \\
        \bottomrule
    \end{tabular}
    }
    \vspace{-5mm}
    \label{tab:benchmark_comparison}
\end{table*}

To this end, we introduce \textbf{RAGRouter-Bench}, the first dataset and benchmark for adaptive RAG routing. Grounded in query–corpus compatibility, RAGRouter-Bench systematically constructs a comprehensive dataset with fine-grained, quantifiable indicators from both query and corpus perspectives, along with effectiveness and efficiency measurements across representative RAG paradigms. Each $(\text{query}, \text{corpus}, \text{paradigm}, \text{performance})$ tuple define a concrete query–corpus combination evaluated under a specific paradigm. Furthermore, we introduce a suite of query–corpus routers to benchmark this dataset, validating the absence of a one-size-fits-all solution and highlighting the role of query–corpus compatibility. Our results show that adaptive routing achieves improved effectiveness–efficiency trade-offs, offering actionable insights toward next-generation RAG systems. Our detailed contributions can be summarized as follows:

$\circ$\; \textbf{Standardized RAG Abstraction.} We formulate RAG paradigms under a unified retriever abstraction, enabling fair comparisons across various real-world RAG applications.
    
$\circ$\;  \textbf{Dual-View Compatibility Framework.} We propose a hierarchical framework to characterize query–corpus compatibility. Specifically, we curate and augment three canonical query types—factual, reasoning, and summarization, while we design multi-dimensional indicators to capture semantic and structural properties, informing how query and corpus, both individually and interactively, influence routing decisions.
    
$\circ$\;  \textbf{Effectiveness-Efficiency Protocol.} We establish a unified evaluation protocol, combining quantitative metrics and LLM-as-a-judge evaluation for response quality, and construction and inference token costs for efficiency.
    
$\circ$\;  \textbf{Systematic Benchmarking.} We implement standardized RAG paradigms (NaiveRAG, GraphRAG, HybridRAG, IterativeRAG) with different backbone LLMs (e.g., LLaMA-3.1-8B, DeepSeek-V3) across diverse datasets, facilitating a comprehensive benchmark that characterizes effectiveness-efficiency trade-offs over query-corpus combinations.
    
$\circ$\; \textbf{Validation and Insights.} We benchmark a range of query–corpus routers on RAGRouter-Bench, validating the practical operability of query-corpus compatibility and demonstrating its potential for next-generation adaptive RAG systems.

\section{Related Work}

\textbf{RAG Benchmarks and Evaluation.}
Despite the growing diversity of RAG paradigms, no existing benchmark is designed to evaluate adaptive routing across paradigms under effectiveness--efficiency trade-offs. Current benchmarks primarily target specific retrieval challenges such as multi-hop reasoning \citep{tang2024multihop, yang2018hotpotqa}, entity-centric QA \citep{yih2016value}, graph-augmented retrieval \citep{xiang2025whentouse, xiao2025graphragbench}, and long-document understanding \citep{qian2024memorag}. Automated evaluation frameworks \citep{es2023ragas, saadfalcon2023ares} and unified toolkits \citep{jin2024flashrag} enable scalable paradigm comparison, while recent work introduces corpus-level diagnostics \citep{facco2017estimating, radovanovic2010hubs} and cost-aware evaluation \citep{lee2025stronger}. However, these efforts evaluate paradigms in isolation under fixed routing without jointly considering cost-effectiveness trade-offs, leaving no systematic platform for benchmarking adaptive RAG routing decisions.

\textbf{Adaptive RAG Routing.}
Adaptive RAG strategies fall into three categories: query-complexity-driven paradigm selection \citep{jeong2024adaptive_rag, autothinkrag2026}, selective retrieval triggering based on model confidence or external signals \citep{jiang2023active_retrieval_augmented_generation, su2024dragin, trivedi2023ircot, yao2024seakr, marina2025llm_independent_adaptive_rag, cheng2024uar}, and multi-source routing across retrieval strategies or knowledge types \citep{wu2025self_routing_rag, bai2025learning_to_route_hybrid_source_rag, li2025lara}, often under cost-aware objectives \citep{wang2024corag, su2025fast_or_better}. A common limitation across all three directions is the reliance on query-side or generation-side signals alone, without explicitly modeling how corpus structural and semantic properties shape paradigm effectiveness. This is the gap that our query--corpus compatibility framework addresses.


\begin{figure*}[t]
  \centering
  \vspace{-6mm}
  \caption{\label{fig:overall}
    \textbf{Overview of the RAGRouter-Bench framework.} \textbf{Left:} Query types with representative examples. \textbf{Center:} Five RAG paradigms as routing targets. \textbf{Right:} Multi-domain corpora with structural and semantic characterization. \textbf{Bottom:} Dual-axis evaluation covering response quality and resource efficiency.
    }
    \vspace{-2mm}
  \includegraphics[width=0.9\textwidth]{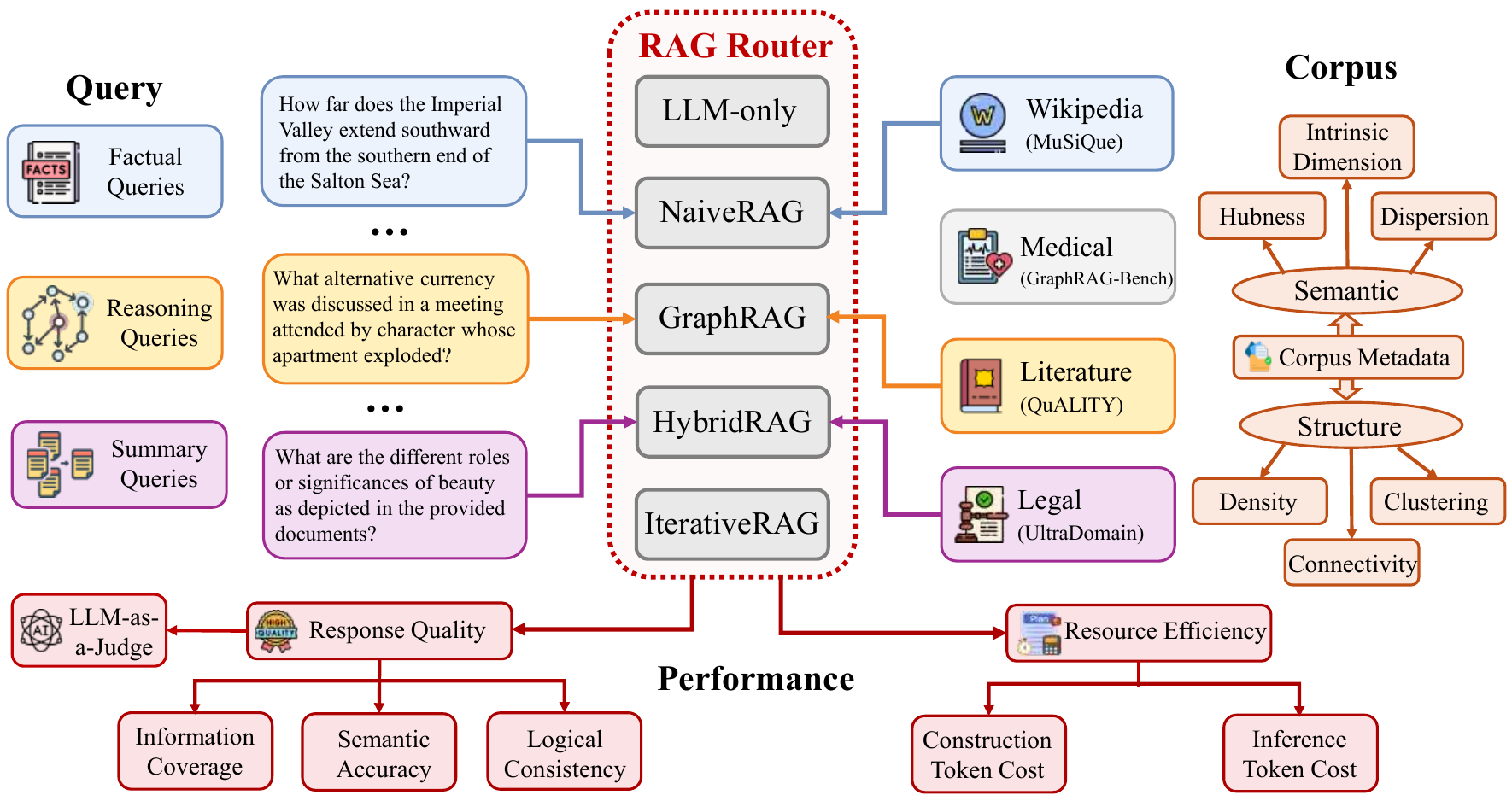} 
  \vspace{-3mm}
\end{figure*}

\section{The RAGRouter-Bench}

We construct \textsc{RAGRouter-Bench} to systematically investigate the joint influence of query and corpus characteristics on RAG paradigm selection, as shown in Figure \ref{fig:overall}. 
Formally, given a query--corpus pair $(q, \mathcal{C})$ and a candidate paradigm space $\Pi$ , our objective is to identify the optimal paradigm $\pi^* = \arg\max_{\pi \in \Pi} \mathcal{U}(\pi; q, \mathcal{C})$, where the utility function $\mathcal{U}$ captures task-specific objectives such as response quality, token consumption, or their trade-offs. 
This formulation frames RAG routing as context-dependent paradigm selection, positioning query-corpus compatibility as the central determinant of adaptive RAG routing.


\subsection{RAG Paradigm Instantiation}

To enable principled cross-paradigm comparison, we revisit real-world RAG pipelines and standardize two retriever primitives that capture common design patterns, from which we instantiate five representative RAG paradigms, as illustrated in Figure~\ref{fig:rag_paradigms} in Appendix~\ref{sec:app_implementation}.

\textbf{Base Retrievers.}
We decompose RAG pipelines into atomic retrievers for fair comparison while preserving real-world fidelity. Specifically, we define two primitives: 
(i) \text{NaiveRetriever}, which performs dense vector retrieval by encoding queries and chunks and selecting top-$K$ segments \citep{karpukhin2020dense}. (ii) \text{GraphRetriever}, which operates over knowledge graphs by extracting seed entities, propagating relevance via Personalized PageRank (PPR), and retrieving text from high-scoring nodes \citep{edge2024graphrag}.

\textbf{RAG Paradigm Instances.} 
Building on these retrievers, we define five paradigms  to accommodate a broad spectrum of RAG designs: 
(i) \text{LLM-only}, a retrieval-free baseline that directly prompts the LLM \citep{petroni2019language};
(ii) \text{NaiveRAG}, which invokes NaiveRetriever once and concatenates retrieved chunks as context \citep{lewis2020rag};
(iii) \text{GraphRAG}, which applies GraphRetriever to extract relevant triplets and enhance generation \citep{edge2024graphrag};
(iv) \text{HybridRAG}, which merges both retrievers via Reciprocal Rank Fusion \citep{cormack2009reciprocal, sarmah2024hybridrag};
(v) \text{IterativeRAG}, which adopts a retrieve–generate–evaluate loop for iterative query decomposition \citep{asai2024selfrag, trivedi2022interleaving}. We focus on these structured pipelines to enable controlled evaluation across paradigms, removing confounding factors from heterogeneous and implementation-specific RAG methods. Implementation details are provided in Appendix~\ref{sec:app_implementation}.

\subsection{Data Curation and Augmentation}
\label{sec:data_curation}

\begin{figure*}[t]
  \vspace{-4mm}
  \centering
  \caption{\label{fig:data_profile}
  \textbf{Query type taxonomy and dataset composition in RAGRouter-Bench.} \textbf{Left:} Three query types definition. \textbf{Right:} Query type distribution across four datasets.}
  \includegraphics[width=0.95\textwidth]{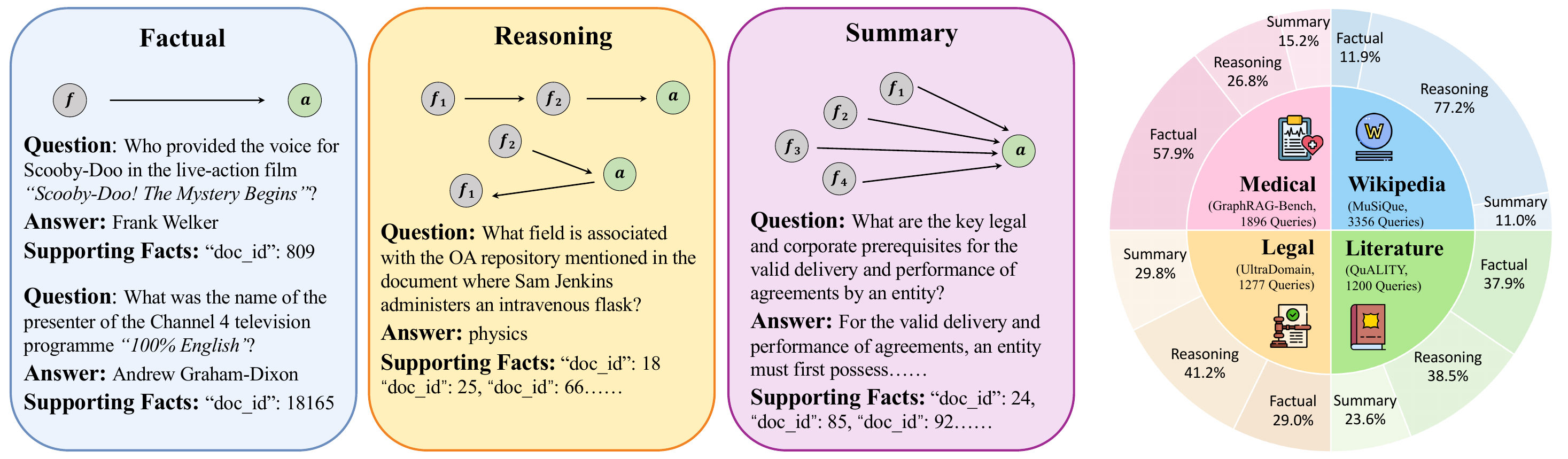} 
  \vspace{-4mm}
\end{figure*}

As shown in Figure~\ref{fig:data_profile}, RAGRouter-Bench is built through systematic corpus composition and query generation to enable comprehensive analysis of query–corpus interactions.


\textbf{Corpus Sourcing.}
We integrate datasets spanning four domains: encyclopedic knowledge from Wikipedia (MuSiQue \citep{trivedi2022musique}), literature (QuALITY \citep{pang2022quality}), legal documentation (UltraDomain\_legal \citep{qian2024memorag}), and medical textbooks (GraphRAGBench\_medical \citep{xiao2025graphragbench}), totaling 21,460 documents.

\textbf{Query Generation.}
We incorporate three canonical query types across all corpora: (i) \text{factual queries} for single-segment entity retrieval; (ii) \text{reasoning queries} for multi-hop inference; and (iii) \text{summary queries} for multi-source information aggregation. However, existing benchmarks exhibit highly skewed query distributions; for example, MuSiQue consists exclusively of reasoning queries, while QuALITY contains over 90\% factual queries \citep{trivedi2022musique, pang2022quality}. To address this issue, we apply \text{query data augmentation} via LLM-based structure-aware expansion \citep{xiao2025graphragbench}, and validate it with a \text{verify-then-filter protocol} \citep{chen2023benchmarking}, where human verification on a stratified sample (N=200) achieves 94\% agreement with automated judgments.
The resulting dataset comprises 7,727 queries, including 4,086 reasoning (52.9\%), 2,320 factual (30.0\%), and 1,321 summary queries (17.1\%), with all corpora covering all query types to enable controlled cross-type comparison. Details are provided in Appendix~\ref{sec:app_preprocess}.


\subsection{Dual-View Analysis}
\label{sec:dual_view_analysis} 

RAG performance should be governed by query–corpus interactions rather than query complexity alone. We have considered three canonical query types for coverage (Section \ref{sec:data_curation}), while the corpus defines the retrieval environment and constrains retrieval feasibility. Accordingly, we characterize corpus properties along two complementary axes.
(i) \text{Structural topology metrics} capture properties that govern graph-based retrieval. Specifically, largest connected component ratio (LCC ratio) measures global connectivity, where low values indicate fragmentation that impedes multi-hop reasoning \citep{newman2010networks}. Average degree and maximum degree centrality reflect connectivity strength, where excessive sparsity limits relational bridging \citep{sun2024think}. Relation type diversity captures edge semantic richness for precise traversal and clustering coefficient measures local cohesiveness that facilitates evidence aggregation \citep{watts1998collective}.
(ii) \text{Semantic space metrics} characterize embedding-space properties that affect vector-based retrieval. In particular, intrinsic dimension measures effective dimensionality, where high values degrade distance-based similarity \citep{facco2017estimating}; dispersion quantifies embedding uniformity, where low values cause semantic crowding and reduce discriminability \citep{wang2020alignment}; and hubness captures retrieval bias, where high values skew results toward frequent but potentially irrelevant passages \citep{radovanovic2010hubs}.
Together, these metrics provide a hierarchical representation of corpus characteristics for routing decisions. Detailed definitions are provided in Appendix~\ref{sec:app_preprocess}.

\subsection{Evaluation Protocol for RAG Paradigms}
\label{sec:evaluation_protocol}
RAGRouter-Bench introduces a unified evaluation protocol for effectiveness–efficiency trade-offs central to RAG routing decisions \citep{jin2024flashrag}.
(i) \text{Generation Quality.} We evaluate generation quality using three complementary metrics: Semantic F1 for token-level accuracy \citep{zhang2019bertscore}, coverage for key information completeness \citep{reimers2019sentence}, Faithfulness for factual grounding in retrieved content \citep{es2023ragas}. In addition, we use LLM-as-a-Judge for holistic correctness via ternary classification \citep{zheng2023judging}.
(ii) \text{Resource Consumption.} We decompose token usage into construction cost, which captures one-time preprocessing overhead (e.g., knowledge graph construction) amortized over queries, and inference cost, which measures per-query overhead including retrieval and generation. Generation input is truncated to 8k tokens to respect LLM context limits. Formal definitions are provided in Appendix~\ref{sec:app_evaluation}.

\subsection{Query-Corpus Router Design}
\label{sec:router_design}
To validate the practical operability of query–corpus compatibility, we formalize paradigm routing as a multi-class classification task over the established dataset. Given a query–corpus pair, the router integrates query-side features with corpus indicators to predict the optimal RAG paradigm. As defined in Section~\ref{sec:dual_view_analysis}, the router encodes two complementary signal sources: query-side features capturing the information need, and corpus-side features characterizing the retrieval environment, including structural and semantic properties. These signals are projected into a shared representation space to enable joint modeling of query–corpus interactions.
Following Section~\ref{sec:evaluation_protocol}, the optimal paradigm for each query–corpus pair can be determined via a utility-driven mapping aligned with practical RAG objectives, where quantifiable evaluation metrics, LLM-as-a-Judge signals, or their combinations serve as references for label assignment. Crucially, when no paradigm yields a satisfactory response, the pair is assigned a dedicated ``\textit{cannot answer}'' label. This allows the router to recognize inherent task boundaries and system limitations, rather than forcing a suboptimal routing decision. Implementation details on feature dimensions, encoders, and training procedures are provided in Section~\ref{sec:router_model} and Appendix~\ref{sec:app_router}.


\begin{table*}[t]
\centering
\vspace{-3mm}
\renewcommand{\arraystretch}{1.25}
\caption{
\textbf{Main evaluation results across RAG paradigms, datasets, and backbone LLMs.} Each paradigm reports LLM-as-a-Judge accuracy (\%) by query type (Factual, Reasoning, Summary) and overall average, along with average token consumption per query. \colorbox{rankfirst}{Green} indicates best performance; \colorbox{ranksecond}{Blue} indicates second best.}
\resizebox{\textwidth}{!}{%
\begin{tabular}{l | ccccc | ccccc | ccccc | ccccc}
\toprule

\multirow{2}{*}{\textbf{Method}} & 
\multicolumn{5}{c|}{\textbf{MuSiQue}} & 
\multicolumn{5}{c|}{\textbf{QuALITY}} & 
\multicolumn{5}{c|}{\textbf{Legal}} & 
\multicolumn{5}{c}{\textbf{Medical}} \\
\cmidrule(lr){2-6} \cmidrule(lr){7-11} \cmidrule(lr){12-16} \cmidrule(lr){17-21}

 & Fac. & Rea. & Sum. & \textbf{Avg.} & Tok. 
 & Fac. & Rea. & Sum. & \textbf{Avg.} & Tok. 
 & Fac. & Rea. & Sum. & \textbf{Avg.} & Tok. 
 & Fac. & Rea. & Sum. & \textbf{Avg.} & Tok. \\
\midrule

\rowcolor{modeldeepseek}
\multicolumn{21}{c}{\textit{\textbf{Model: DeepSeek-V3}}} \\
\midrule
NaiveRAG & 11.1 & \cellcolor{ranksecond}\textbf{28.3} & \cellcolor{ranksecond}\textbf{29.4} & 26.4 & 13k & \cellcolor{rankfirst}\textbf{83.7} & \cellcolor{rankfirst}\textbf{33.8} & \cellcolor{ranksecond}\textbf{17.0} & \cellcolor{rankfirst}\textbf{48.8} & 50k & 54.9 & 11.2 & \cellcolor{rankfirst}\textbf{39.1} & \cellcolor{ranksecond}\textbf{32.2} & 46k & \cellcolor{ranksecond}\textbf{63.1} & 63.5 & 49.1 & 61.1 & 51k \\
GraphRAG & \cellcolor{rankfirst}\textbf{90.2} & 22.5 & 20.6 & \cellcolor{ranksecond}\textbf{30.3} & 9k & 70.7 & \cellcolor{ranksecond}\textbf{20.4} & \cellcolor{rankfirst}\textbf{19.8} & 39.3 & 50k & \cellcolor{ranksecond}\textbf{61.6} & 6.7 & 29.1 & 29.3 & 184k & 55.0 & 56.0 & 43.6 & 53.5 & 38k \\
HybridRAG & \cellcolor{ranksecond}\textbf{83.7} & \cellcolor{rankfirst}\textbf{32.8} & \cellcolor{rankfirst}\textbf{30.2} & \cellcolor{rankfirst}\textbf{38.6} & 22k & \cellcolor{ranksecond}\textbf{80.0} & \cellcolor{ranksecond}\textbf{20.4} & 14.8 & \cellcolor{ranksecond}\textbf{41.6} & 99k & \cellcolor{rankfirst}\textbf{72.2} & \cellcolor{ranksecond}\textbf{11.8} & \cellcolor{ranksecond}\textbf{34.6} & \cellcolor{rankfirst}\textbf{36.1} & 230k & \cellcolor{rankfirst}\textbf{67.8} & \cellcolor{ranksecond}\textbf{64.0} & \cellcolor{ranksecond}\textbf{54.0} & \cellcolor{rankfirst}\textbf{64.7} & 74k \\
IterativeRAG & 10.3 & 21.8 & 21.2 & 20.4 & 20k & 67.0 & 17.1 & 16.2 & 35.8 & 21k & 49.5 & \cellcolor{rankfirst}\textbf{12.4} & 30.4 & 28.5 & 20k & 62.1 & \cellcolor{rankfirst}\textbf{67.8} & \cellcolor{rankfirst}\textbf{56.1} & \cellcolor{ranksecond}\textbf{62.7} & 7k \\

\midrule
\rowcolor{modelllama}
\multicolumn{21}{c}{\textit{\textbf{Model: LLaMA-3.1-8B}}} \\
\midrule
NaiveRAG & 10.3 & 7.9 & \cellcolor{ranksecond}\textbf{12.0} & 8.6 & 13k & \cellcolor{ranksecond}\textbf{69.2} & 10.9 & \cellcolor{ranksecond}\textbf{1.4} & \cellcolor{ranksecond}\textbf{30.7} & 50k & 50.3 & \cellcolor{ranksecond}\textbf{10.7} & \cellcolor{ranksecond}\textbf{22.8} & \cellcolor{ranksecond}\textbf{25.8} & 46k & \cellcolor{ranksecond}\textbf{52.1} & 37.9 & 30.1 & \cellcolor{ranksecond}\textbf{44.9} & 51k \\
GraphRAG & \cellcolor{rankfirst}\textbf{84.4} & \cellcolor{ranksecond}\textbf{9.7} & 11.4 & \cellcolor{ranksecond}\textbf{18.7} & 9k & 44.7 & 9.5 & \cellcolor{rankfirst}\textbf{2.5} & 21.2 & 50k & \cellcolor{ranksecond}\textbf{55.4} & 7.4 & 19.7 & 25.0 & 184k & 48.1 & 33.6 & \cellcolor{ranksecond}\textbf{31.8} & 41.7 & 38k \\
HybridRAG & \cellcolor{ranksecond}\textbf{79.9} & \cellcolor{rankfirst}\textbf{12.1} & \cellcolor{rankfirst}\textbf{13.9} & \cellcolor{rankfirst}\textbf{20.3} & 22k & \cellcolor{rankfirst}\textbf{70.3} & \cellcolor{rankfirst}\textbf{15.6} & \cellcolor{rankfirst}\textbf{2.5} & \cellcolor{rankfirst}\textbf{33.2} & 99k & \cellcolor{rankfirst}\textbf{63.8} & \cellcolor{rankfirst}\textbf{13.1} & \cellcolor{rankfirst}\textbf{24.2} & \cellcolor{rankfirst}\textbf{31.1} & 230k & \cellcolor{rankfirst}\textbf{55.7} & \cellcolor{ranksecond}\textbf{39.9} & \cellcolor{rankfirst}\textbf{33.9} & \cellcolor{rankfirst}\textbf{48.2} & 74k \\
IterativeRAG & 12.3 & 6.0 & 6.8 & 6.8 & 20k & 62.3 & \cellcolor{ranksecond}\textbf{11.5} & \cellcolor{ranksecond}\textbf{1.4} & 28.4 & 21k & 48.7 & 10.1 & 12.6 & 22.0 & 20k & 47.5 & \cellcolor{rankfirst}\textbf{44.6} & 27.3 & 43.6 & 7k \\

\bottomrule
\end{tabular}%
}
\vspace{-3mm}
\label{tab:main_results_full}
\end{table*}

\section{Experiments}


\subsection{Experimental Setup}

\textbf{RAG Paradigms.}
We standardize infrastructure across all paradigms for fair comparison. We use DeepSeek-V3 \citep{deepseek2024v3} and LLaMA-3.1-8B \citep{dubey2024llama} as generators, and text-embedding-3-small for vectorization \citep{openai2024embeddings}, with a unified 8k token context budget. For retrieval, NaiveRAG retrieves top-100 chunks via cosine similarity \citep{karpukhin2020dense}; GraphRAG extracts 20 seed entities and propagates relevance via PPR ($ \alpha = 0.85 $) \citep{haveliwala2002topic} to retrieve top-100 nodes \citep{edge2024graphrag}; HybridRAG combines both retrievers \citep{sarmah2024hybridrag}; IterativeRAG performs up to 3 retrieve-generate-evaluate iterations \citep{trivedi2022interleaving}. Details are provided in Appendix~\ref{sec:app_hyperparams}.

\textbf{Evaluation.} To support fine-grained corpus modeling, we compute structural metrics (LCC ratio, density, clustering coefficient) \citep{newman2010networks, sun2024think, watts1998collective} and semantic metrics (intrinsic dimension, dispersion, hubness) \citep{facco2017estimating, wang2020alignment, radovanovic2010hubs}. We then evaluate all settings following the protocol in Section~\ref{sec:evaluation_protocol}.
For generation quality, we measure semantic F1 \citep{zhang2019bertscore}, coverage, faithfulness \citep{es2023ragas}, and LLM-as-a-Judge accuracy using GPT-4o as the evaluator \citep{achiam2023gpt4}. For resource efficiency, we track token consumption decomposed into retrieval and generation costs \citep{jin2024flashrag}.    These results form the dataset of our RAGRouter-Bench. Implementation details are provided in Appendix~\ref{sec:app_evaluation}.

\textbf{Router Models.}
\label{sec:router_model}
We primarily evaluate two categories of router architectures on the constructed dataset. Traditional ML routers include XGBoost \citep{chen2016xgboost} and MLP \citep{rumelhart1986learning}, which take query type one-hot encodings and 6-dimensional corpus indicators as input.
Neural routers include DNN, TwoTower, Adaptive-RAG \citep{jeong2024adaptive}, MBA-RAG \citep{xu2024mba}, and RouterDC \citep{chen2024routerdc}. We encode query text using BERT \citep{devlin2019bert} and BGE \citep{xiao2024bge}, projecting representations to 64 dimensions. Corpus indicators are discretized into bins and embedded into 64-dimensional vectors, ensuring balanced representation across query and corpus signals.
Additionally, we include three naive baselines to establish lower and upper bounds: Random uniformly samples from all classes; Best-fixed always selects the globally most frequent optimal paradigm; and Oracle selects the ground-truth optimal paradigm for each query.
For label construction, we first identify paradigms judged as correct by LLM-as-a-Judge, and then select the one with the highest semantic F1 as the label. All routers are trained and evaluated using 5-fold stratified cross-validation, with average results reported.
We report three metrics to assess routing performance: routing accuracy measures the rate of selecting the optimal paradigm, macro F1 captures balanced performance across classes, and correct rate measures the LLM-as-a-Judge correctness of the selected paradigm.

\subsection{Results and Analysis}

\begin{table*}[t]
\centering
\vspace{-4mm}
\renewcommand{\arraystretch}{1.25}
\caption{\textbf{Router evaluation under joint query-corpus features.} R.A: routing accuracy (\%), F1: macro F1 (\%), C.R: correct rate (\%). Neural routers are evaluated with BERT and BGE encoders. \colorbox{rankfirst}{Green}: best per column; \colorbox{ranksecond}{Blue}: second best.}
\resizebox{\textwidth}{!}{%
\begin{tabular}{l | ccc | ccc | ccc | ccc | ccc}
\toprule
\multirow{2}{*}{\textbf{Method}} & 
\multicolumn{3}{c|}{\textbf{Avg}} & \multicolumn{3}{c|}{\textbf{MuSiQue}} & \multicolumn{3}{c|}{\textbf{QuALITY}} & \multicolumn{3}{c|}{\textbf{Legal}} & \multicolumn{3}{c}{\textbf{Medical}} \\
\cmidrule(lr){2-4} \cmidrule(lr){5-7} \cmidrule(lr){8-10} \cmidrule(lr){11-13} \cmidrule(lr){14-16}
 & R.A & F1 & C.R & R.A & F1 & C.R & R.A & F1 & C.R & R.A & F1 & C.R & R.A & F1 & C.R \\
\midrule

\rowcolor{modeldeepseek}
\multicolumn{16}{c}{\textit{\textbf{Model: DeepSeek-V3}}} \\
\midrule
\rowcolor{gray!10} Random & 20.06 & 18.90 & 39.06 & 20.04 & -- & 28.40 & 19.96 & -- & 42.84 & 20.05 & -- & 31.61 & 20.16 & -- & 60.58 \\
\rowcolor{gray!10} Best-fixed & 20.91 & 6.92 & 39.32 & 15.46 & -- & 26.37 & 30.05 & -- & 48.75 & 18.72 & -- & 32.18 & 26.27 & -- & 61.08 \\
\midrule
MLP & \cellcolor{rankfirst}46.78 & 19.92 & 42.93 & \cellcolor{rankfirst}54.82 & 23.89 & 35.31 & \cellcolor{rankfirst}54.61 & 24.68 & 48.52 & 51.38 & 19.72 & 31.89 & 24.51 & 10.02 & 60.31 \\
XGBoost & 45.76 & 20.58 & 42.93 & 52.78 & 24.08 & 35.42 & 51.45 & 24.82 & 48.73 & 52.76 & 21.95 & 32.01 & 25.04 & 10.78 & 59.92 \\
\midrule
TwoTower (BGE) & 43.29 & 25.41 & 42.28 & 49.71 & 29.02 & 32.33 & 45.97 & 23.48 & 48.30 & 52.49 & 25.58 & 33.19 & 24.02 & 20.12 & 62.22 \\
TwoTower (BERT) & 44.95 & 25.42 & 42.81 & 51.44 & 28.26 & 33.67 & 49.13 & 24.49 & 48.21 & \cellcolor{rankfirst}53.28 & 26.10 & 33.57 & 25.21 & 20.51 & 61.80 \\
Adap-RAG (BGE) & 43.93 & 26.24 & 42.43 & 49.98 & 29.55 & 33.22 & 47.55 & \cellcolor{ranksecond}25.04 & \cellcolor{ranksecond}48.88 & 51.79 & 25.10 & 32.64 & 25.65 & 21.92 & 61.24 \\
Adap-RAG (BERT) & \cellcolor{ranksecond}46.19 & 23.20 & 42.87 & 53.01 & 26.68 & 34.27 & \cellcolor{ranksecond}51.90 & 24.75 & \cellcolor{rankfirst}48.99 & 53.07 & 25.03 & 33.36 & 25.87 & 14.81 & 60.64 \\
MBA-RAG (BGE) & 43.30 & 25.04 & 41.61 & 48.48 & 26.67 & 30.73 & 47.65 & 23.59 & 48.72 & 50.06 & \cellcolor{rankfirst}27.51 & 32.98 & \cellcolor{rankfirst}26.82 & 21.40 & 62.17 \\
MBA-RAG (BERT) & 45.17 & 21.48 & 42.66 & 51.49 & 25.15 & 33.72 & 50.23 & 23.43 & 48.83 & 52.51 & 22.58 & 32.62 & 25.85 & 13.02 & 61.33 \\
DNN (BGE) & 43.92 & \cellcolor{rankfirst}29.76 & 43.37 & 50.95 & \cellcolor{rankfirst}35.15 & 35.01 & 44.77 & 24.91 & 46.92 & 51.03 & \cellcolor{rankfirst}27.51 & 32.99 & \cellcolor{ranksecond}26.15 & \cellcolor{rankfirst}24.78 & \cellcolor{rankfirst}62.90 \\
DNN (BERT) & 45.48 & 27.77 & \cellcolor{ranksecond}43.40 & \cellcolor{ranksecond}53.42 & 33.61 & \cellcolor{ranksecond}35.49 & 48.28 & 24.77 & 47.95 & 52.83 & 26.64 & \cellcolor{rankfirst}33.91 & 24.71 & 20.09 & 60.93 \\
RouterDC (BGE) & 43.84 & \cellcolor{ranksecond}28.82 & 42.84 & 50.20 & 34.56 & 34.18 & 44.77 & 23.80 & 47.08 & 52.91 & \cellcolor{ranksecond}27.49 & 32.70 & 25.87 & \cellcolor{ranksecond}22.72 & \cellcolor{ranksecond}62.32 \\
RouterDC (BERT) & 45.97 & 28.80 & \cellcolor{rankfirst}43.69 & 52.64 & \cellcolor{ranksecond}35.01 & \cellcolor{rankfirst}35.55 & 51.62 & \cellcolor{rankfirst}27.88 & 48.57 & \cellcolor{ranksecond}53.22 & 26.82 & \cellcolor{ranksecond}33.79 & 25.71 & 19.71 & 61.69 \\
\midrule
\rowcolor{gray!10} Oracle & 100.00 & 100.00 & 60.83 & 100.00 & -- & 51.61 & 100.00 & -- & 59.93 & 100.00 & -- & 50.51 & 100.00 & -- & 84.65 \\

\midrule
\rowcolor{modelllama}
\multicolumn{16}{c}{\textit{\textbf{Model: LLaMA-3.1-8B}}} \\
\midrule
\rowcolor{gray!10} Random & 24.95 & 21.87 & 26.57 & 24.92 & -- & 16.62 & 25.18 & -- & 26.58 & 24.94 & -- & 26.65 & 24.86 & -- & 44.13 \\
\rowcolor{gray!10} Best-fixed & 14.42 & 6.30 & 25.78 & 12.81 & -- & 18.71 & 9.35 & -- & 21.20 & 13.47 & -- & 24.98 & 21.10 & -- & 41.72 \\
\midrule
MLP & 62.63 & 30.15 & 27.07 & \cellcolor{ranksecond}78.32 & 35.74 & 18.41 & 66.89 & 31.98 & 28.21 & 58.55 & 26.18 & 24.01 & 34.92 & 21.78 & 43.72 \\
XGBoost & 62.45 & 29.03 & 26.72 & \cellcolor{rankfirst}78.48 & 35.91 & 18.52 & \cellcolor{ranksecond}67.12 & 32.41 & \cellcolor{ranksecond}28.38 & 58.98 & 27.43 & 23.89 & 33.45 & 15.81 & 42.08 \\
\midrule
TwoTower (BGE) & 59.91 & 29.91 & 27.06 & 73.23 & 30.72 & 18.61 & 62.83 & 35.15 & 26.22 & 61.28 & 32.51 & 25.10 & 33.57 & 23.43 & \cellcolor{ranksecond}43.88 \\
TwoTower (BERT) & 61.74 & 31.16 & 26.66 & 74.96 & 33.65 & 18.28 & 63.76 & 34.32 & 27.49 & \cellcolor{ranksecond}61.95 & 32.41 & 24.73 & 36.92 & 23.93 & 42.26 \\
Adap-RAG (BGE) & 60.25 & 30.40 & 26.88 & 73.56 & 31.32 & 18.67 & 62.53 & 31.43 & 26.06 & 61.11 & 34.23 & 25.14 & 34.69 & 25.56 & 43.11 \\
Adap-RAG (BERT) & \cellcolor{ranksecond}62.66 & 30.75 & 27.22 & 75.39 & 32.27 & 18.73 & \cellcolor{rankfirst}67.83 & 34.75 & \cellcolor{rankfirst}28.83 & 61.49 & 34.14 & 25.02 & \cellcolor{ranksecond}37.64 & 23.24 & 42.73 \\
MBA-RAG (BGE) & 59.18 & 29.16 & 27.14 & 72.76 & 29.74 & 18.50 & 62.48 & 31.81 & 27.24 & 58.48 & 32.31 & \cellcolor{ranksecond}25.66 & 33.53 & 24.32 & 43.37 \\
MBA-RAG (BERT) & \cellcolor{rankfirst}62.85 & 30.49 & 27.15 & 75.77 & 31.36 & 18.73 & 66.87 & 33.79 & 27.72 & \cellcolor{rankfirst}62.44 & 35.07 & 25.46 & \cellcolor{rankfirst}37.73 & 23.80 & 42.83 \\
DNN (BGE) & 58.33 & \cellcolor{ranksecond}34.61 & \cellcolor{rankfirst}27.40 & 72.14 & 37.50 & 18.70 & 60.43 & 35.41 & 27.56 & 58.35 & \cellcolor{ranksecond}35.26 & 25.09 & 32.53 & \cellcolor{rankfirst}28.54 & \cellcolor{rankfirst}44.27 \\
DNN (BERT) & 61.42 & \cellcolor{rankfirst}34.65 & \cellcolor{ranksecond}27.38 & 74.67 & \cellcolor{ranksecond}38.06 & 18.79 & 65.23 & \cellcolor{ranksecond}35.93 & 27.94 & 61.80 & \cellcolor{rankfirst}35.68 & \cellcolor{rankfirst}25.68 & 35.30 & \cellcolor{ranksecond}27.10 & 43.37 \\
RouterDC (BGE) & 60.01 & 33.81 & 27.09 & 73.98 & \cellcolor{rankfirst}38.20 & \cellcolor{rankfirst}18.92 & 60.49 & 32.91 & 26.65 & 59.49 & 34.97 & 25.62 & 35.32 & 25.81 & 42.83 \\
RouterDC (BERT) & 61.30 & 34.52 & 27.20 & 73.77 & 37.51 & \cellcolor{ranksecond}18.86 & 65.56 & \cellcolor{rankfirst}38.68 & 28.03 & 61.37 & 34.81 & 25.31 & 36.48 & 26.42 & 42.73 \\
\midrule
\rowcolor{gray!10} Oracle & 100.00 & 100.00 & 40.57 & 100.00 & -- & 26.76 & 100.00 & -- & 39.73 & 100.00 & -- & 41.19 & 100.00 & -- & 65.14 \\
\bottomrule
\end{tabular}%
}
\label{tab:router_result}
\vspace{-5mm}
\end{table*}

\begin{figure}[t] 
\vspace{-3mm}
  \centering
    \caption{
    \textbf{Paradigm performance across datasets and query types.} Each panel shows one RAG paradigm's LLM-as-a-Judge accuracy (correct\%), with rows as query types and columns as datasets. Asterisk (*) marks the best-performing paradigm for each combination.
  }
  \includegraphics[width=\columnwidth]{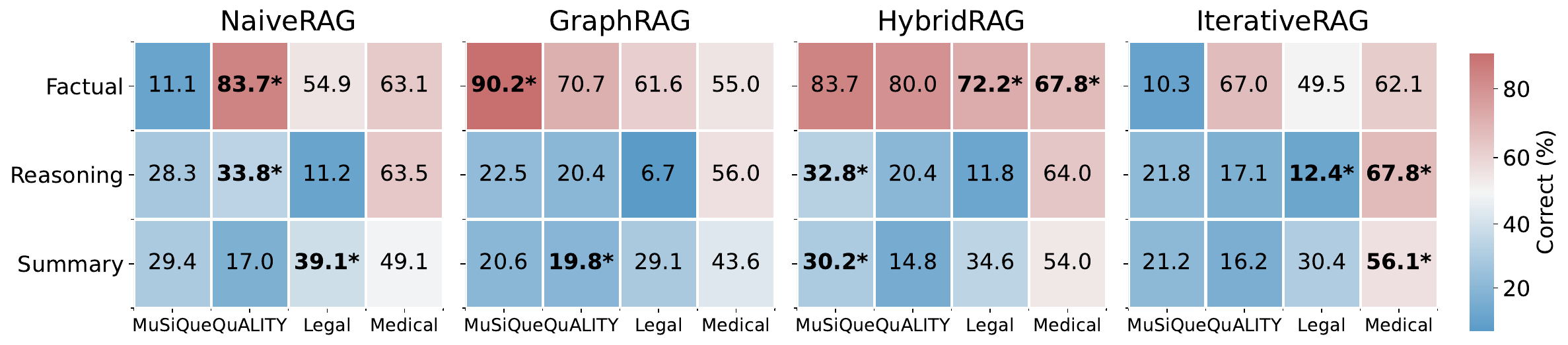}
  \label{fig:rq1_heatmap}
  \vspace{-3mm}
\end{figure}

\textbf{Comparative Paradigm Analysis.}
Table~\ref{tab:main_results_full} and Figure~\ref{fig:rq1_heatmap} summarize performance across datasets and query types, yielding these insights: (i) No paradigm is universally optimal across corpora. Performance varies with corpus environment: \text{GraphRAG} achieves 90.2\% on MuSiQue factual queries but drops to 70.7\% on QuALITY, while \text{NaiveRAG} excels on QuALITY (83.7\%) but fails on MuSiQue (11.1\%). These reversals align with corpus structure: MuSiQue’s high reachability favors graph traversal, whereas QuALITY’s semantic crowding degrades vector retrieval.
(ii) Paradigm effectiveness depends on query type. EWithin the same corpus, the optimal paradigm shifts with task complexity: on MuSiQue, from \text{GraphRAG} for factual (90.2\%) to \text{HybridRAG} for reasoning (32.8\%) and summarization (30.2\%); in the Medical domain, \text{IterativeRAG} dominates reasoning (67.8\%), while \text{HybridRAG} remains best for factual extraction (67.8\%). (iii) Together, these results show that RAG paradigm selection is inherently a multi-criteria decision problem. No single paradigm can accommodate the diverse query–corpus landscape, motivating adaptive routing grounded in both query characteristics and corpus properties.

\textbf{Router Benchmarking Analysis.}
Table~\ref{tab:router_result} presents the performance of all routers under joint query–corpus features with following insights:
(i) Neural routers outperform traditional ML in balanced performance. Neural routers consistently achieve higher F1 and correct rate, while traditional ML methods remain competitive in routing accuracy but exhibit substantially lower F1, indicating strong majority-class bias. Within the same architecture, BERT-based encodings enable more balanced paradigm selection than BGE embeddings.
(ii) Stronger LLMs make routing harder but improve outcome quality. LLaMA-3.1-8B achieves higher routing accuracy than DeepSeek-V3, yet lower correct rate. This counterintuitive result arises because weaker generation tends to concentrate labels toward ``cannot answer'', simplifying the classification task but limiting downstream correctness. In contrast, stronger models produce more diverse label distributions, making routing more challenging but yielding higher correct rate when correct paradigms are selected. This suggests that routing difficulty is intrinsically coupled with generation quality, rather than purely a modeling issue.
(iii) Routing provides measurable gains but remains far from optimal. The best router reaches 43.69\% correct rate, outperforming Random (39.06\%) and Best-fixed (39.32\%), confirming the value of adaptive routing. However, a substantial gap remains to Oracle (60.83\%), indicating that current models are insufficient to fully capture query–corpus compatibility. This also highlights that adaptive RAG routing remains a largely open problem, with significant room for improvement in router modeling.


\begin{figure*}[t]
  \centering
  \caption{\textbf{Routing performance under different input signals across datasets.} Average R.A (routing accuracy) and C.R (correct rate, \%) across all nine routers for query (Q), corpus (C) feature configuration, with Random and Oracle as lower and upper bounds.}
  \includegraphics[width=0.95\textwidth]{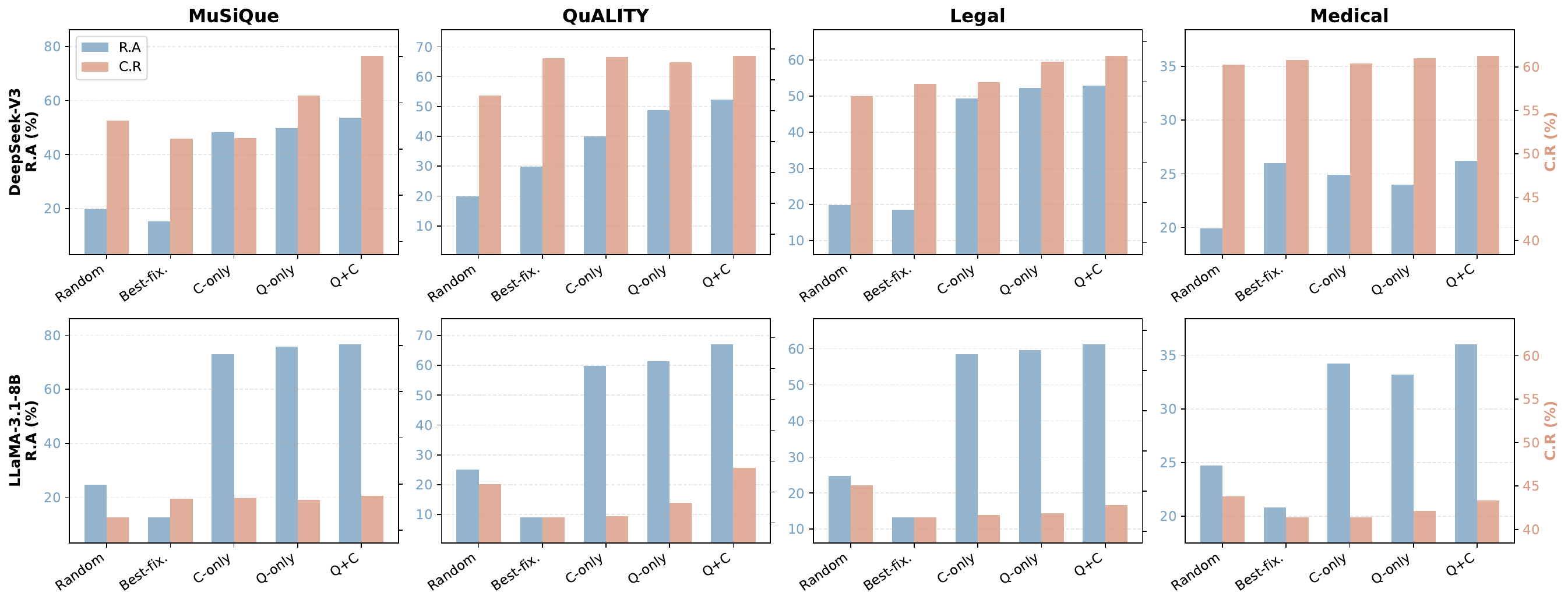}
  \label{fig:router_bar}
  \vspace{-3mm}
\end{figure*}

\begin{wrapfigure}{r}{0.5\textwidth}
  \centering
  \caption{
    \textbf{Corpus features across structural and semantic dimensions.} \textbf{Left:} Graph topology metrics capturing knowledge graph properties. \textbf{Right:} Embedding space metrics characterizing semantic distribution.
  }
  \includegraphics[width=0.5\columnwidth]{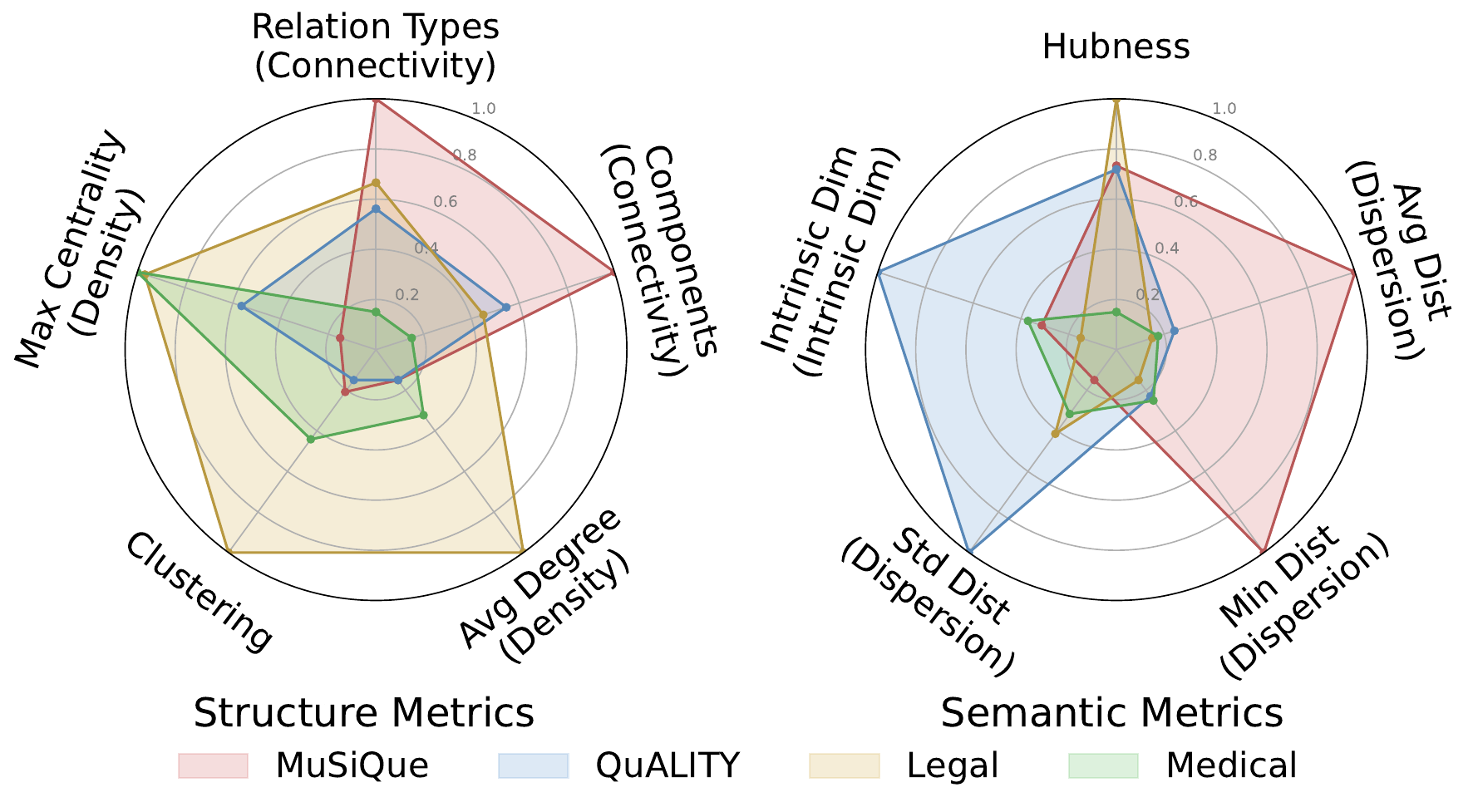}
  \label{fig:corpus_fingerprints}
  \vspace{-3mm}
\end{wrapfigure}
\textbf{Query-Corpus Compatibility Analysis.}
Figure~\ref{fig:corpus_fingerprints} illustrates the distinct structural and semantic profiles across domains, while Figure~\ref{fig:router_bar} presents routing performance across different feature configurations, from which our observations include: 
(i) Joint query–corpus features are necessary for effective routing. The joint configuration consistently outperforms query-only and corpus-only across all dataset–model settings in both routing accuracy and correct rate. 
For example, on MuSiQue, routing accuracy improves from 48.4\% to 55.3\% and correct rate from 26.4\% to 35.8\%, demonstrating the complementary nature of query and corpus signals.
(ii) Single-sided signals are fundamentally insufficient. Corpus-only features yield nearly identical performance across routers due to their dataset-level invariance, essentially collapsing routing into a fixed per-dataset mapping. While query-only features (Q-only) provide more granular signals, their effectiveness is inconsistent across diverse domains, further justifying the necessity of a unified dual-view representation.
(iii) Corpus structure and semantics shape paradigm effectiveness. Structural and semantic indicators explain systematic performance differences across corpora. For instance, The Legal corpus exhibits a dense graph structure (high connectivity) but a congested embedding space (high hubness), explaining why HybridRAG (72.2\%) significantly outperforms NaiveRAG (54.9\%). Conversely, QuALITY's fragmented structural profile (low connectivity) hinders graph-based methods, allowing NaiveRAG (83.7\%) to dominate via direct semantic matching. These fine-grained metrics provide the computable signals necessary for context-aware routing decisions.

\begin{figure*}[t]
  \centering
  \vspace{-2mm}
  \caption{\textbf{Cost-effectiveness trade-off across RAG paradigms and router (DeepSeek-V3).} Each point represents a strategy on one dataset, with token cost per query on the x-axis and LLM-as-a-Judge correct rate on the y-axis. Router (Q+C) achieves comparable correct rate to the best fixed paradigm at substantially lower token cost across most datasets.}
  \includegraphics[width=\textwidth]{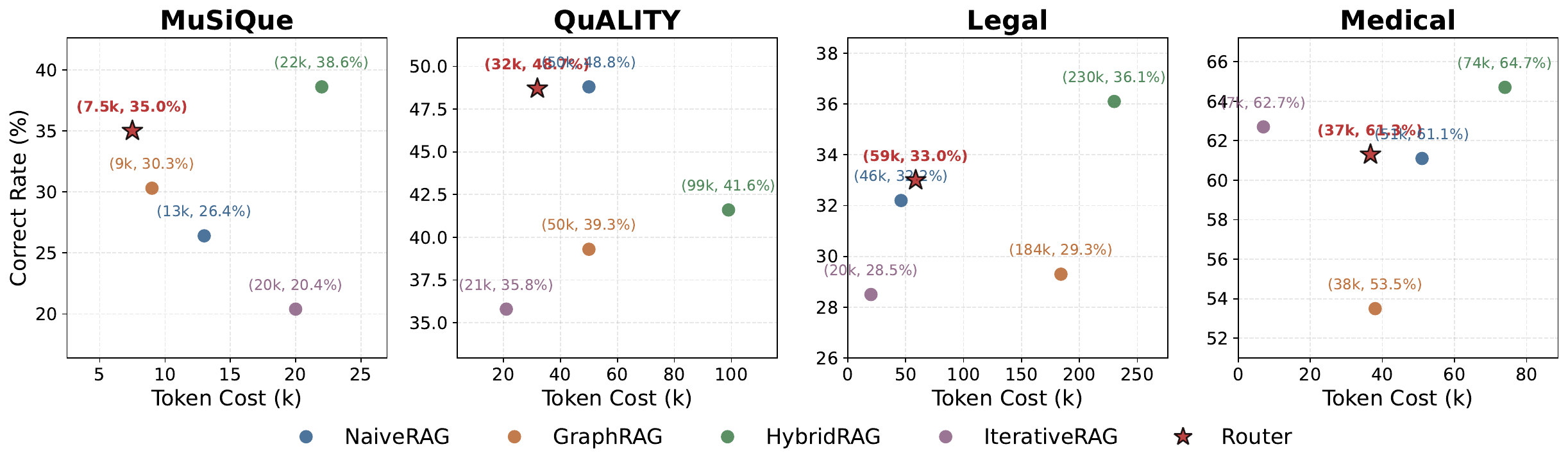}
  \label{fig:router_scatter}
  \vspace{-3mm}
\end{figure*}

\textbf{Cost-Performance Trade-off Analysis.}
Table~\ref{tab:main_results_full} and Figure~\ref{fig:router_scatter} present the cost–effectiveness landscape across RAG paradigms and the router. Specifically, (i) No fixed paradigm is Pareto-optimal across datasets. While HybridRAG delivers the highest correct rate on most datasets, it incurs a prohibitive computational burden (e.g., 230k tokens per query on Legal). In contrast, lower-cost paradigms may reduce token usage but incur accuracy degradation. As shown in Figure~\ref{fig:router_scatter}, no single fixed strategy simultaneously occupies the upper-left corner across all four datasets, confirming that optimal RAG selection is a multi-criteria decision governed by specific deployment constraints.
(ii) The router achieves near-optimal effectiveness at significantly lower cost. For example, on MuSiQue, the router attains 35.0\% correct rate at 7.5k tokens, compared to HybridRAG’s 38.6\% at 22k (66\% cost reduction). On Legal, it reaches 33.0\% at 58.6k versus 36.1\% at 230k (75\% reduction). Across datasets, router solutions consistently lie closer to the Pareto frontier, indicating improved cost–effectiveness trade-offs.
(iii) Cost variation is substantial and fundamentally shapes paradigm selection. Token consumption spans over an order of magnitude across paradigms, making deployment constraints a primary factor in selecting RAG strategies. RAGRouter-Bench explicitly captures this trade-off by jointly modeling effectiveness and efficiency, enabling cost-aware routing decisions beyond accuracy-only optimization.

\section{Conclusion}

We introduce RAGRouter-Bench, the first benchmark for adaptive RAG routing grounded in query–corpus compatibility. By unifying multi-domain datasets, canonical query types, fine-grained structural and semantic corpus indicators, and a dual-view evaluation protocol, it provides a principled testbed for analyzing effectiveness–efficiency trade-offs across RAG paradigms.
Our results show that (i) no single paradigm is universally optimal across query–corpus settings; (ii) paradigm effectiveness is jointly determined by query characteristics and corpus properties; and (iii) adaptive routing consistently achieves better cost–effectiveness trade-offs than fixed strategies. These findings position RAG paradigm selection as a context-dependent decision problem rather than a static pipeline choice.
Despite these advances, a significant gap remains between current routers and the oracle upper bound, indicating substantial room for improving query–corpus representations and routing mechanisms. 
We hope RAGRouter-Bench will serve as a foundation for future research on adaptive, interpretable, and cost-aware RAG systems, and further establish query–corpus compatibility as a core principle for next-generation RAG. Limitations of this work are discussed in Appendix~\ref{sec:limitations}.

\bibliography{colm2026_conference}
\bibliographystyle{colm2026_conference}

\appendix
\appendix

\clearpage
\vspace{1em}
\begin{center}
{\LARGE \textbf{Appendix}}
\end{center}
\vspace{1em}



\section{Data Construction \& Corpus Analysis}
\label{sec:app_preprocess}

\subsection{Corpus Statistics \& Preprocessing.}
\label{sec:app_corpus}

\paragraph{Data Overview.}
To establish a benchmark encompassing diverse retrieval environments, we integrate four representative datasets spanning encyclopedic knowledge (MuSiQue, 21,100 Wikipedia articles), long-form narratives (QuALITY, 265 Gutenberg novels), specialized legal corpora (UltraDomain\_Legal, 94 contract documents), and medical literature (GraphRAGBench\_Medical, a single comprehensive textbook). As shown in Table \ref{tab:raw_stats}, these datasets exhibit extreme disparities in both document count (ranging from 1 to 21,100) and average length (107.9 to 221,495 tokens), thereby serving as an ideal testbed for evaluating the adaptability of RAG paradigms across distinct scales and structural settings. The complete preprocessing pipeline is formalized in Algorithm~\ref{alg:corpus_preprocess}.

\begin{table}[H]
\centering
\small
\renewcommand{\arraystretch}{1.2}
\setlength{\tabcolsep}{8pt}
\begin{tabular}{l l c c c}
\toprule
\textbf{Dataset} & \textbf{Domain} & \textbf{Num. Docs} & \textbf{Avg. Tokens} & \textbf{Total Tokens} \\
\midrule
MuSiQue & Wikipedia & 21,100 & 107.9 & 2,276,013 \\
QuALITY & Narrative & 265    & 5,741.1 & 1,521,395 \\
UltraDomain\_legal   & Legal & 94     & 50,785.0 & 4,773,793 \\
GraphRAGBench\_medical & Medical Textbook & 1      & 221,495.0 & 221,495 \\
\bottomrule
\end{tabular}
\caption{\textbf{Raw Corpus Statistics.} Overview of the source documents illustrating extreme variations in document length (Avg. Tokens) and scale (Num. Docs), ranging from fragmented Wikipedia articles to monolithic textbooks.}
\label{tab:raw_stats}
\end{table}

\paragraph{Chunking Strategy.}
We employ a sliding window chunking strategy to process the raw corpora (see Table~\ref{tab:preprocessing_params} for all hyperparameters). Specifically, utilizing the \texttt{cl100k\_base} encoder from \texttt{tiktoken}, each document (concatenated title and content) is segmented into fixed-size chunks with a window size of 512 tokens and a 100-token overlap to preserve contextual coherence. This configuration strikes a balance between retrieval granularity and contextual completeness. Following this segmentation, the four datasets yield distinct chunk inventories: 21,153 (MuSiQue), 3,822 (QuALITY), 11,632 (Legal), and 538 (Medical).

\begin{table}[htbp]
    \centering
    \small
    \setlength{\tabcolsep}{15pt}
    \renewcommand{\arraystretch}{1.1}
    \begin{tabular}{l l l l}
    \toprule
    \textbf{Stage} & \textbf{Parameter} & \textbf{Value} & \textbf{Description} \\
    \midrule
    \multirow{3}{*}{Chunking} 
      & Size & 512 & Fixed-size segment \\
      & Overlap & 100 & Sliding window \\
      & Tokenizer & cl100k\_base & OpenAI encoding \\
    \midrule
    \multirow{5}{*}{Extraction} 
      & LLM & DeepSeek-V3 & Base Model \\
      & Temp. & 0.0 & Deterministic \\
      & Concur. & 15 & Parallel requests \\
      & Timeout & 60s & API limit \\
      & Retries & 3 & Fault tolerance \\
    \midrule
    Graph & Directed & False & Undirected edges \\
    \midrule
    \multirow{3}{*}{Embedding} 
      & Model & text-emb-3-small & OpenAI Model \\ 
      & Dim. & 1536 & Vector size \\
      & Batch & 30 & API batch size \\
    \bottomrule
    \end{tabular}
    \caption{Hyperparameters for Corpus Preprocessing.}
    \label{tab:preprocessing_params}
\end{table}

\begin{figure}[H] 
    \centering
    \includegraphics[width=0.95\textwidth]{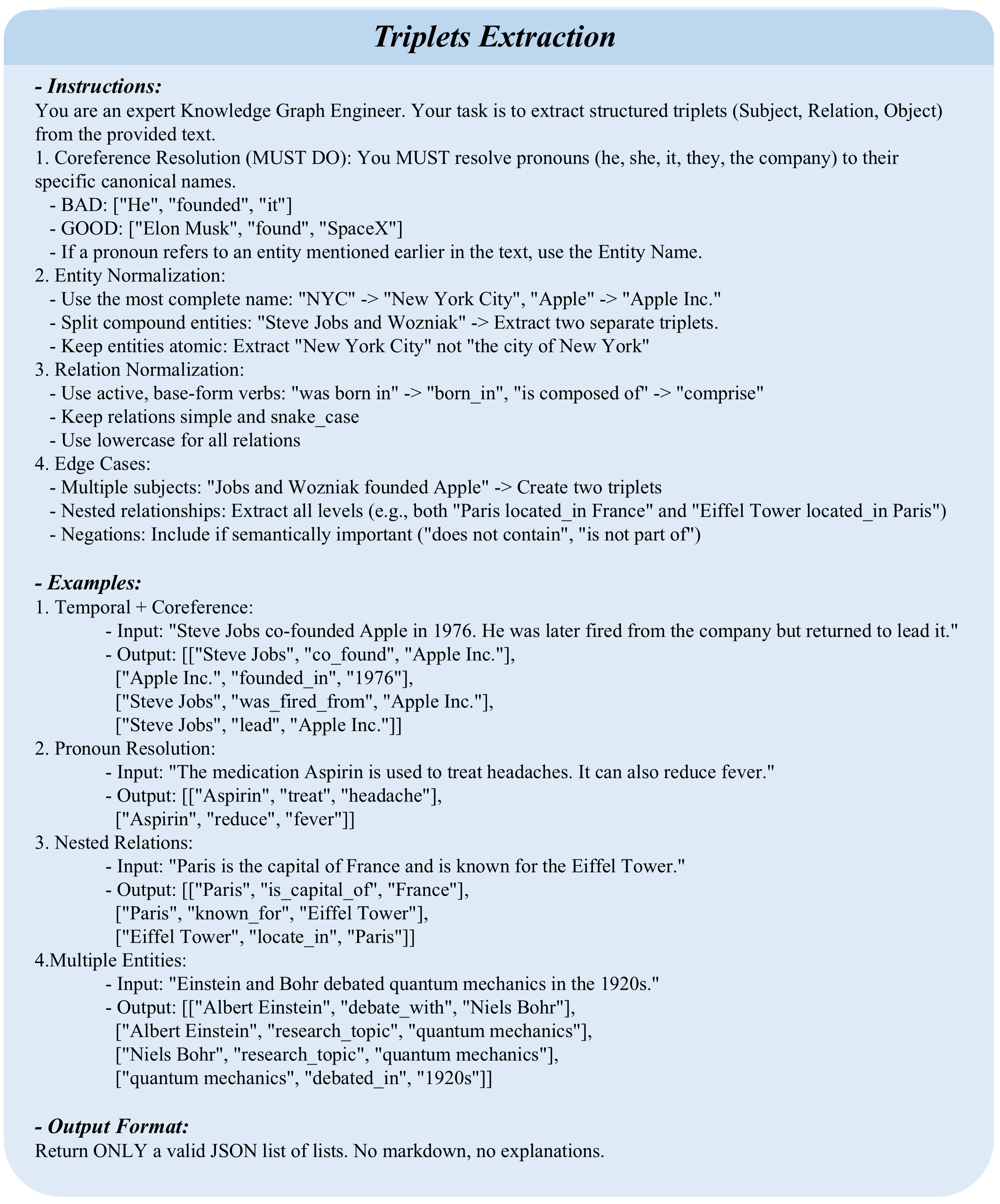}
    \caption{The prompt template for extracting structured knowledge triplets from text chunks.}
    \label{fig:triplets_extraction}
\end{figure}

\paragraph{Knowledge Graph Construction.}
To facilitate the structure-aware retrieval required by GraphRAG and HybridRAG, we extract knowledge triplets from each individual text chunk. We employ DeepSeek-V3 as the underlying extraction engine, with hyperparameters detailed in Table~\ref{tab:preprocessing_params}, setting the temperature to 0.0 to ensure deterministic generation. To optimize processing throughput, we implement an asynchronous parallelization strategy configured with a maximum concurrency of 15, a 60-second request timeout, and a retry mechanism allowing up to 3 attempts with exponential backoff. As illustrated in Figure~\ref{fig:triplets_extraction}, the extraction prompt instructs the model to identify \texttt{(Subject, Relation, Object)} triplets within the input text and format the output as a JSON array. The scale of the resulting knowledge graphs is detailed in Table~\ref{tab:struct_full}. Specifically, MuSiQue comprises 206,738 entity nodes and 276,898 edges; QuALITY contains 90,088 nodes and 120,611 edges; the Legal corpus yields 135,231 nodes and 261,207 edges; and the Medical textbook generates 14,712 nodes and 21,480 edges. Graph density varies significantly from $6.0 \times 10^{-6}$ (MuSiQue) to $9.9 \times 10^{-5}$ (Medical), reflecting inherent disparities in structural sparsity across domains. All constructed graphs are represented as undirected graphs to facilitate bidirectional traversal. 

\begin{algorithm}[H]
\caption{Unified Corpus Preprocessing Pipeline.}
\small
\label{alg:corpus_preprocess}
\begin{algorithmic}[1]
\Require Raw Corpus $\mathcal{C} = \{d_1, d_2, ..., d_n\}$, where each $d_i = (\text{title}, \text{text})$
\Ensure Knowledge Graph $\mathcal{G}=(V, E)$, Dense Vector Index $\mathcal{I}_{\text{vec}}$, Entity Embeddings $\mathbf{E}_{\text{ent}}$
\Statex \textbf{Hyperparameters:} Chunk Size $L=512$, Overlap $O=100$, Embedding Model $\mathcal{M}_{\text{emb}}$, LLM $\mathcal{M}_{\text{gen}}$
\Statex \textcolor{gray}{\textit{// Phase 1: Sliding Window Chunking}}
\State $\mathcal{S}_{\text{chunks}} \gets \emptyset$
\For{each document $d \in \mathcal{C}$}
    \State $\text{Tokens} \gets \text{Tokenize}(d.\text{title} \oplus d.\text{text})$ \Comment{Using \texttt{tiktoken} (cl100k\_base)}
    \State $ptr \gets 0$
    \While{$ptr < \text{len}(\text{Tokens})$}
        \State $c_{\text{raw}} \gets \text{Tokens}[ptr : ptr + L]$
        \State $c_{\text{text}} \gets \text{Decode}(c_{\text{raw}})$
        \State $\mathcal{S}_{\text{chunks}} \gets \mathcal{S}_{\text{chunks}} \cup \{(d.\text{id}, c_{\text{text}})\}$
        \State $ptr \gets ptr + (L - O)$
    \EndWhile
\EndFor
\Statex \textcolor{gray}{\textit{// Phase 2: Graph Construction \& Entity Extraction}}
\State Initialize $V \gets \emptyset, E \gets \emptyset$
\For{each chunk $c \in \mathcal{S}_{\text{chunks}}$}
    \State $\mathcal{T} \gets \mathcal{M}_{\text{gen}}(\text{Prompt}_{\text{extract}}, c)$ \Comment{Extract triplets $(s, r, o)$ via DeepSeek}
    \For{each triplet $(s, r, o) \in \mathcal{T}$}
        \State $V \gets V \cup \{s, o\}$
        \State $E \gets E \cup \{(s, r, o)\}$
    \EndFor
\EndFor
\Statex \textcolor{gray}{\textit{// Phase 3: Vectorization \& Indexing}}
\State $\mathbf{X}_{\text{chunks}} \gets \emptyset$
\For{each chunk $c \in \mathcal{S}_{\text{chunks}}$}
    \State $\mathbf{v}_c \gets \mathcal{M}_{\text{emb}}(c)$ \Comment{Dimension $d=1536$}
    \State $\mathbf{X}_{\text{chunks}}.\text{append}(\mathbf{v}_c)$
\EndFor
\State $\mathcal{I}_{\text{vec}} \gets \text{FAISS.Index}(\mathbf{X}_{\text{chunks}})$ \Comment{Build dense retrieval index}
\State $\mathbf{E}_{\text{ent}} \gets \text{EmbedEntities}(V)$ \Comment{Embed unique entities for GraphRAG}
\State \Return $\mathcal{G}, \mathcal{I}_{\text{vec}}, \mathbf{E}_{\text{ent}}$
\end{algorithmic}
\end{algorithm}

\paragraph{Vectorization \& Indexing.}
To facilitate the dense retrieval mechanism of NaiveRAG, we utilize the OpenAI \texttt{text-embedding-3-small} model to encode each text chunk into a 1,536-dimensional vector representation. We employ a batch size of 30 to mitigate API rate-limiting constraints. All generated vectors undergo L2 normalization and are subsequently indexed using \texttt{FAISS} (\texttt{IndexFlatIP}) to enable efficient retrieval based on cosine similarity. Concurrently, to support the entity-level retrieval required by GraphRAG, we generate distinct embeddings for all unique entities within the knowledge graphs, specifically, 206,738 for MuSiQue, 90,088 for QuALITY, 135,231 for Legal, and 14,712 for Medical. These entity embeddings serve as the foundation for precise entity matching and the selection of seed nodes for graph traversal during the query execution phase.

\subsection{Query Generation Pipeline}
\label{sec:app_query}

\paragraph{Generation Overview.}
To construct a query set encompassing varying degrees of cognitive complexity, we devise three distinct query generation strategies, as detailed in Algorithm~\ref{alg:query_gen}. All generation processes utilize DeepSeek-V3 as the backbone model, with a temperature setting of 0.7 to strike a balance between diversity and quality.

\begin{figure}[H]
    \centering
    \includegraphics[width=0.95\textwidth]{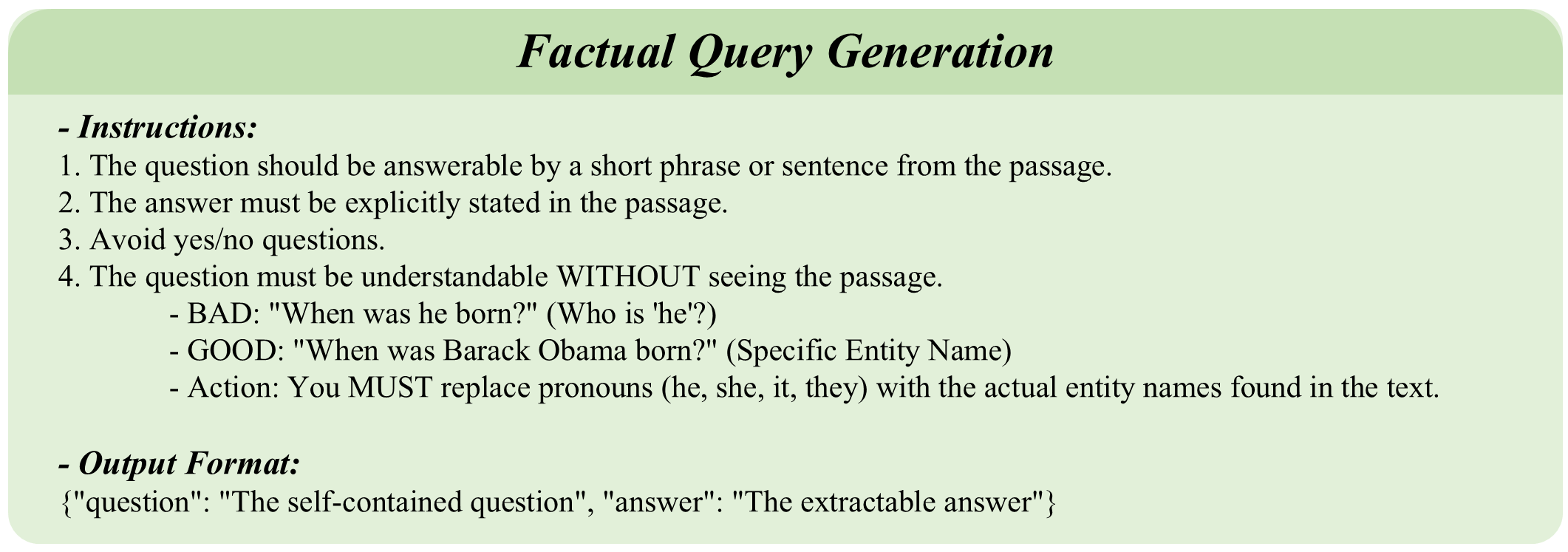} 
    \caption{The prompt template for generating Factual queries.}
    \label{fig:factual_query_generation}
\end{figure}

\textit{\noindent\textbf{Factual Queries:}} We perform uniform random sampling of text segments from the corpus and employ the prompt illustrated in Figure~\ref{fig:factual_query_generation} to steer the LLM in generating factual QA pairs. Crucially, the prompt enforces a \textit{self-contained} constraint, ensuring that questions are semantically independent of context and devoid of ambiguous pronominal references.

\begin{figure}[H]
\vspace{-2mm}
    \centering
    \includegraphics[width=0.95\textwidth]{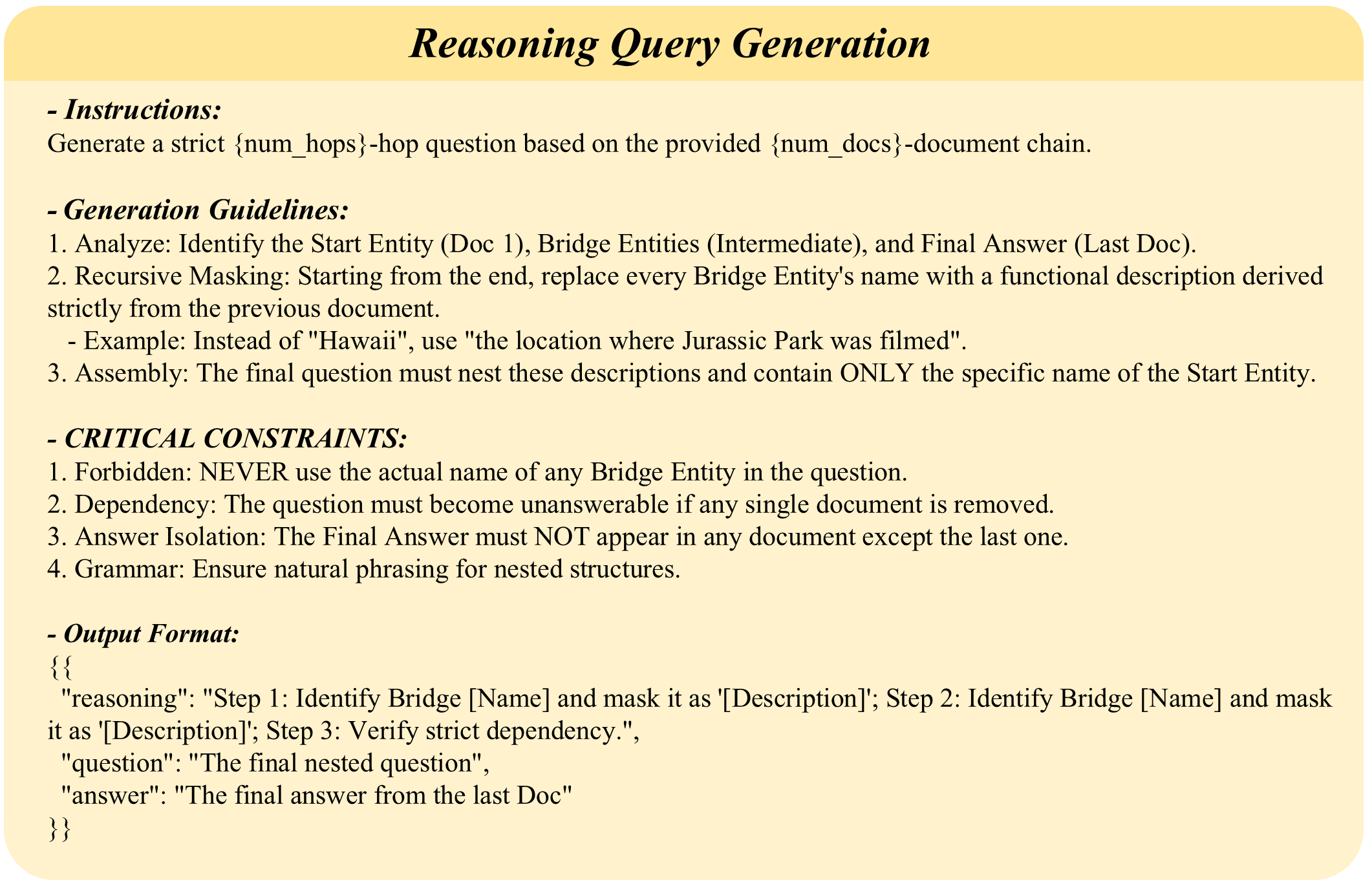} 
    \caption{The instruction set for synthesizing Reasoning queries.}
    \label{fig:reasoning_query_generation}
\end{figure}

\textit{\noindent\textbf{Reasoning Queries:}} Leveraging the topological structure of the knowledge graph, we identify document chains linked via shared entities. Specifically, for $k$-hop inquiries, we execute random walks on the graph to locate $k$ documents connected by bridge entities. Figure~\ref{fig:reasoning_query_generation} depicts the generation prompt, which centers on a Reverse Substitution strategy: starting from the target answer, bridge entities are iteratively replaced with functional descriptions derived from preceding documents. This mechanism ensures that the resulting questions necessitate traversing the complete reasoning chain for resolution. We generate reasoning queries at both 2-hop and 3-hop complexity levels.

\begin{figure}[H]
    \centering
    \includegraphics[width=0.95\textwidth]{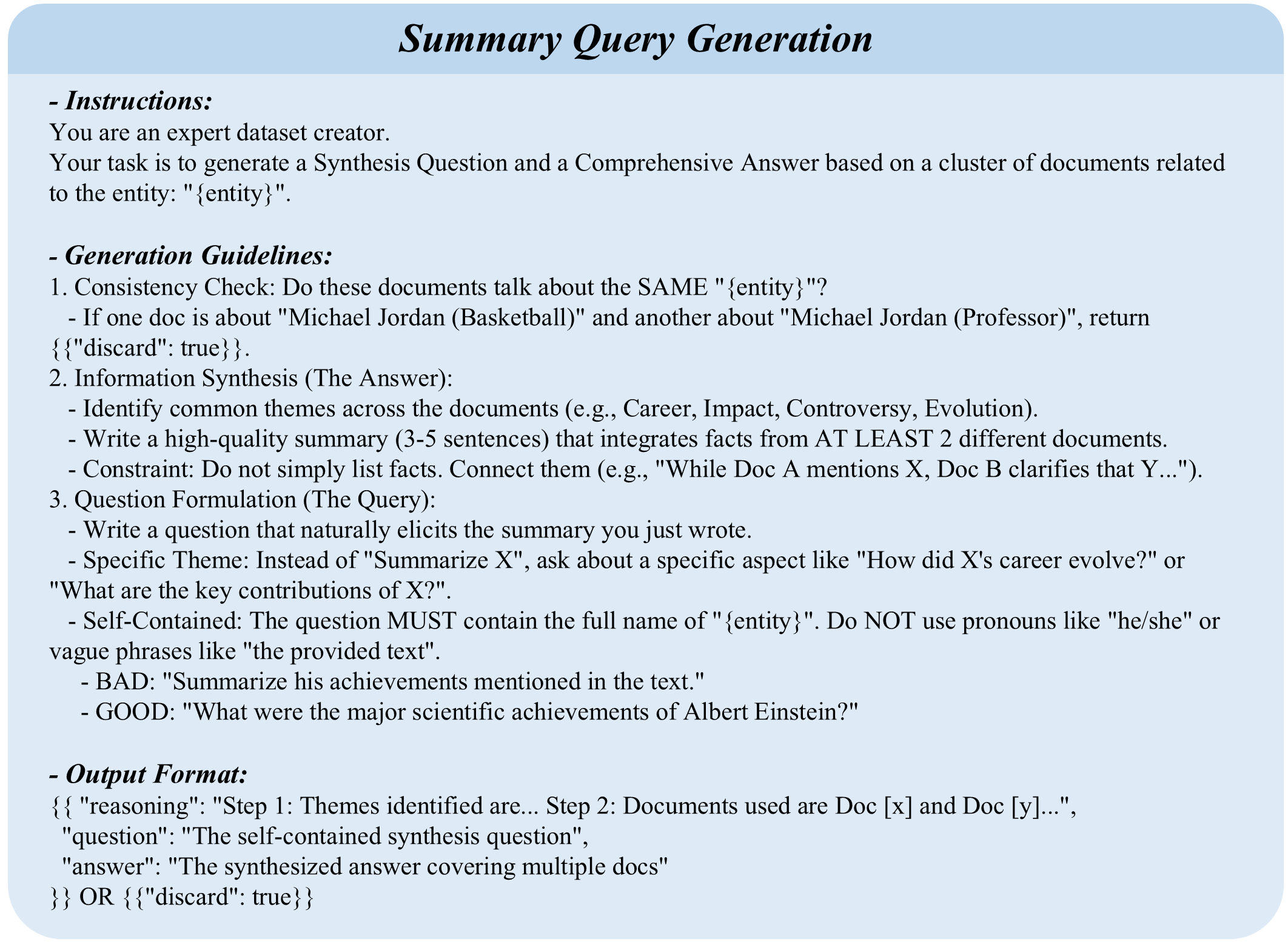} 
    \caption{The prompt designed for Summary query generation.}
    \label{fig:summary_query_generation}
\end{figure}

\textit{\noindent\textbf{Summary Queries:}} We cluster documents by entity, selecting those with high connectivity within the graph as summarization targets. As shown in Figure~\ref{fig:summary_query_generation}, the prompt mandates an initial \textit{consistency check}, verifying that multiple documents refer to the same entity rather than homonyms, followed by the synthesis of information from at least two documents. This process yields summarization questions that explicitly require cross-document integration for a complete response.

\paragraph{Verify-then-Filter Validation.}
Raw queries undergo a rigorous three-tiered verification protocol before inclusion in the final benchmark (Algorithm~\ref{alg:query_gen}, Phase 2).

\textit{\noindent\textbf{Grounding Check:}} We task an LLM with generating answers derived exclusively from the supporting facts. An LLM-as-a-Judge then validates the semantic consistency between the generated response and the expected answer. This step guarantees that the question is rigorously \textit{answerable} given the provided documents.

\textit{\noindent\textbf{Shortcut Detection:}} We evaluate whether any single supporting fact suffices to answer the question in isolation. If an individual document yields the correct answer, the query is identified as containing a ``shortcut'', violating the intrinsic requirement of multi-hop reasoning, and is subsequently filtered out.

\textit{\noindent\textbf{Knowledge Leakage Check:}} We screen for two forms of information leakage: (1) Lexical Leakage, where the answer appears as a substring within the question itself; and (2) Parametric Leakage, where the LLM can answer correctly relying solely on pre-trained knowledge. The latter is assessed via a closed-book test; if the model succeeds without retrieval context, the query is deemed ineffective for evaluating retrieval capabilities.

\textit{\noindent\textbf{Human Verification:}} To address potential LLM self-validation bias in the above automated checks, we conduct manual verification on a stratified sample of 50 queries per corpus (N=200 total). Two annotators independently assess answerability and information leakage, achieving 91\% inter-annotator agreement (Cohen's $\kappa$=0.85). The human-LLM agreement rate of 94\% confirms the reliability of the automated filtering pipeline.

\paragraph{Validation Statistics.}
Table~\ref{tab:validation_stats} presents the validation statistics across the constituent datasets. We generated a total of 7,444 candidate queries, of which 3,414 were retained following the \textit{Verify-then-Filter} process, yielding an overall acceptance rate of 45.9\%.

\begin{algorithm}[H]
\caption{Query Generation \& Validation Pipeline}
\small
\label{alg:query_gen}
\begin{algorithmic}[1]
\Require Corpus $\mathcal{C}$, Knowledge Graph $\mathcal{G}$, Target Counts $N_{\text{fact}}, N_{\text{hop}}, N_{\text{sum}}$
\Ensure Validated Query Set $\mathcal{Q}_{\text{final}}$

\Statex \textbf{Hyperparameters:} Generator $\mathcal{M}_{\text{gen}}$, Evaluator $\mathcal{M}_{\text{eval}}$

\Statex \textcolor{gray}{\textit{// Phase 1: Diversity-Driven Generation}}
\State $\mathcal{Q}_{\text{raw}} \gets \emptyset$

\State \textbf{(1) Factual Queries (Single-hop):}
\For{$i \gets 1$ \textbf{to} $N_{\text{fact}}$}
    \State $c \sim \text{Uniform}(\mathcal{C})$ \Comment{Sample random chunk}
    \State $(q, a) \gets \mathcal{M}_{\text{gen}}(\text{Prompt}_{\text{fact}}, c)$
    \State $\mathcal{Q}_{\text{raw}} \gets \mathcal{Q}_{\text{raw}} \cup \{(q, a, \{c\}, \text{"factual"})\}$
\EndFor

\State \textbf{(2) Reasoning Queries (Multi-hop):}
\State $\mathcal{P}_{\text{chains}} \gets \text{RandomWalk}(\mathcal{G}, \text{len}=2)$ \Comment{Find connected doc pairs via bridge entities}
\For{$i \gets 1$ \textbf{to} $N_{\text{hop}}$}
    \State $D_{\text{chain}} \gets \mathcal{P}_{\text{chains}}[i]$
    \State $(q, a, \text{reasoning}) \gets \mathcal{M}_{\text{gen}}(\text{Prompt}_{\text{hop}}, D_{\text{chain}})$
    \State $\mathcal{Q}_{\text{raw}} \gets \mathcal{Q}_{\text{raw}} \cup \{(q, a, D_{\text{chain}}, \text{"multi\_hop"})\}$
\EndFor

\State \textbf{(3) Summary Queries (Global):}
\For{$i \gets 1$ \textbf{to} $N_{\text{sum}}$}
    \State $e \sim \text{PageRank}(\mathcal{G})$ \Comment{Sample important entity}
    \State $D_{\text{cluster}} \gets \text{GetNeighbors}(e, \mathcal{G})$ \Comment{Retrieve ego-graph documents}
    \State $(q, a) \gets \mathcal{M}_{\text{gen}}(\text{Prompt}_{\text{sum}}, D_{\text{cluster}})$
    \State $\mathcal{Q}_{\text{raw}} \gets \mathcal{Q}_{\text{raw}} \cup \{(q, a, D_{\text{cluster}}, \text{"summary"})\}$
\EndFor

\Statex \textcolor{gray}{\textit{// Phase 2: The "Verify-then-Filter" Validation Loop}}
\State $\mathcal{Q}_{\text{final}} \gets \emptyset$
\For{each query instance $\mathbf{x} = (q, a, D_{\text{supp}}, \text{type}) \in \mathcal{Q}_{\text{raw}}$}
    \State $\text{valid} \gets \textbf{True}$

    \State \textcolor{gray}{\textit{Check 1: Grounding (Answerable from Context?)}}
    \State $\hat{a} \gets \mathcal{M}_{\text{eval}}(\text{Prompt}_{\text{qa}}, q, D_{\text{supp}})$
    \If{$\text{Sim}(\hat{a}, a) < \tau_{\text{strict}}$} $\text{valid} \gets \textbf{False}$ \EndIf

    \State \textcolor{gray}{\textit{Check 2: Shortcut Detection (Multi-hop Only)}}
    \If{$\text{type} == \text{"multi\_hop"}$}
        \For{$d \in D_{\text{supp}}$}
            \If{$\mathcal{M}_{\text{eval}}(q, \{d\}) \approx a$} \Comment{Can single doc answer it?}
                \State $\text{valid} \gets \textbf{False}$; \textbf{break}
            \EndIf
        \EndFor
    \EndIf

    \State \textcolor{gray}{\textit{Check 3: Knowledge Leakage (LLM Prior Knowledge)}}
    \State $a_{\text{prior}} \gets \mathcal{M}_{\text{eval}}(\text{Prompt}_{\text{closed\_book}}, q)$ \Comment{Ask without context}
    \If{$\text{Sim}(a_{\text{prior}}, a) > \tau_{\text{leak}}$} 
        \State $\text{valid} \gets \textbf{False}$ \Comment{Reject if LLM already knows the answer}
    \EndIf

    \If{$\text{valid}$} $\mathcal{Q}_{\text{final}} \gets \mathcal{Q}_{\text{final}} \cup \{\mathbf{x}\}$ \EndIf
\EndFor

\State \Return $\mathcal{Q}_{\text{final}}$
\end{algorithmic}
\end{algorithm}

\begin{table}
\centering
\small
\setlength{\tabcolsep}{8pt}
\renewcommand{\arraystretch}{1.1}
\begin{tabular}{l ccc ccc ccc}
\toprule
\multirow{2}{*}{\textbf{Type}} &
\multicolumn{3}{c}{\textbf{MuSiQue}} &
\multicolumn{3}{c}{\textbf{QuALITY}} &
\multicolumn{3}{c}{\textbf{Legal}} \\
\cmidrule(lr){2-4}\cmidrule(lr){5-7}\cmidrule(lr){8-10}
& \textbf{Gen.} & \textbf{Pass} & \textbf{Rate}
& \textbf{Gen.} & \textbf{Pass} & \textbf{Rate}
& \textbf{Gen.} & \textbf{Pass} & \textbf{Rate} \\
\midrule
Single-hop & 700 & 398 & 56.9\% & 500 & 454 & 90.8\% & 400 & 370 & 92.5\% \\
2-hop      & 400 & 84  & 21.0\% & 561 & 212 & 37.8\% & 784 & 238 & 30.4\% \\
3-hop      & 400 & 89  & 22.2\% & 600 & 249 & 41.5\% & 800 & 288 & 36.0\% \\
Summary    & 527 & 368 & 69.8\% & 789 & 283 & 35.9\% & 983 & 381 & 38.8\% \\
\textit{Dataset Total} & \textit{2,027} & \textit{939} & \textit{46.3\%}
& \textit{2,450} & \textit{1,198} & \textit{48.9\%}
& \textit{2,967} & \textit{1,277} & \textit{43.0\%} \\
\midrule
\multicolumn{7}{l}{\textbf{Overall Total}} & \textbf{7,444} & \textbf{3,414} & \textbf{45.9\%} \\
\bottomrule
\end{tabular}
\caption{Verify-then-Filter validation statistics.}
\label{tab:validation_stats}
\end{table}

Pass rates exhibit significant variance across distinct query types. Single-hop queries achieve pass rates exceeding 90\% on QuALITY and Legal datasets, yet only 56.9\% on MuSiQue. This disparity is primarily attributed to MuSiQue's foundation in Wikipedia, where extensive factual overlap with the LLM's pre-training corpus frequently triggers the Knowledge Leakage filter. Multi-hop queries register the lowest pass rates (21\%--42\%), with the vast majority discarded by Shortcut Detection, underscoring the inherent challenge in generating questions that genuinely necessitate multi-step reasoning. Summary queries exhibit pass rates ranging from 36\% to 70\%, with failures predominantly stemming from the Grounding Check, specifically, semantic deviations between the response derived from the provided document set and the expected gold standard.

The final benchmark comprises validated queries from MuSiQue (939), QuALITY (1,198), and Legal (1,277). Integrating the Medical dataset from GraphRAGBench (1,896 pre-annotated questions), the final corpus totals 5,310 high-quality queries, spanning four domains and three levels of cognitive complexity.

\subsection{Corpus Evaluation Metrics}
\label{sec:app_metrics}

Table~\ref{tab:corpus_metrics} summarizes the core metrics employed to characterize the retrieval environment. By quantifying corpus properties along the dual dimensions of \textit{structural topology} and \textit{semantic space}, these metrics provide a quantitative foundation for delineating the applicability boundaries of distinct RAG paradigms.

\begin{table}[H]
    \centering
    \small
    \setlength{\tabcolsep}{2.5pt}
    \begin{tabularx}{\textwidth}{l l l >{\raggedright\arraybackslash}X}
        \toprule
        \textbf{Dimension} & \textbf{Metric} & \textbf{Physical Interpretation} & \textbf{Constraint Mechanism on Retrieval} \\
        \midrule
        \multirow{3}{*}{\textbf{Structural}} 
        & LCC Ratio & Global Reachability & Fragmentation into isolated components severs reasoning paths for multi-hop retrieval. \\
        & Density & Edge Saturation & Sparse graphs lack sufficient relational bridges; overly dense graphs introduce noise. \\
        & Clustering Coeff. & Local Coherence & Facilitates evidence aggregation within topical neighborhoods. \\
        \midrule
        \multirow{3}{*}{\textbf{Semantic}} 
        & Intrinsic Dim. & Effective Degrees of Freedom & High dimensionality exacerbates the curse of dimensionality, degrading similarity metrics. \\
        & Dispersion & Semantic Spread & Low dispersion causes semantic crowding, hindering distinction of hard negatives. \\
        & Hubness & Retrieval Interference & Hub embeddings dominate neighbor lists, causing systematic retrieval bias. \\
        \bottomrule
    \end{tabularx}
    \caption{Key corpus metrics characterizing the retrieval environment.}
    \label{tab:corpus_metrics}
\end{table}

\paragraph{Structural Topology Metrics.}
We employ three graph-theoretic metrics to evaluate the topological structure of the knowledge graphs (Table~\ref{tab:struct_full}):

\begin{table}[H]
    \centering
    \small
    \renewcommand{\arraystretch}{1.2}
    \setlength{\tabcolsep}{8pt}
    \resizebox{\textwidth}{!}{
    \begin{tabular}{l c c c c c c c c}
    \toprule
    \textbf{Dataset} & \textbf{Nodes} & \textbf{Edges} & \textbf{Density} & \textbf{Rel. Types} & \textbf{Avg. Deg.} & \textbf{Comp.} & \textbf{LCC Ratio} & \textbf{Cluster. Coeff.} \\
    \midrule
    MuSiQue   & 206,738 & 276,898 & 6.00e-06 & 44,766 & 2.68 & 7,722 & 0.882 & 0.0213 \\
    QuALITY   & 90,088  & 120,611 & 1.50e-05 & 23,828 & 2.68 & 3,997 & 0.883 & 0.0177 \\
    Legal     & 135,231 & 261,207 & 1.40e-05 & 28,799 & 3.86 & 3,204 & 0.933 & 0.0701 \\
    Medical   & 14,712  & 21,480  & 9.90e-05 & 4,169  & 2.92 & 741   & 0.861 & 0.0357 \\
    \bottomrule
\end{tabular}%
}
\caption{\textbf{Full Structural Statistics.} Detailed graph topology metrics including node/edge counts, graph density, number of unique relation types, average node degree, number of connected components (Comp.), ratio of the largest connected component (LCC Ratio), and average clustering coefficient.}
\label{tab:struct_full}
\end{table}

\textit{\noindent\textbf{LCC Ratio (Largest Connected Component Ratio):}} This metric measures global reachability. It is defined as the ratio of nodes in the largest connected component to the total number of nodes:
\begin{equation}
    \text{LCC Ratio} = \frac{|V_{\text{LCC}}|}{|V|}
\end{equation}
where $V_{\text{LCC}}$ denotes the node set of the largest connected component and $V$ represents the total node set. A lower ratio indicates severe graph fragmentation, increasing the risk that multi-hop reasoning paths are physically severed.

\textit{\noindent\textbf{Density:}} This metric measures edge saturation. For an undirected graph, it is defined as:
\begin{equation}
    D = \frac{2|E|}{|V|(|V|-1)}
\end{equation}
where $|E|$ is the number of edges and $|V|$ is the number of nodes. Excessively low density implies a lack of sufficient relational bridges between entities, while excessively high density introduces noise, thereby degrading graph traversal efficiency.

\textit{\noindent\textbf{Clustering Coefficient:}} This metric measures local cohesiveness. The clustering coefficient for a node $v$ is defined as the ratio of actual edges between its neighbors to the number of possible edges:
\begin{equation}
    C_v = \frac{2 \cdot |\{e_{jk} : v_j, v_k \in N(v), e_{jk} \in E\}|}{k_v(k_v - 1)}
\end{equation}
where $N(v)$ is the neighborhood set of node $v$, and $k_v = |N(v)|$ represents the node degree. The global clustering coefficient is the average over all nodes. A high coefficient indicates the presence of tight-knit thematic communities, facilitating local evidence aggregation.

\paragraph{Semantic Space Metrics}
We employ three vector space metrics to assess embedding quality (Table~\ref{tab:semantic_full}):

\begin{table}[H]
    \centering
    \small
    \renewcommand{\arraystretch}{1.2}
    \setlength{\tabcolsep}{10pt}
    \resizebox{\textwidth}{!}{%
    \begin{tabular}{l c c c c c c c}
    \toprule
    \textbf{Dataset} & \textbf{Chunks} & \textbf{Int. Dim.} & \textbf{Hubness} & \textbf{Avg. Dist.} & \textbf{Std. Dist.} & \textbf{Min. Dist.} & \textbf{Max. Dist.} \\
    \midrule
    MuSiQue   & 21,153 & 8.17  & 1.27 & 0.708 & 0.049 & 0.552 & 0.924 \\
    QuALITY   & 3,822  & 10.75 & 1.26 & 0.345 & 0.119 & 0.186 & 0.805 \\
    Legal     & 11,632 & 7.56  & 1.46 & 0.300 & 0.071 & 0.147 & 0.792 \\
    Medical   & 538    & 8.39  & 0.86 & 0.312 & 0.063 & 0.196 & 0.700 \\
    \bottomrule
    \end{tabular}%
    }
    \caption{\textbf{Full Semantic Statistics.} Metrics covering vector space properties: total number of text chunks, intrinsic dimension (Int. Dim.), Hubness score (interference), and centroid distance statistics (Average, Standard Deviation, Minimum, Maximum).}
    \label{tab:semantic_full}
\end{table}

\textit{\noindent\textbf{Intrinsic Dimension:}} This metric quantifies the effective degrees of freedom within the embedding space. We estimate it using the \textbf{TwoNN} algorithm: for each data point, we calculate the distance to its nearest neighbor ($r_1$) and second-nearest neighbor ($r_2$). Letting $\mu = r_2 / r_1$, the intrinsic dimension is defined as:
\begin{equation}
    d_{\text{int}} = \frac{1}{\mathbb{E}[\ln \mu]}
\end{equation}
High intrinsic dimensionality exacerbates the \textit{curse of dimensionality}, rendering distance-based similarity metrics ineffective.

\textit{\noindent\textbf{Dispersion:}} This metric measures the uniformity of semantic distribution. We compute the cosine distance of each embedding vector from the global centroid:
\begin{equation}
    \text{dist}(x_i) = 1 - \frac{x_i \cdot \bar{x}}{\|x_i\| \|\bar{x}\|}
\end{equation}
where $\bar{x} = \frac{1}{n}\sum_{i=1}^{n} x_i$ denotes the centroid vector. Table~\ref{tab:semantic_full} reports the mean, standard deviation, minimum, and maximum of these distances. Low dispersion results in \textit{semantic crowding}, hindering the retriever's ability to distinguish between semantically similar yet factually unrelated ``hard negatives.''

\textit{\noindent\textbf{Hubness:}} This metric quantifies the extent of retrieval interference. Defining $N_k(i)$ as the number of times point $i$ appears in the $k$-nearest neighbor lists of all other points ($k$-occurrence), Hubness is calculated as the skewness of this distribution:
\begin{equation}
    S_k = \frac{\mathbb{E}[(N_k - \mu_{N_k})^3]}{\sigma_{N_k}^3}
\end{equation}
Positive skewness indicates the presence of ``hub'' embeddings, vectors that frequently appear in the nearest neighbor lists of others. This phenomenon causes a systematic bias in retrieval results towards these hubs, thereby reducing both retrieval diversity and accuracy.

\section{RAG Paradigm Implementation}
\label{sec:app_implementation}

\begin{figure*}[t]
  \centering
  \caption{\textbf{Overview of the five RAG paradigms evaluated in RAGRouter-Bench.} \textbf{(a) Input:} Query and corpus shared across all paradigms. \textbf{(b) Retrieval:} Paradigm-specific pipelines. \textbf{(c) Generation:} Retrieved context combined with query as prompt to LLM. \textbf{(d) Output:} Final response.}
  \includegraphics[width=0.95\textwidth]{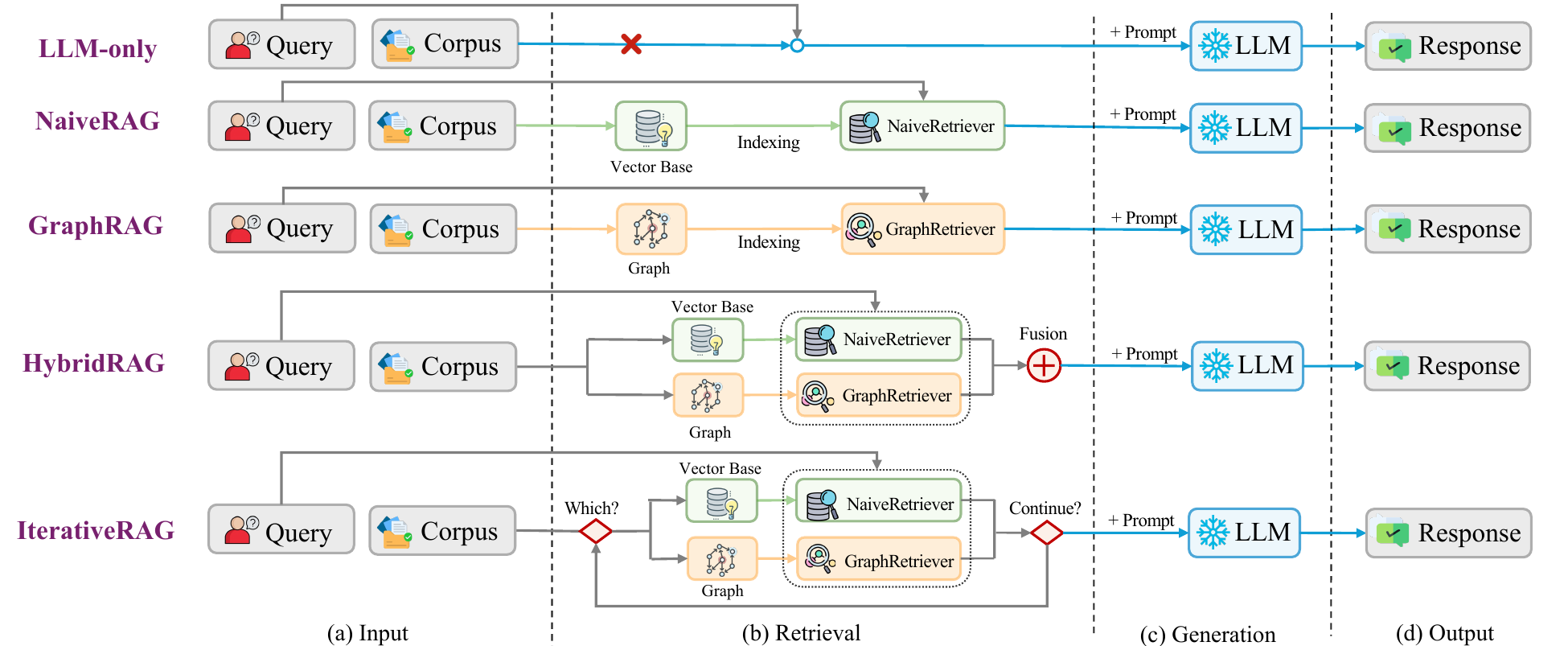}
  \label{fig:rag_paradigms}
\end{figure*}

The architectural overview is illustrated in Figure~\ref{fig:rag_paradigms}.

\subsection{Unified Hyperparameters}
\label{sec:app_hyperparams}

To guarantee a fair comparison across distinct RAG paradigms, we standardize the core hyperparameter configurations. Table~\ref{tab:shared_params} enumerates the foundational settings shared across all methodologies: DeepSeek-V3 serves as the uniform generation backbone, with the temperature set to 0.3 to elicit stable outputs. All retrieval-augmented approaches are constrained by a shared context budget of 8,000 tokens, employ cosine similarity as the distance metric, and enforce a minimum relevance threshold of 0.4 to filter out low-quality evidence.

\begin{table}
    \centering
    \small
    \setlength{\tabcolsep}{15pt}
    \renewcommand{\arraystretch}{1.2}
    \begin{tabular}{l l l l}
    \toprule
    \textbf{Category} & \textbf{Parameter} & \textbf{Value} & \textbf{Description} \\
    \midrule
    \multirow{4}{*}{LLM}
      & Model & DeepSeek-V3 & Generation backbone \\
      & Temperature & 0.3 & Focused generation \\
      & Max Tokens & 1000 & Response length limit \\
      & Timeout & 120s & API request limit \\
    \midrule
    \multirow{3}{*}{Retrieval}
      & Token Budget & 8000 & Context length limit \\
      & Similarity & Cosine & Distance metric \\
      & Min Threshold & 0.4 & Relevance filter \\
    \bottomrule
    \end{tabular}
    \caption{Shared hyperparameters across all RAG paradigms.}
    \label{tab:shared_params}
\end{table}

Table~\ref{tab:paradigm_params} details the paradigm-specific parameters. LLM-only, serving as the retrieval-free baseline, utilizes a higher temperature (0.7) to encourage the model to fully leverage its internal parametric knowledge. NaiveRAG retrieves a maximum of 100 text chunks, truncating the selection based on similarity ranking to fit within the token budget. GraphRAG initiates from 20 seed entities and expands the subgraph using Personalized PageRank ($\alpha=0.85$), retaining a maximum of 100 nodes and 500 triplets. IterativeRAG employs GraphRAG as the base retriever, executing up to 3 rounds of iterative refinement, with the evaluator operating at a low temperature (0.1) to ensure decisional consistency. Finally, HybridRAG inherits the parameter settings of both NaiveRAG and GraphRAG, fusing their respective retrieval results.

\begin{table}
    \centering
    \small
    \setlength{\tabcolsep}{15pt}
    \renewcommand{\arraystretch}{1.2}
    \begin{tabular}{l l l l}
    \toprule
    \textbf{Paradigm} & \textbf{Parameter} & \textbf{Value} & \textbf{Description} \\
    \midrule
    \multirow{2}{*}{LLM-only}
      & Temperature & 0.7 & Creative generation \\
      & Max Tokens & 1000 & Response length \\
    \midrule
    NaiveRAG & Top-K & 100 & Max chunks retrieved \\
    \midrule
    \multirow{4}{*}{GraphRAG}
      & Seed Entities & 20 & Initial anchors \\
      & PPR Alpha & 0.85 & Damping factor \\
      & PPR Max Nodes & 100 & Subgraph size limit \\
      & Max Triplets & 500 & Serialization limit \\
    \midrule
    \multirow{3}{*}{IterativeRAG}
      & Base Retriever & GraphRAG & Underlying method \\
      & Max Iterations & 3 & Reasoning loop limit \\
      & Eval Temp. & 0.1 & Evaluator setting \\
    \bottomrule
    \end{tabular}
    \caption{Paradigm-specific hyperparameters.}
    \label{tab:paradigm_params}
\end{table}

\subsection{Retrieval Paradigm Implementation}
\label{sec:app_paradigms}

\begin{figure}[H]
    \centering
    \includegraphics[width=1.0\textwidth]{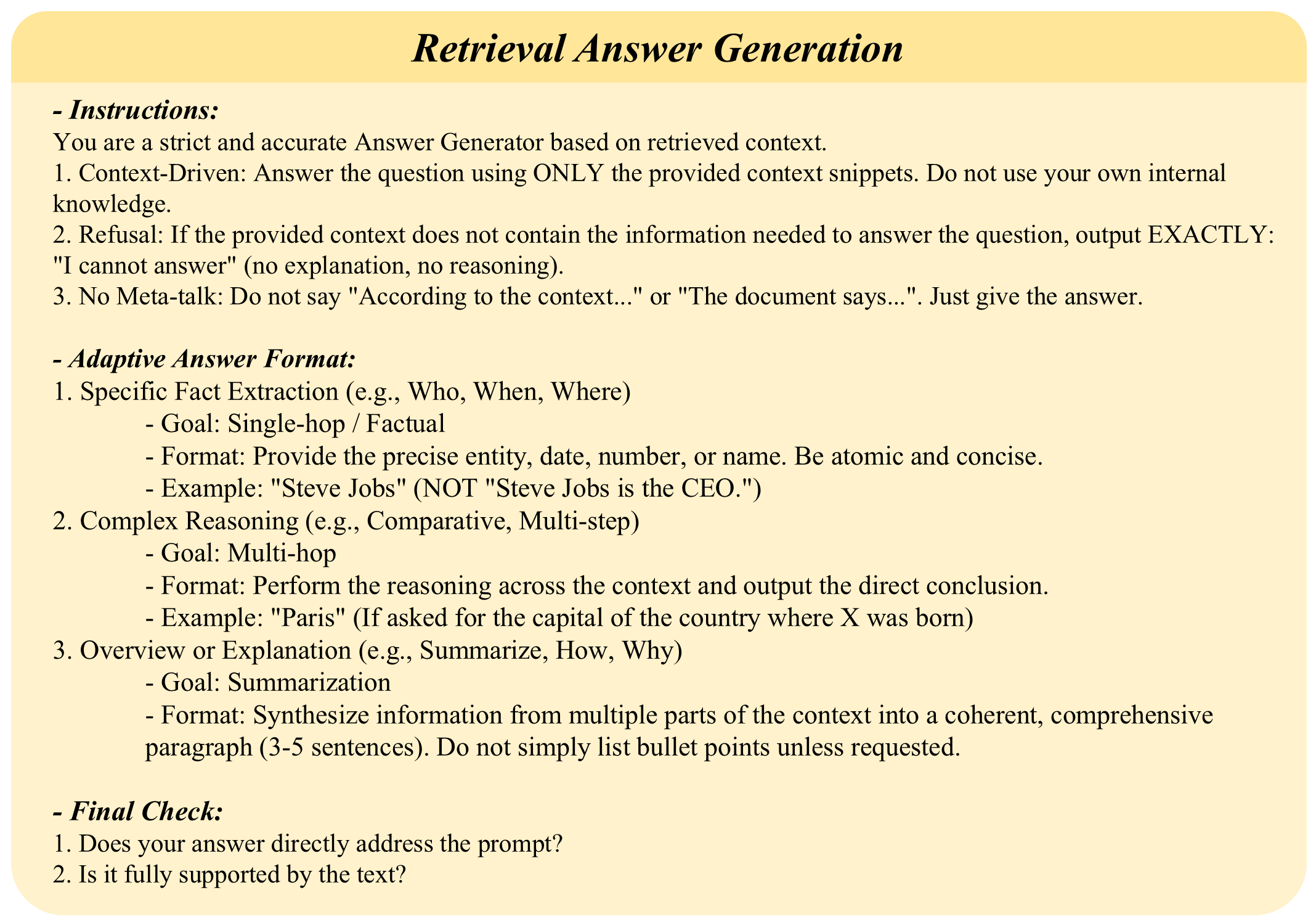} 
    \caption{The Context-Aware Generation prompt.}
    \label{fig:retrieval_answer_generation}
\end{figure}

Table~\ref{tab:rag_paradigm_intro} contrasts the five paradigms from a methodological perspective, characterizing their retrieval substrates, information granularity, and search mechanisms. This section elaborates on the specific implementation details of each paradigm. To ensure a fair comparison, all methodologies employ a uniform generation prompt, as illustrated in Figure~\ref{fig:retrieval_answer_generation}.

\begin{table}[H]
    \centering
    \scriptsize
    \setlength{\tabcolsep}{4pt}
    \renewcommand{\arraystretch}{1.2}
    \setlength{\tabcolsep}{1pt}

    \begin{tabularx}{\textwidth}{ 
        l 
        >{\raggedright\arraybackslash}p{0.18\textwidth}
        >{\raggedright\arraybackslash}p{0.15\textwidth}
        >{\raggedright\arraybackslash}p{0.18\textwidth}
        >{\raggedright\arraybackslash}X 
    }
        \toprule
        \textbf{Paradigm} & 
        \textbf{Retrieval Substrate} & 
        \textbf{Info. Granularity} & 
        \textbf{Search Mechanism} & 
        \textbf{Optimal Use Cases} \\
        \midrule
        
        \textbf{LLM-only} & 
        Parametric Weights \newline (Implicit) & 
        Internal Knowledge & 
        Next-token Prediction & 
        General chit-chat; Creative writing; Tasks requiring no external facts \citep{mallen2023trust}. \\
        \addlinespace[4pt]
        
        \textbf{NaiveRAG} & 
        Flat Vector Space & 
        Coarse-grained \newline (Passage/Chunk) & 
        Semantic Similarity \newline (Dense Retrieval) & 
        Explicit fact retrieval; Simple QA; Queries with high semantic overlap \citep{karpukhin2020dense}. \\
        \addlinespace[4pt]
        
        \textbf{HybridRAG} & 
        Dual-Pathway Space \newline (Dense + Sparse) & 
        Multi-granular \newline (Keyword + Chunk) & 
        Hybrid Fusion \newline (BM25 + Vector) & 
        Precision-critical search; Low-frequency entity lookup; Exact matching \citep{thakur2021beir}. \\
        \addlinespace[4pt]
        
        \textbf{GraphRAG} & 
        Graph Topology \newline (Knowledge Graph) & 
        Fine-grained \newline (Entity/Relation) & 
        Structure-aware \newline Traversal & 
        Multi-hop reasoning \citep{yasunaga2021qa}; Global summarization \citep{edge2024graphrag}; Connecting disparate information. \\
        \addlinespace[4pt]
        
        \textbf{IterativeRAG} & 
        Dynamic Context & 
        Adaptive \newline (Coarse to Fine) & 
        Multi-step Feedback \newline (Reasoning Loop) & 
        Ambiguous queries; Complex research; Tasks needing progressive clarification \citep{trivedi2022interleaving, asai2024selfrag}. \\
        
        \bottomrule
    \end{tabularx}
    
    \caption{\textbf{A Methodology Perspective on RAG Paradigms.} We categorize existing paradigms by their \textit{Retrieval Substrate} (data structure), \textit{Information Granularity}, and \textit{Search Mechanism}. Each paradigm imposes different trade-offs between retrieval cost and reasoning capability, highlighting that no single strategy fits all scenarios.}
    \label{tab:rag_paradigm_intro}
\end{table}

\begin{figure}[H]
    \centering
    \includegraphics[width=0.95\textwidth]{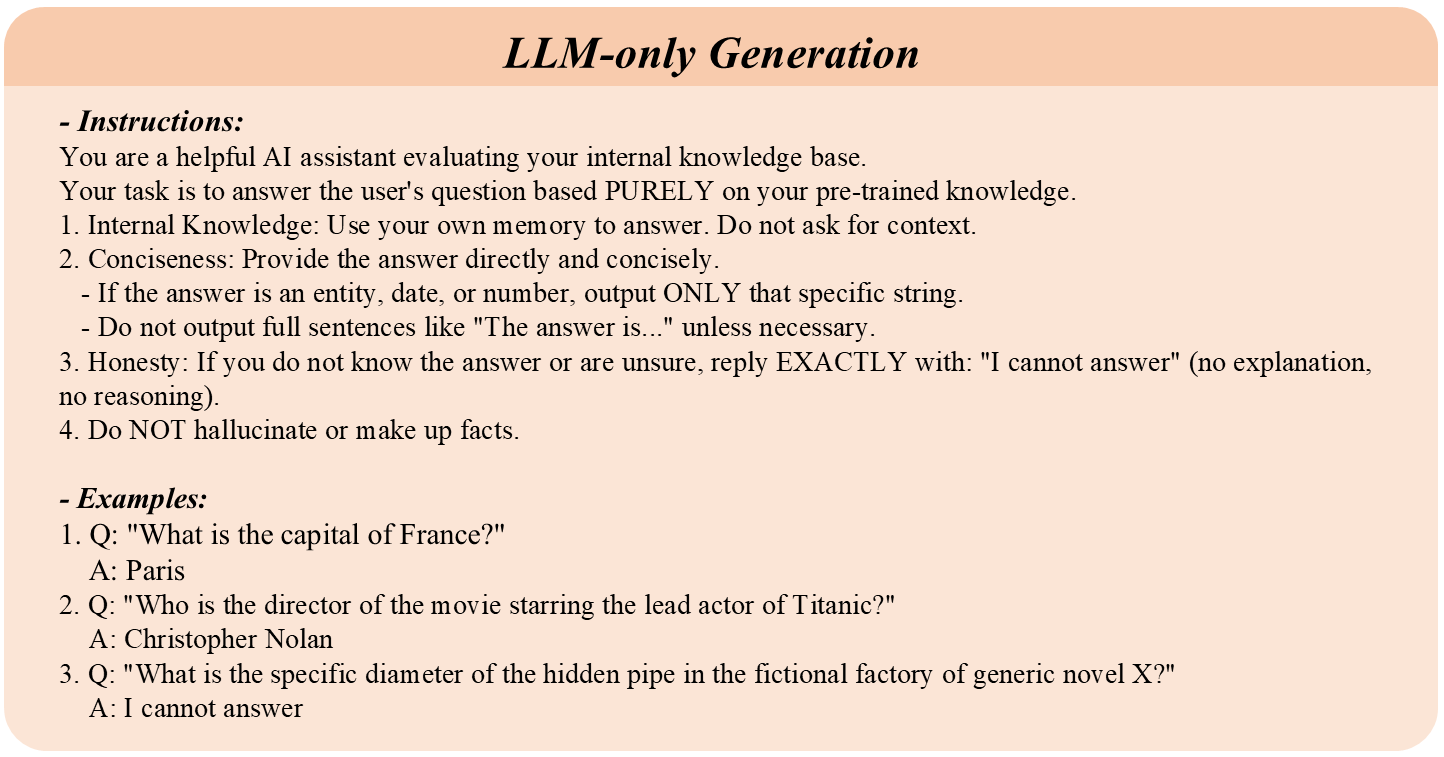} 
    \caption{The prompt for Direct Generation (LLM-only).}
    \label{fig:llm_direct_generation}
\end{figure}

\afterpage{
\paragraph{LLM-only and NaiveRAG.}
LLM-only operates as the retrieval-free baseline, generating responses by directly querying the model and thereby relying exclusively on its internal parametric knowledge (Figure~\ref{fig:llm_direct_generation}). NaiveRAG follows the standard dense retrieval protocol, retrieving the top-100 semantically similar text chunks and subsequently truncating the concatenated context to adhere to the 8,000-token budget limit.}

\begin{figure}[H]
    \centering
    \includegraphics[width=0.95\textwidth]{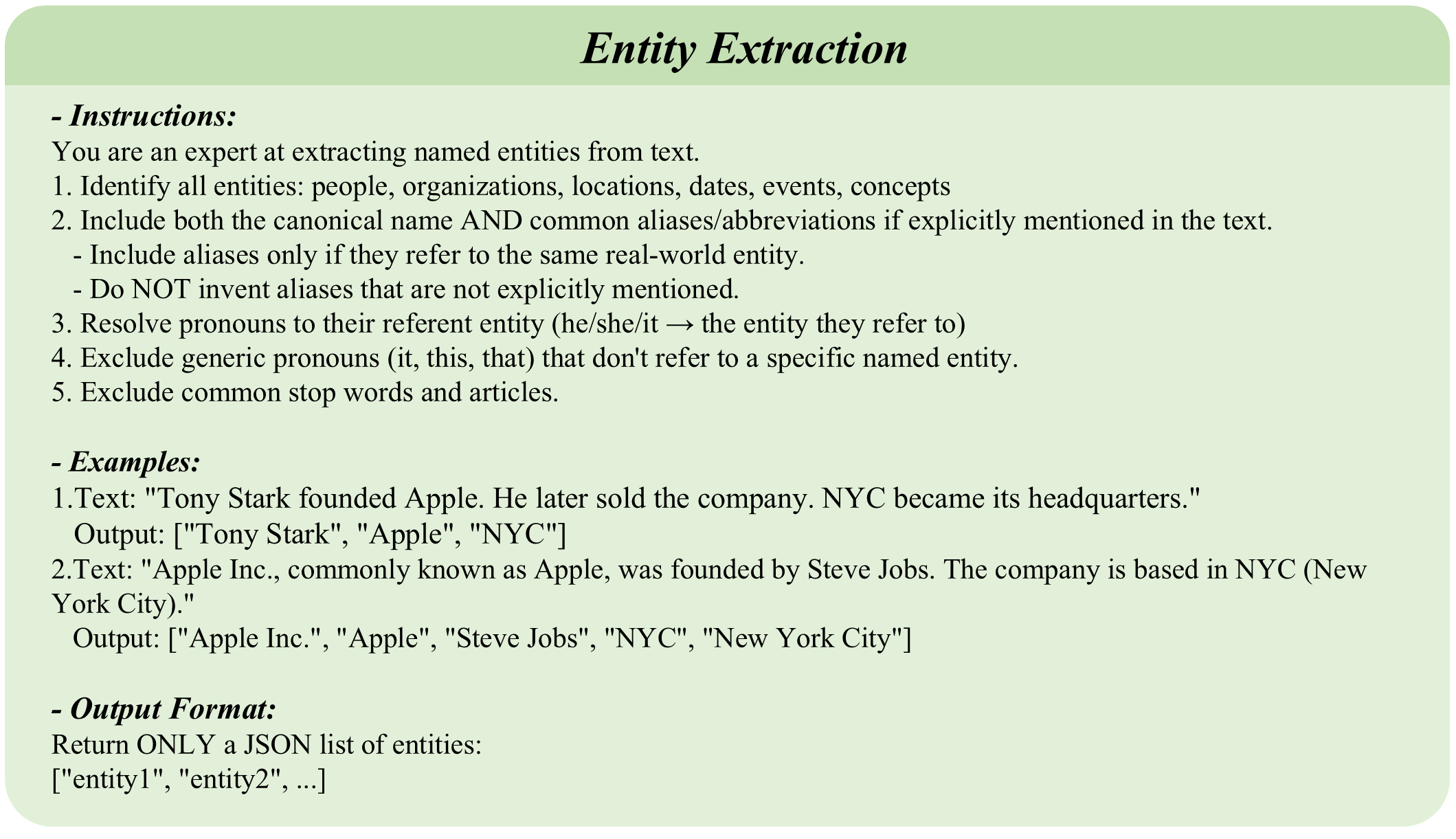} 
    \caption{The prompt template for Entity Extraction.}
    \label{fig:entity_extractionn}
\vspace{-1mm}
\end{figure}

\paragraph{GraphRAG.}
GraphRAG leverages knowledge graphs to perform structure-aware retrieval (Algorithm~\ref{alg:graphrag_pipeline}). The process initiates by extracting query entities via an LLM (Figure~\ref{fig:entity_extractionn}) and linking them to graph entities to establish a seed node set $S$. Subsequently, Personalized PageRank (PPR) is executed over the global graph topology. We construct the personalization vector $\mathbf{p}$ based on semantic similarity to the seeds:
\begin{equation}
    \mathbf{p}[v] = \frac{\text{sim}(v)}{\sum_{u \in S} \text{sim}(u)}, \quad v \in S
\end{equation}
The PPR iterative update rule is defined as:
\begin{equation}
    \boldsymbol{\pi}^{(t+1)} = \alpha \cdot \mathbf{A} \cdot \boldsymbol{\pi}^{(t)} + (1-\alpha) \cdot \mathbf{p}
\end{equation}
where $\mathbf{A}$ denotes the column-normalized adjacency matrix of the graph, and $\alpha=0.85$ serves as the damping factor. Upon convergence, we identify the top-100 nodes with the highest PPR scores to construct a salient subgraph. The associated triplets are then extracted and mapped back to their original textual source to serve as the generation context.

\paragraph{HybridRAG.}
HybridRAG integrates the retrieval outputs from both NaiveRAG (vector-based) and GraphRAG (graph-based). Following the independent acquisition of ranked lists from both pathways, we employ Reciprocal Rank Fusion (RRF) to merge the rankings:
\begin{equation}
    \text{RRF}(d) = \sum_{r \in \mathcal{R}} \frac{1}{k + \text{rank}_r(d)}
\end{equation}
where $\mathcal{R} = \{\text{Naive}, \text{Graph}\}$ represents the set of retrievers, $\text{rank}_r(d)$ denotes the rank position of document $d$ within retriever $r$, and $k=60$ serves as the smoothing constant. Post-fusion, documents are sorted in descending order of their RRF scores. We subsequently remove duplicates and truncate the sequence to adhere to the strict 8,000-token context budget.

\begin{figure}[H]
\vspace{-5mm}
    \centering
    \includegraphics[width=0.95\textwidth]{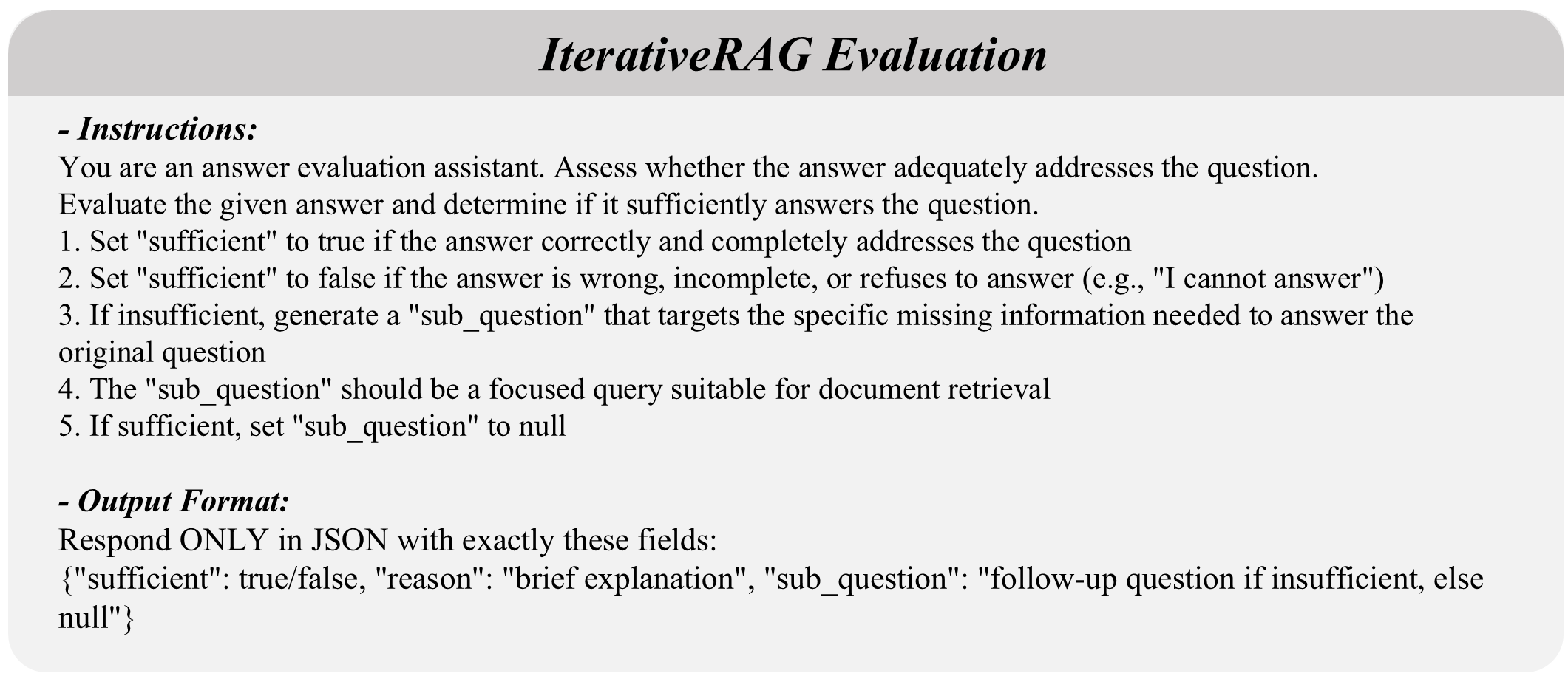} 
    \caption{The Self-Evaluation prompt for Iterative RAG.}
    \label{fig:iterativerag_evaluation}
\end{figure}

\paragraph{IterativeRAG.}
IterativeRAG implements a ``Retrieve-Generate-Evaluate'' feedback loop, as detailed in Algorithm~\ref{alg:iterativerag}. In the initial phase (Round 0), the system attempts a direct response using the LLM. If the evaluator (Figure~\ref{fig:iterativerag_evaluation}) deems this response insufficient, it generates targeted sub-questions to trigger the retrieval cycle. In each subsequent iteration, newly retrieved evidence is aggregated with the cumulative context to synthesize an updated answer, which is then re-evaluated. This cycle persists until one of the following termination criteria is met: (i) the answer is judged sufficient; (ii) the maximum iteration count ($T=3$) is reached; (iii) no new sub-questions are generated; or (iv) generated sub-questions are repetitive. We instantiate the framework using either NaiveRAG or GraphRAG as the underlying base retriever.

\begin{algorithm}[H]
\caption{IterativeRAG (Multi-Round Retrieval with Self-Evaluation)}
\small
\label{alg:iterativerag}
\begin{algorithmic}[1]
\Require Question $q$, Base Retriever $\mathcal{R}$ (NaiveRAG or GraphRAG), Max Iterations $T$
\Ensure Final Answer $a$

\Statex \textbf{Initialization}
\State $\mathcal{Q}_{\text{history}} \gets \{q\}$ \Comment{\textcolor{gray}{Track all queries to prevent loops}}
\State $\mathcal{C}_{\text{accum}} \gets \emptyset$ \Comment{\textcolor{gray}{Accumulated retrieved chunks}}
\State $H \gets []$ \Comment{\textcolor{gray}{Reasoning trace}}

\vspace{0.5em}
\Statex \textbf{Round 0: Direct LLM Answer (No Retrieval)}
\State $a_0 \gets \text{LLM}(\text{Prompt}_{\text{QA}}, q)$ \Comment{\textcolor{gray}{Answer without context}}
\State $\text{eval}_0 \gets \text{LLM}(\text{Prompt}_{\text{Eval}}, q, a_0)$ \Comment{\textcolor{gray}{\{sufficient, reason, sub\_question\}}}
\State $H.\text{append}((0, q, a_0, \text{eval}_0))$

\If{$\text{eval}_0.\text{sufficient} \textbf{ is True}$}
    \State \Return $a_0$ \Comment{\textcolor{gray}{LLM already knows the answer}}
\EndIf

\State $q_{\text{curr}} \gets \text{eval}_0.\text{sub\_question} \textbf{ or } q$ \Comment{\textcolor{gray}{Get refined query}}

\vspace{0.5em}
\Statex \textbf{Round 1+: Iterative Retrieval Loop}
\For{$t \gets 1$ \textbf{to} $T$}
    
    \State \textcolor{gray}{\textit{// Step 1: Retrieve new chunks}}
    \State $\mathcal{C}_{\text{new}} \gets \text{Retrieve}(\mathcal{R}, q_{\text{curr}})$
    
    \State \textcolor{gray}{\textit{// Step 2: Merge and Deduplicate}}
    \State $\mathcal{C}_{\text{accum}} \gets \mathcal{C}_{\text{accum}} \cup \mathcal{C}_{\text{new}}$
    
    \State \textcolor{gray}{\textit{// Step 3: Apply Token Budget}}
    \State $\mathcal{C}_{\text{ctx}} \gets \text{ApplyTokenBudget}(\mathcal{C}_{\text{accum}}, B=8000)$
    \State $\text{ctx} \gets \text{Concatenate}(\mathcal{C}_{\text{ctx}})$
    
    \State \textcolor{gray}{\textit{// Step 4: Generate answer with accumulated context}}
    \State $a_t \gets \text{LLM}(\text{Prompt}_{\text{RAG}}, \text{ctx}, q)$ \Comment{\textcolor{gray}{Always answer ORIGINAL question}}
    
    \State \textcolor{gray}{\textit{// Step 5: Evaluate answer sufficiency}}
    \State $\text{eval}_t \gets \text{LLM}(\text{Prompt}_{\text{Eval}}, q, a_t, \text{ctx})$
    \State $H.\text{append}((t, q_{\text{curr}}, a_t, \text{eval}_t))$
    
    \State \textcolor{gray}{\textit{// Step 6: Check termination conditions}}
    \If{$\text{eval}_t.\text{sufficient} \textbf{ is True}$}
        \State \Return $a_t$ \Comment{\textcolor{gray}{Answer is sufficient}}
    \EndIf
    
    \State $q_{\text{next}} \gets \text{eval}_t.\text{sub\_question}$
    
    \If{$q_{\text{next}} \textbf{ is null}$}
        \State \Return $a_t$ \Comment{\textcolor{gray}{No further refinement possible}}
    \EndIf
    
    \If{$q_{\text{next}} \in \mathcal{Q}_{\text{history}}$}
        \State \Return $a_t$ \Comment{\textcolor{gray}{Prevent query loop}}
    \EndIf
    
    \State \textcolor{gray}{\textit{// Step 7: Update for next iteration}}
    \State $\mathcal{Q}_{\text{history}} \gets \mathcal{Q}_{\text{history}} \cup \{q_{\text{next}}\}$
    \State $q_{\text{curr}} \gets q_{\text{next}}$

\EndFor

\State \Return $H[-1].\text{answer}$ \Comment{\textcolor{gray}{Return last answer if max iterations reached}}
\end{algorithmic}
\end{algorithm}

\begin{algorithm}[H]
\caption{GraphRAG Retrieval \& Generation Pipeline}
\small
\label{alg:graphrag_pipeline}
\begin{algorithmic}[1]
\Require Question $q$, Knowledge Graph $G=(V, E)$, Entity Index $\mathcal{I}_E$, Token Budget $B$
\Ensure Generated Answer $a$

\Statex \textbf{Hyperparameters:} Entity Threshold $\tau_{\text{entity}}=0.4$, PPR Threshold $\tau_{\text{ppr}}=1e^{-5}$, Damping $\alpha=0.85$

\vspace{0.5em}
\Statex \textcolor{gray}{\textit{// Phase 1: Seed Entity Retrieval}}
\State $E_{\text{query}} \gets \text{LLM}(\text{Prompt}_{\text{extract}}, q)$
\If{$E_{\text{query}} = \emptyset$} 
    \State $E_{\text{query}} \gets \{q\}$ \Comment{Fallback: use entire question}
\EndIf

\State $S \gets \emptyset$ \Comment{Initialize seed set mapping: $id \to score$}
\For{\textbf{each} $e \in E_{\text{query}}$}
    \State $v_e \gets \text{Embed}(e)$
    \State $\mathcal{K} \gets \text{FAISS\_Search}(\mathcal{I}_E, v_e, k=20)$
    \For{\textbf{each} $(id, \text{sim}) \in \mathcal{K}$}
        \If{$\text{sim} > \tau_{\text{entity}}$}
            \State $S[id] \gets \max(S[id], \text{sim})$ \Comment{Max pooling for duplicates}
        \EndIf
    \EndFor
\EndFor
\State $S \gets \text{TopK}(S, k=20)$

\vspace{0.5em}
\Statex \textcolor{gray}{\textit{// Phase 2: PPR-Based Subgraph Expansion}}
\State $\mathbf{p} \gets \text{Zeros}(|V|)$ \Comment{Initialize personalization vector}
\State $Z \gets \sum_{(id, \text{sim}) \in S} \text{sim}$
\For{\textbf{each} $(id, \text{sim}) \in S$}
    \State $\mathbf{p}[id] \gets \text{sim} / Z$ \Comment{Normalize to probability distribution}
\EndFor

\State $\boldsymbol{\pi} \gets \text{PageRank}(G, \text{personalization}=\mathbf{p}, \alpha=\alpha, \text{iter}=100)$

\State $V_{\text{sub}} \gets \{ v \mid \boldsymbol{\pi}[v] \ge \tau_{\text{ppr}} \}$
\State $V_{\text{sub}} \gets \text{TopK}(V_{\text{sub}}, k=100) \cup \text{Keys}(S)$ \Comment{Keep top-100 expanded nodes + seeds}

\vspace{0.5em}
\Statex \textcolor{gray}{\textit{// Phase 3: Context Construction}}
\State $T \gets \{ (u, r, v) \mid u,v \in V_{\text{sub}}, (u,v) \in E \}$ \Comment{Extract triplets from induced subgraph}
\State $\mathcal{C} \gets \emptyset$
\State Sort $T$ by $\max(\text{sim}(u), \text{sim}(v))$ descending \Comment{Prioritize relevance}

\State $ctx \gets \text{``''}, \quad \text{count} \gets 0$
\For{\textbf{each} $(u, r, v) \in T$}
    \State $\text{sents} \gets \text{TripletSourceMap}[(u, r, v)]$
    \For{\textbf{each} $s \in \text{sents}$}
        \If{$\text{count} + \text{Len}(s) > B$} \textbf{break} \EndIf
        \State $ctx \gets ctx \oplus s$
        \State $\text{count} \gets \text{count} + \text{Len}(s)$
    \EndFor
\EndFor

\vspace{0.5em}
\Statex \textcolor{gray}{\textit{// Phase 4: Answer Generation}}
\State $a \gets \text{LLM}(\text{Prompt}_{\text{RAG}}, ctx, q)$
\State \Return $a$
\end{algorithmic}
\end{algorithm}

\begin{figure*}[!htbp] 
    \centering
    \includegraphics[width=0.95\textwidth]{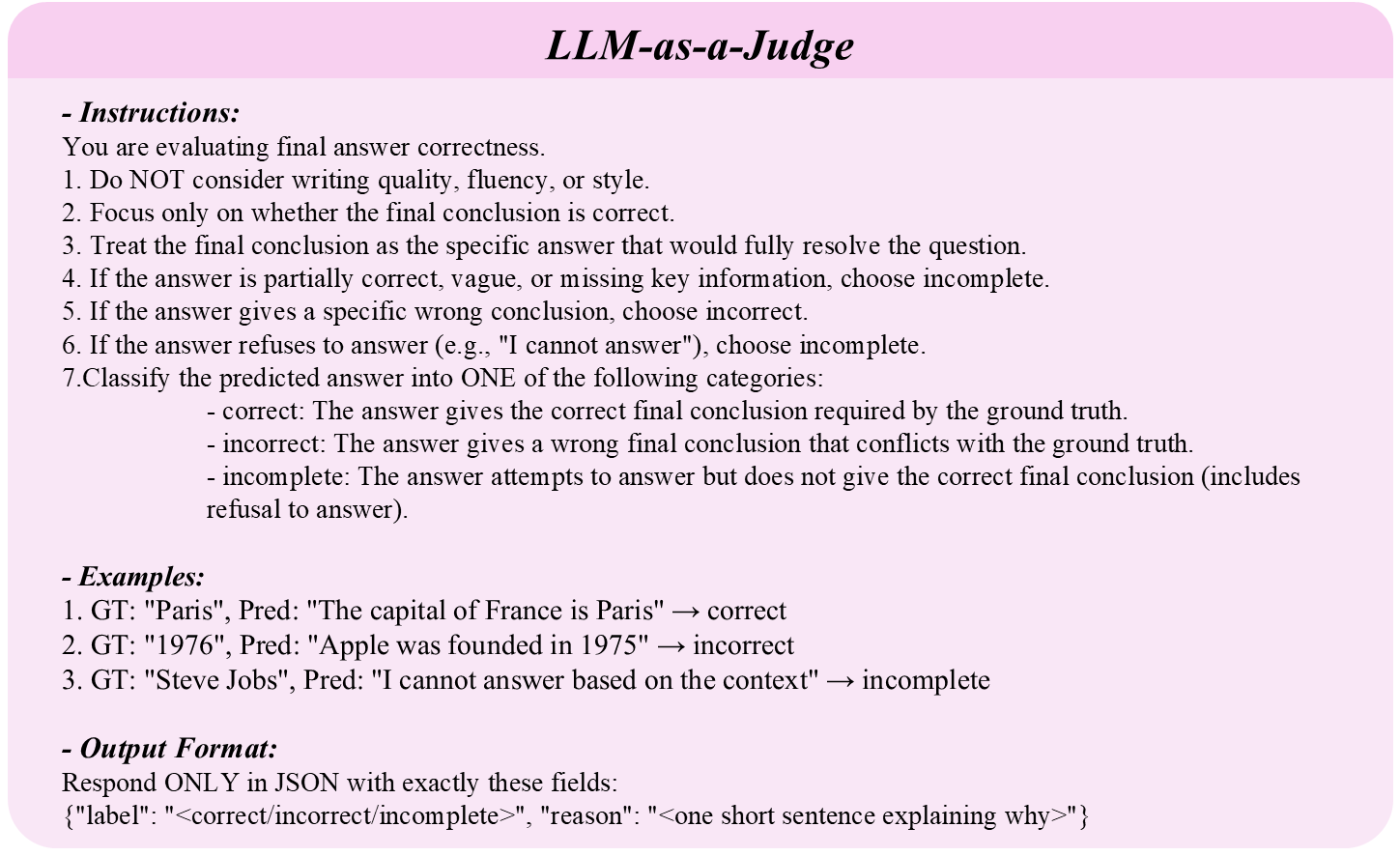} 
    \caption{The LLM-as-a-Judge instruction template used for automated evaluation.}
    \label{fig:llm_as_judge}
\end{figure*}

\begin{table*}[!htbp]
\centering
\scriptsize 
\renewcommand{\arraystretch}{1.1}
\setlength{\tabcolsep}{3pt}
\begin{tabular}{ll|ccc|ccc|ccc|ccc}
\toprule
\multirow{2}{*}{\textbf{Method}} & \multirow{2}{*}{\textbf{Type}} &
\multicolumn{3}{c|}{\textbf{MuSi.}} &
\multicolumn{3}{c|}{\textbf{QuAL.}} &
\multicolumn{3}{c|}{\textbf{Legal}} &
\multicolumn{3}{c}{\textbf{Med.}} \\
\cmidrule(lr){3-5}\cmidrule(lr){6-8}\cmidrule(lr){9-11}\cmidrule(lr){12-14}
& & \textbf{Cor} & \textbf{Inc} & \textbf{No-A}
& \textbf{Cor} & \textbf{Inc} & \textbf{No-A}
& \textbf{Cor} & \textbf{Inc} & \textbf{No-A}
& \textbf{Cor} & \textbf{Inc} & \textbf{No-A} \\
\midrule
\rowcolor{modeldeepseek}
\multicolumn{14}{c}{\textit{\textbf{Model: DeepSeek-V3}}} \\
\midrule
\multirow{3}{*}{NaiveRAG}
& Multi-hop  & 28.3 & 24.0 & 47.7 & 33.8 & 48.8 & 17.4 & 11.2 & 13.3 & 75.5 & 63.5 & 5.1 & 31.4 \\
& Single-hop & 11.1 & 5.3  & 83.7 & \textbf{83.7} & 13.2 & 3.1  & 54.9 & 17.3 & 27.8 & 63.1 & 6.2 & 30.7 \\
& Summary    & 29.4 & 27.5 & 43.2 & 17.0 & 30.7 & 52.3 & 39.1 & 11.6 & 49.3 & 49.1 & 1.0 & 49.8 \\
\cmidrule{1-14}
\multirow{3}{*}{GraphRAG}
& Multi-hop  & 22.5 & 20.6 & 56.9 & 20.4 & 49.7 & 29.9 & 6.7 & 10.7 & 82.7 & 56.0 & 3.9 & 40.1 \\
& Single-hop & \textbf{90.2} & 0.5 & 9.3 & 70.7 & 19.8 & 9.5 & 61.6 & 13.5 & 24.9 & 55.0 & 6.2 & 38.8 \\
& Summary    & 20.7 & 20.4 & 59.0 & 19.8 & 24.0 & 56.2 & 29.1 & 8.4 & 62.5 & 43.6 & 2.4 & 54.0 \\
\cmidrule{1-14}
\multirow{3}{*}{HybridRAG}
& Multi-hop  & 32.8 & 22.7 & 44.5 & 20.4 & 32.1 & 47.5 & 11.8 & 14.8 & 73.4 & 64.1 & 4.1 & 31.8 \\
& Single-hop & 83.7 & 1.5 & 14.8 & 80.0 & 13.7 & 6.4 & \textbf{72.2} & 15.1 & 12.7 & \textbf{67.8} & 5.7 & 26.5 \\
& Summary    & 30.2 & 29.6 & 40.2 & 14.8 & 35.3 & 49.8 & 34.7 & 13.1 & 52.2 & 54.0 & 1.7 & 44.3 \\
\cmidrule{1-14}
\multirow{3}{*}{IterativeRAG}
& Multi-hop  & 21.8 & 21.9 & 56.3 & 17.1 & 37.3 & 45.6 & 12.4 & 13.9 & 73.8 & \textbf{67.8} & 4.3 & 27.9 \\
& Single-hop & 10.3 & 24.1 & 65.6 & 67.0 & 24.5 & 8.6 & 49.5 & 24.6 & 25.9 & 62.1 & 12.5 & 25.4 \\
& Summary    & 21.2 & 34.2 & 44.6 & 16.3 & 38.2 & 45.6 & 30.5 & 10.8 & 58.8 & \textbf{56.1} & 3.8 & 40.1 \\
\midrule
\rowcolor{modelllama}
\multicolumn{14}{c}{\textit{\textbf{Model: LLaMA-3.1-8B}}} \\
\midrule
\multirow{3}{*}{NaiveRAG}
& Multi-hop  & 7.9 & 9.3 & 82.7 & 10.9 & 20.4 & 68.8 & 10.7 & 21.3 & 68.1 & 37.9 & 8.1 & 54.0 \\
& Single-hop & 10.3 & 5.8 & 83.9 & \textbf{69.2} & 13.9 & 17.0 & 50.3 & 24.6 & 25.1 & 52.1 & 12.3 & 35.6 \\
& Summary    & 12.0 & 47.8 & 40.2 & 1.4 & 65.4 & 33.2 & 22.8 & 27.8 & 49.3 & 30.1 & 14.5 & 55.4 \\
\cmidrule{1-14}
\multirow{3}{*}{GraphRAG}
& Multi-hop  & 9.7 & 13.6 & 76.7 & 9.5 & 12.8 & 77.7 & 7.4 & 12.2 & 80.4 & 33.6 & 6.3 & 60.1 \\
& Single-hop & \textbf{84.4} & 1.3 & 14.3 & 44.7 & 13.0 & 42.3 & 55.4 & 15.7 & 28.9 & 48.1 & 8.5 & 43.4 \\
& Summary    & 11.4 & 37.5 & 51.1 & 2.5 & 67.8 & 29.7 & 19.7 & 26.5 & 53.8 & 31.8 & 9.7 & 58.5 \\
\cmidrule{1-14}
\multirow{3}{*}{HybridRAG}
& Multi-hop  & 12.1 & 14.9 & 73.0 & 15.6 & 21.5 & 62.9 & 13.1 & 19.0 & 67.9 & 39.9 & 6.1 & 54.0 \\
& Single-hop & 79.9 & 2.3 & 17.8 & 70.3 & 12.1 & 17.6 & \textbf{63.8} & 20.8 & 15.4 & \textbf{55.7} & 11.2 & 33.1 \\
& Summary    & 13.9 & 48.6 & 37.5 & 2.5 & 68.6 & 29.0 & 24.2 & 27.0 & 48.8 & 33.9 & 10.4 & 55.7 \\
\cmidrule{1-14}
\multirow{3}{*}{IterativeRAG}
& Multi-hop  & 6.0 & 12.9 & 81.1 & 11.5 & 14.8 & 73.8 & 10.1 & 20.2 & 69.8 & 44.6 & 8.8 & 46.6 \\
& Single-hop & 12.3 & 32.9 & 54.8 & 62.3 & 16.1 & 21.6 & 48.6 & 25.7 & 25.7 & 47.4 & 20.5 & 32.1 \\
& Summary    & 6.8 & 44.8 & 48.4 & 1.4 & 61.5 & 37.1 & 12.6 & 27.3 & 60.1 & 27.3 & 12.5 & 60.2 \\
\bottomrule
\end{tabular}
\caption{Comprehensive LLM-as-a-Judge evaluation comparison between DeepSeek-V3 and Llama-3-8B across four datasets. Results report Accuracy (Cor), Incorrectness (Inc), and No-Answer rates (No-A).}
\label{tab:llm_eval_comparison}
\end{table*}

\section{Evaluation \& Analysis Details}
\label{sec:app_evaluation}

\begin{table*}[!htbp]
    \centering
    \small
    \setlength{\tabcolsep}{3.5pt}
    \caption{Multi-dimensional metrics for evaluating RAG generation quality.}
    \label{tab:eval_metrics}
    \begin{tabularx}{\textwidth}{l l l >{\raggedright\arraybackslash}X}
        \toprule
        \textbf{Category} & \textbf{Metric} & \textbf{Focus} & \textbf{Description \& Rationale} \\
        \midrule
        \multirow{3}{*}{\textbf{Answer Quality}} 
        & LLM-as-a-Judge & Answer Correctness & LLM classifies answers as \textit{correct}, \textit{incorrect}, or \textit{incomplete}, providing human-aligned judgment. \\
        & Semantic F1 & Reference Similarity & BERTScore-based token-level semantic similarity between prediction and ground truth, robust to paraphrase variations. \\
        & Soft Coverage & Completeness & Maximum cosine similarity between GT embedding and any prediction sentence, measuring information recall. \\
        \midrule
        \multirow{2}{*}{\textbf{Grounding}} 
        & Faithfulness (Hard) & Hallucination & Fraction of answer sentences with retrieval support above threshold ($\tau=0.7$), detecting unsupported claims. \\
        & Faithfulness (Soft) & Support Strength & Mean of max similarities between answer sentences and retrieval content, measuring grounding degree. \\
        \bottomrule
    \end{tabularx}
\end{table*}

\begin{table*}[!htbp]
\centering
\scriptsize 
\renewcommand{\arraystretch}{1.1} 
\setlength{\tabcolsep}{2.5pt}
\resizebox{\textwidth}{!}{%
\begin{tabular}{llccccc|ccccc}
\toprule
\multirow{2}{*}{\textbf{Dataset}} & \multirow{2}{*}{\textbf{Method}} &
\multicolumn{5}{c|}{\textbf{DeepSeek-V3}} &
\multicolumn{5}{c}{\textbf{Llama 3 8B}} \\
\cmidrule(lr){3-7}\cmidrule(lr){8-12}
& & \textbf{Sem-F1} & \textbf{COV} & \textbf{Faith-H} & \textbf{Faith-S} & \textbf{LLM-Cor\%}
& \textbf{Sem-F1} & \textbf{COV} & \textbf{Faith-H} & \textbf{Faith-S} & \textbf{LLM-Cor\%} \\
\midrule
\multirow{6}{*}{MuSiQue} 
& LLM-only & - & - & - & - & 0.0 & - & - & - & - & 0.0 \\
& NaiveRAG & 0.503 & 0.362 & 0.143 & 0.486 & 26.4 & 0.249 & 0.161 & 0.194 & 0.540 & 8.6 \\
& GraphRAG & 0.510 & 0.386 & 0.114 & 0.439 & 30.3 & 0.374 & 0.262 & 0.185 & 0.494 & 18.7 \\
& HybridRAG & 0.613 & 0.472 & - & - & 38.6 & 0.406 & 0.284 & 0.582 & 0.178 & 20.3 \\
& Iterative (Naive) & 0.469 & 0.320 & - & - & 20.4 & 0.289 & 0.169 & - & - & 6.8 \\
& Iterative (Graph) & 0.000 & 0.400 & - & - & 29.2 & 0.032 & 0.271 & - & - & 16.1 \\
\midrule
\multirow{6}{*}{QuALITY} 
& LLM-only & - & - & - & - & 0.0 & - & - & - & - & 0.0 \\
& NaiveRAG & 0.858 & 0.627 & 0.009 & 0.404 & 48.7 & 0.620 & 0.475 & 0.011 & 0.414 & 30.7 \\
& GraphRAG & 0.794 & 0.546 & 0.009 & 0.377 & 39.3 & 0.501 & 0.376 & 0.014 & 0.385 & 21.2 \\
& HybridRAG & 0.738 & 0.553 & 0.438 & 0.146 & 41.7 & 0.643 & 0.493 & 0.283 & 0.201 & 33.2 \\
& Iterative (Naive) & 0.724 & 0.506 & - & - & 35.8 & 0.583 & 0.435 & - & - & 28.4 \\
& Iterative (Graph) & 0.657 & 0.446 & - & - & 28.7 & 0.528 & 0.378 & - & - & 20.4 \\
\midrule
\multirow{6}{*}{Legal} 
& LLM-only & - & - & - & - & 0.0 & - & - & - & - & 0.0 \\
& NaiveRAG & 0.568 & 0.469 & 0.145 & 0.537 & 32.2 & 0.587 & 0.473 & 0.164 & 0.531 & 25.8 \\
& GraphRAG & 0.530 & 0.443 & 0.094 & 0.510 & 29.3 & 0.536 & 0.445 & 0.146 & 0.508 & 25.0 \\
& HybridRAG & 0.617 & 0.520 & 0.589 & 0.326 & 36.1 & 0.622 & 0.512 & 0.627 & 0.349 & 31.1 \\
& Iterative (Naive) & 0.572 & 0.466 & - & - & 28.5 & 0.577 & 0.453 & - & - & 22.0 \\
& Iterative (Graph) & 0.534 & 0.439 & - & - & 26.0 & 0.013 & 0.444 & - & - & 21.0 \\
\midrule
\multirow{6}{*}{Medical} 
& LLM-only & - & - & - & - & 0.0 & - & - & - & - & 0.0 \\
& NaiveRAG & 0.770 & 0.599 & 0.207 & 0.583 & 61.1 & 0.732 & 0.574 & 0.175 & 0.575 & 44.9 \\
& GraphRAG & 0.691 & 0.541 & 0.250 & 0.588 & 53.5 & 0.673 & 0.532 & 0.214 & 0.572 & 41.7 \\
& HybridRAG & 0.792 & 0.620 & 0.767 & 0.358 & 64.7 & 0.759 & 0.600 & 0.813 & 0.374 & 48.2 \\
& Iterative (Naive) & 0.826 & 0.595 & - & - & 62.7 & 0.802 & 0.582 & - & - & 43.6 \\
& Iterative (Graph) & 0.801 & 0.575 & - & - & 59.8 & 0.818 & 0.595 & - & - & 43.6 \\
\bottomrule
\end{tabular}%
}
\caption{Evaluation results on DeepSeek-V3 and Llama 3 8B across all datasets and RAG paradigms. Sem-F1: Semantic F1 (BERTScore-based), COV: Coverage, Faith-H/S: Faithfulness Hard/Soft, LLM-Cor\%: LLM-as-a-Judge correct rate. ``-'' indicates metric not applicable or not computed.}
\label{tab:model_results}
\end{table*}

\subsection{Metric Implementation}
\label{sec:app_eval_metrics}

\paragraph{Metric Categories.} 
To comprehensively assess the generation quality of RAG systems, we devise a multi-dimensional evaluation framework (Table~\ref{tab:eval_metrics}). The metrics are categorized into three distinct classes: (1) Answer Quality Metrics, which quantify the semantic similarity and informational completeness of the generated response relative to the gold standard; (2) Faithfulness Metrics, which verify whether the response is strictly grounded in the retrieved context, serving as a mechanism to identify hallucinations; and (3) LLM-as-a-Judge, which provides a holistic assessment of correctness that aligns with human judgment.

\paragraph{Answer Quality Metrics.}
\noindent Semantic F1 calculates the token-level semantic similarity between the generated response and the ground truth, derived from BERTScore. Let $\hat{y} = \{\hat{x}_1, \dots, \hat{x}_m\}$ denote the predicted answer and $y = \{x_1, \dots, x_n\}$ denote the reference answer. We first extract contextual embeddings utilizing a pre-trained language model, specifically \texttt{microsoft/deberta-xlarge-mnli}, and subsequently compute the precision ($P_{\text{BERT}}$), recall ($R_{\text{BERT}}$), and F1 score:

\begin{equation}
    P_{\text{BERT}} = \frac{1}{|\hat{y}|} \sum_{\hat{x}_i \in \hat{y}} \max_{x_j \in y} \cos(\mathbf{h}_{\hat{x}_i}, \mathbf{h}_{x_j})
\end{equation}

\begin{equation}
    R_{\text{BERT}} = \frac{1}{|y|} \sum_{x_j \in y} \max_{\hat{x}_i \in \hat{y}} \cos(\mathbf{h}_{\hat{x}_i}, \mathbf{h}_{x_j})
\end{equation}

\begin{equation}
    \text{Semantic F1} = 2 \cdot \frac{P_{\text{BERT}} \cdot R_{\text{BERT}}}{P_{\text{BERT}} + R_{\text{BERT}}}
\end{equation}

Here, $\mathbf{h}$ represents the contextual embedding vector of a token, and $\cos(\cdot, \cdot)$ signifies cosine similarity. This metric exhibits robustness against synonym substitution and paraphrastic variations.

\noindent Soft Coverage quantifies the extent to which the generated response encapsulates the information present in the gold standard. We utilize \texttt{SentenceTransformer} (\texttt{all-MiniLM-L6-v2}) to segment both the reference and the prediction into individual sentences and encode them into sentence-level embeddings. For each sentence $s_i^{gt}$ in the ground truth, we compute its maximum similarity with respect to all sentences in the prediction:

\begin{equation}
    \text{Coverage} = \frac{1}{|S^{gt}|} \sum_{s_i^{gt} \in S^{gt}} \max_{s_j^{pred} \in S^{pred}} \cos(\mathbf{e}_{s_i^{gt}}, \mathbf{e}_{s_j^{pred}})
\end{equation}

where $S^{gt}$ and $S^{pred}$ denote the sentence sets of the ground truth and the prediction, respectively, and $\mathbf{e}$ represents the sentence embedding. A higher coverage value indicates that the generated response has successfully captured a greater proportion of the critical information contained in the reference.

\paragraph{Faithfulness Metrics.}  
The Faithfulness metric evaluates whether the generated response is faithful to the retrieved context, serving as a primary mechanism for detecting hallucinations. We segment the generated response into individual sentences and calculate the semantic support for each sentence against the retrieved content.

\noindent Faithfulness (Hard) utilizes a strict threshold to determine the proportion of sentences in the answer that are supported by the retrieval:

\begin{equation}
    \text{Faith}_{\text{hard}} = \frac{1}{|S^{ans}|} \sum_{s_i \in S^{ans}} \mathbb{1}\left[\max_{c_j \in C} \cos(\mathbf{e}_{s_i}, \mathbf{e}_{c_j}) \geq \tau \right]
\end{equation}

where $S^{ans}$ denotes the set of answer sentences, $C$ denotes the set of retrieved context sentences, $\tau = 0.7$ serves as the similarity threshold, and $\mathbb{1}[\cdot]$ is the indicator function. This metric strictly quantifies the fraction of the response that possesses explicit grounding within the retrieved results.

\textit{\noindent\textbf{Faithfulness (Soft).}} employs a continuous calculation to measure the average support strength between the answer and the retrieved content:

\begin{equation}
    \text{Faith}_{\text{soft}} = \frac{1}{|S^{ans}|} \sum_{s_i \in S^{ans}} \max_{c_j \in C} \cos(\mathbf{e}_{s_i}, \mathbf{e}_{c_j})
\end{equation}

The Soft version provides a more granular measure of grounding, reflecting partial support even when the strict threshold is not met.

It is important to note that Faithfulness metrics are calculated exclusively for NaiveRAG and GraphRAG. We exclude HybridRAG and IterativeRAG from this specific evaluation, as their complex retrieval formats, involving multi-turn interactions or hybrid sources, may introduce bias into the direct calculation.

\paragraph{LLM-as-a-Judge.}
Complementing automated metrics, we employ the LLM-as-a-Judge methodology to conduct human-aligned correctness evaluation. As illustrated in Figure~\ref{fig:llm_as_judge}, we devise a structured evaluation prompt instructing the evaluator model (GPT-4o-mini) to classify generated responses into three mutually exclusive categories. (1) Correct: The response is logically accurate and encapsulates the core information of the ground truth. (2) Incorrect: The response contains erroneous information that contradicts the ground truth. (3)Incomplete: The response is partially correct yet lacks critical details, or the model refuses to generate an answer.

This evaluation paradigm mitigates the limitations of similarity-based metrics, which often fail to capture logical inconsistencies, thereby providing a quality assessment that more closely aligns with human judgment.

\paragraph{Evaluation Results.}
Table~\ref{tab:model_results} presents the comprehensive evaluation results for DeepSeek-V3 and Llama 3 8B across the four constituent datasets. The key findings are summarized as follows:

\textit{\noindent\textbf{Overall Performance:}} HybridRAG achieves optimal or near-optimal performance across the majority of datasets, particularly excelling in Semantic F1 and Coverage metrics. For instance, on the Medical dataset, DeepSeek-V3 combined with HybridRAG attains an LLM accuracy of 64.7\%, surpassing both NaiveRAG (61.1\%) and GraphRAG (53.5\%).

\textit{\noindent\textbf{Model Disparity:}} DeepSeek-V3 significantly outperforms Llama 3 8B. Taking the MuSiQue dataset as an example, DeepSeek-V3 with HybridRAG achieves an LLM accuracy of 38.6\%, whereas Llama 3 8B reaches only 20.3\%, a substantial performance gap of 18.3 percentage points.

\textit{\noindent\textbf{Faithfulness Analysis:}} Faithfulness scores for NaiveRAG and GraphRAG are generally low (mostly below 0.25), indicating that even with retrieval augmentation, models continue to generate content unsupported by the retrieved context. Conversely, the Medical dataset exhibits relatively higher faithfulness (GraphRAG reaches 0.250 Hard / 0.588 Soft), potentially attributable to the medical domain's strict reliance on retrieved factual evidence for answer formulation.

Table~\ref{tab:llm_eval_comparison} further provides a breakdown of LLM evaluation results by query type. Key observations include:

\textit{\noindent\textbf{Factual Dominance:}} All methods yield their best performance on single-hop queries. GraphRAG achieves accuracies of 90.2\% (DeepSeek) and 84.4\% (Llama) on MuSiQue's single-hop questions, significantly outperforming other query types. This aligns with GraphRAG's entity-based retrieval mechanism, where single-entity queries facilitate the precise localization of relevant information.

\textit{\noindent\textbf{Reasoning Challenge:}} Multi-hop queries prove universally challenging across all methods, with accuracy generally remaining below 35\%. The Legal dataset is particularly demanding, where the best-performing method (IterativeRAG) attains only 12.4\% accuracy on multi-hop tasks.

\textit{\noindent\textbf{Summary Dilemma:}} Summary-type queries exhibit a high rate of incorrect responses (exceeding 30\% for most methods on QuALITY), suggesting a model tendency to generate summaries that are either over-generalized or deviate from the source text.

\subsection{Cost Calculation}
\label{sec:app_cost}

\begin{table*}[!htbp]
\centering
\centering
\scriptsize 
\renewcommand{\arraystretch}{1.1}
\setlength{\tabcolsep}{6pt}
\begin{tabular}{llrrrrrrrr}
\toprule
\textbf{Dataset} & \textbf{Method} & \textbf{N} & \textbf{Avg-Ctx} & \textbf{Ret-In} & \textbf{Ret-Out} & \textbf{Gen-In} & \textbf{Gen-Out} & \textbf{Total} \\
\midrule
\multirow{5}{*}{MuSiQue}
& LLM-only & 3,356 & 0 & 0 & 0 & 0.07 & 0.03 & 0.10 \\
& NaiveRAG & 3,356 & 13,139 & 0 & 0 & 44.17 & 0.05 & 44.22 \\
& GraphRAG & 3,356 & 8,472 & 2.35 & 0.24 & 28.51 & 0.04 & 31.14 \\
& HybridRAG & 3,356 & 21,602 & 2.35 & 0.24 & 72.57 & 0.06 & 75.22 \\
& IterativeRAG & 3,357 & 6,364 & 43.24 & 0.72 & 21.44 & 0.05 & 65.46 \\
\midrule
\multirow{5}{*}{QuALITY}
& LLM-only & 1,198 & 0 & 0 & 0 & 0.03 & 0.00 & 0.04 \\
& NaiveRAG & 1,198 & 49,444 & 0 & 0 & 59.27 & 0.04 & 59.31 \\
& GraphRAG & 1,198 & 48,502 & 1.56 & 0.07 & 58.14 & 0.04 & 59.81 \\
& HybridRAG & 1,198 & 97,789 & 1.56 & 0.07 & 117.19 & 0.03 & 118.85 \\
& IterativeRAG & 1,198 & 6,871 & 16.69 & 0.25 & 8.27 & 0.03 & 25.24 \\
\midrule
\multirow{5}{*}{Legal}
& LLM-only & 1,277 & 0 & 0 & 0 & 0.05 & 0.01 & 0.06 \\
& NaiveRAG & 1,277 & 46,273 & 0 & 0 & 59.14 & 0.05 & 59.19 \\
& GraphRAG & 1,277 & 179,728 & 4.82 & 0.13 & 229.56 & 0.04 & 234.55 \\
& HybridRAG & 1,277 & 225,571 & 4.82 & 0.13 & 288.10 & 0.05 & 293.10 \\
& IterativeRAG & 1,278 & 6,460 & 16.81 & 0.29 & 8.30 & 0.05 & 25.45 \\
\midrule
\multirow{5}{*}{Medical}
& LLM-only & 1,896 & 0 & 0 & 0 & 0.03 & 0.06 & 0.09 \\
& NaiveRAG & 1,896 & 50,504 & 0 & 0 & 95.78 & 0.08 & 95.86 \\
& GraphRAG & 1,896 & 37,513 & 0.25 & 0.15 & 71.15 & 0.07 & 71.63 \\
& HybridRAG & 1,896 & 73,628 & 0.25 & 0.15 & 139.63 & 0.08 & 140.11 \\
& IterativeRAG & 1,897 & 2,231 & 8.67 & 0.26 & 4.26 & 0.08 & 13.27 \\
\bottomrule
\end{tabular}
\caption{Token consumption breakdown for retrieval and generation across all datasets and methods. All token counts are in millions (M) except Avg-Ctx (average context tokens per question). Ret: Retrieval, Gen: Generation, In: Input, Out: Output.}
\label{tab:cost_analysis}
\end{table*}

\paragraph{Cost Components.}
We decompose the computational overhead of the RAG system into two primary phases (Table~\ref{tab:cost_analysis}): the Retrieval phase and the Generation phase. The total cost is formally defined as:
\begin{equation}
    C_{\text{total}} = C_{\text{retrieval}} + C_{\text{generation}}
\end{equation}
where the cost for each phase comprises both input and output token consumption:
\begin{equation}
    C_{\text{retrieval}} = T_{\text{ret}}^{\text{in}} + T_{\text{ret}}^{\text{out}}
\end{equation}
\begin{equation}
    C_{\text{generation}} = T_{\text{gen}}^{\text{in}} + T_{\text{gen}}^{\text{out}}
\end{equation}
In the generation phase, the input token volume is predominantly governed by the aggregate context length:
\begin{equation}
    T_{\text{gen}}^{\text{in}} \approx N \times (L_{\text{prompt}} + L_{\text{context}} + L_{\text{query}})
\end{equation}
where $N$ denotes the total number of queries, $L_{\text{prompt}}$ represents the fixed length of the system prompt, $L_{\text{context}}$ is the average length of the retrieved context (denoted as \textit{Avg-Ctx} in the table), and $L_{\text{query}}$ is the query length.

For GraphRAG and HybridRAG, the retrieval cost incorporates a one-time graph construction overhead, which is amortized over the query set:
\begin{equation}
    C_{\text{retrieval}}^{\text{graph}} = \underbrace{C_{\text{construction}}}_{\text{one-time, amortized}} + \underbrace{C_{\text{entity\_extraction}}}_{\text{per-query}}
\end{equation}
The construction cost encompasses the input tokens required for processing the raw corpus via the LLM ($T_{\text{corpus}}$) and the resulting output tokens for the extracted triplets ($T_{\text{triplets}}$).

\paragraph{Method Comparison.}
Table~\ref{tab:cost_analysis} reveals substantial disparities in computational cost across the five distinct methodologies:

\textit{\noindent\textbf{LLM-only:}} It incurs minimal overhead (0.04--0.10M tokens), as it bypasses retrieval and context injection, consuming tokens solely for the prompt, query input, and response generation.

\textit{\noindent\textbf{NaiveRAG:}} Its cost footprint is predominantly driven by the generation phase ($T_{\text{gen}}^{\text{in}}$ accounts for $>99\%$), necessitated by the inclusion of extensive retrieved chunks within the context window. For instance, on the Medical dataset, the average context length reaches 50,504 tokens, resulting in a total expenditure of 95.86M tokens.

\textit{\noindent\textbf{GraphRAG:}} It generally exhibits lower generation costs compared to NaiveRAG, as graph-based retrieval yields more precise and concise contexts. On MuSiQue, GraphRAG records an Avg-Ctx of 8,472 (vs. 13,139 for NaiveRAG), translating to a total cost of 31.14M (vs. 44.22M), a reduction of approximately 30\%. However, the Legal dataset presents an exception; here, GraphRAG's Avg-Ctx surges to 179,728. This anomaly arises from the dense entity interconnectivity characteristic of legal documents, where graph traversal retrieves a voluminous amount of associated content.

\textit{\noindent\textbf{HybridRAG:}} It incurs the highest computational burden, as it utilizes results from both vector and graph retrieval, resulting in a context length approximating the sum of both. On the Legal dataset, it peaks at 293.10M tokens, marking the maximum consumption across all evaluated methods.

\textit{\noindent\textbf{IterativeRAG:}} It exhibits a distinct cost structure characterized by high retrieval overhead (due to multi-turn LLM invocations for judgment and sub-query generation) but low generation cost (owing to the refined conciseness of the context). On the Medical dataset, despite a substantial $T_{\text{ret}}^{\text{in}}$ of 8.67M, the Avg-Ctx remains merely 2,231, yielding a total cost of 13.27M, the lowest among all RAG paradigms.

\paragraph{Dataset Variation.}
The cost variations across datasets are predominantly governed by document length and corpus scale:

\textit{\noindent\textbf{The Legal dataset:}} It incurs significantly higher costs compared to other benchmarks. Specifically, GraphRAG's total expenditure on Legal (234.55M) is 7.5 times that on MuSiQue (31.14M). This disparity stems from the extensive length and complex entity interrelations inherent in legal documents, which result in: (1) elevated graph construction overhead; and (2) substantially longer contexts yielded by graph traversal (with Avg-Ctx reaching 179,728).

\textit{\noindent\textbf{The Narrative dataset:}} It's costs of NaiveRAG and GraphRAG are comparable (59.31M vs. 59.81M), indicating that graph-based retrieval fails to effectively reduce context length in this setting. This is attributable to QuALITY's long-document characteristic, where each question corresponds to a complete article, resulting in high inter-chunk correlation.

\textit{\noindent\textbf{The Medical dataset:}} Its IterativeRAG demonstrates superior cost-efficiency (13.27M), amounting to merely 14\% of the cost incurred by NaiveRAG (95.86M). Medical QA typically involves explicit information needs, allowing iterative retrieval to rapidly localize critical content.

\paragraph{Cost-Performance Trade-off.}
Synthesizing the cost profiles in Table~\ref{tab:cost_analysis} with the performance metrics in Table~\ref{tab:model_results}, we analyze the cost-performance trade-offs:

\textit{\noindent\textbf{HybridRAG: }}High Cost, High Performance. On MuSiQue, HybridRAG attains an LLM accuracy of 38.6\% at a cost of 75.22M tokens. Compared to NaiveRAG (44.22M, 26.4\%), this represents a 70\% cost increase yielding a 46\% performance gain. Conversely, on \textbf{Medical}, HybridRAG (140.11M, 64.7\%) incurs a 46\% cost hike over NaiveRAG (95.86M, 61.1\%) for a mere 6\% performance improvement, indicating diminishing marginal returns.

\textit{\noindent\textbf{GraphRAG: }}Dataset-Dependent Cost-Efficiency. On MuSiQue, GraphRAG (31.14M, 30.3\%) delivers superior performance at a lower cost than NaiveRAG, emerging as the optimal choice. However, on QuALITY, GraphRAG (59.81M, 39.3\%) offers no advantage, incurring costs comparable to NaiveRAG while yielding inferior performance (48.7\%).

\textit{\noindent\textbf{IterativeRAG: }}Low Cost, Variable Performance. On Medical, IterativeRAG (Naive-base) achieves the highest cost-effectiveness, reaching 62.7\% accuracy at a minimal cost of 13.27M. Yet, on MuSiQue, the same approach yields only 20.4\% accuracy, underperforming other more resource-intensive methods.

\textit{\noindent\textbf{Practical Recommendations:}} (1) For domains with explicit entity relations (e.g., Medical), IterativeRAG offers the best cost-performance ratio; (2) For complex QA requiring the synthesis of multi-source information, HybridRAG delivers optimal performance despite its high cost; and (3) For long-document comprehension tasks (e.g., QuALITY), NaiveRAG remains the most straightforward and effective solution.

\subsection{Case Studies}
\label{sec:app_cases}

\begin{table*}[!htbp]
\centering
\small 
\renewcommand{\arraystretch}{1.3}
\setlength{\tabcolsep}{4pt} 

\begin{tabularx}{\textwidth}{p{0.08\textwidth} p{0.22\textwidth} X p{0.22\textwidth}}
\toprule
\textbf{Case ID} & \textbf{Query \& Gold Standard} & \textbf{Paradigm Comparison} & \textbf{Key Analysis} \\
\midrule

\textbf{Case 1}\newline
\textit{Medical 1244}\newline
(Multi-hop) 
& 
\textbf{Q:} What are the surgical options for early cervical cancer and how do they relate to fertility preservation? \par\medskip
\textbf{Gold:} Cone biopsy and trachelectomy are surgical options for early-stage disease, with trachelectomy being a fertility-sparing procedure.
& 
\textbf{GraphRAG (\textcolor{teal}{Correct}):} Explicitly states that ``fertility-sparing surgical options include cone biopsy or radical trachelectomy,'' accurately capturing both procedures. \par
\textbf{NaiveRAG (\textcolor{orange}{Incomplete}):} Mentions cone biopsy but fails to explicitly cite ``trachelectomy,'' offering generic ``hysterectomy types.'' \par
\textbf{HybridRAG (\textcolor{orange}{Incomplete}):} Similarly omits ``trachelectomy,'' referencing only distinct hysterectomy types. \par
\textbf{IterativeRAG (\textcolor{teal}{Correct}):} Successfully localizes ``cone biopsy, trachelectomy'' through iterative retrieval, fully covering the gold answer.
& 
This case demonstrates that for multi-hop queries necessitating precise medical terminology, GraphRAG's entity-oriented retrieval and IterativeRAG's multi-turn refinement are effective. In contrast, semantic-based methods (Naive/Hybrid) lack the required precision. \\

\midrule

\textbf{Case 2}\newline
\textit{QuALITY 1034}\newline
(Summary) 
& 
\textbf{Q:} Based on the narrative, what are the common types of extreme adversity faced by spacecraft? What are the crew's survival strategies? \par\medskip
\textbf{Gold:} Scenarios include physical trauma (crash/flip), atmospheric entry, and combat. Reactions involve assessing reparability or risking hyperdrive escape.
& 
\textbf{GraphRAG (\textcolor{red}{Incorrect}):} Hallucinates concepts like ``transphasia'' and ``space cafard'' absent from the source text. \par
\textbf{NaiveRAG (\textcolor{red}{Incorrect}):} Lists generic tropes like ``sabotage, alien attacks'' without addressing specific scenarios. \par
\textbf{HybridRAG (\textcolor{red}{Incorrect}):} Erroneously outputs ``James I,'' suggesting irrelevant retrieval. \par
\textbf{IterativeRAG (\textcolor{orange}{Incomplete}):} Closest result, citing ``mechanical failures'' and ``damaged hulls,'' capturing the thematic direction but missing details.
& 
This highlights the challenge of Summary-type queries requiring cross-document synthesis. Even IterativeRAG only achieves Incomplete status, indicating significant room for improvement in long-document summarization tasks. \\

\bottomrule
\end{tabularx}
\caption{\label{tab:case_studies} Qualitative analysis of two representative cases.}
\end{table*}

\paragraph{Comparative Analysis.}
Table \ref{tab:case_studies} presents qualitative analysis of representative cases, comparing paradigm performance on multi-hop reasoning and cross-document summarization tasks.

\paragraph{Error Analysis.}
Drawing upon the aforementioned case studies and the comprehensive evaluation results, we categorize the primary failure modes as follows:

\textit{\noindent\textbf{Retrieval Imprecision:}}
The failure of NaiveRAG and HybridRAG in Case 1 stems from the fact that while the retrieved content possessed thematic relevance (cervical cancer surgery), it failed to precisely hit the pivotal term ``trachelectomy.'' This error is particularly prevalent in specialized domains (e.g., Medical, Legal), where the semantic similarity of domain-specific terminology may be lower than that of generic descriptions. GraphRAG, through its entity matching mechanism, demonstrates superior precision in localizing such specialized terms.

\textit{\noindent\textbf{Context Overload:}}
When retrieval yields a high volume of relevant yet redundant content, the LLM is prone to overlooking critical information. In Case 1, HybridRAG's average context length (Avg-Ctx) reached 73,628 tokens. Such an excessively long context can cause the model to become ``lost in the middle,'' resulting in performance inferior to that of GraphRAG, which utilizes a more concise context (Avg-Ctx: 37,513).

\textit{\noindent\textbf{Hallucination:}}
In Case 2, GraphRAG generated concepts absent from the source documents (e.g., ``transphasia''). This indicates that when the alignment between retrieved content and the query is poor, the model may resort to its internal parametric knowledge to ``complete'' the answer, leading to fabrication. This issue is particularly pronounced in Summary-type queries, as summarization necessitates cross-document synthesis which fragmented retrieval results often fail to fully cover.

\textit{\noindent\textbf{Retrieval Mismatch:}}
In Case 2, HybridRAG output ``James I'', an answer entirely unrelated to spacecraft, suggesting that the retrieval module returned completely irrelevant document fragments. Such severe mismatches likely stem from a semantic gap between the query and the corpus, or from errors in entity extraction within the graph retrieval process.

\textit{\noindent\textbf{Reasoning Chain Failure:}}
For questions necessitating multi-step reasoning (e.g., the multi-hop query in Case 1), even if partially relevant content is retrieved, the model may fail to correctly link the information. IterativeRAG, via its multi-turn ``Retrieve-Generate-Evaluate'' loop, effectively patches the reasoning chain step-by-step, thereby exhibiting relatively stable performance on complex queries.

\section{Router Evaluation Details}
\label{sec:app_router}

\subsection{Router Architecture Details}
\label{sec:app_router_arch}

Table~\ref{tab:router_arch} summarizes the architecture and key hyperparameters of all nine routers evaluated in this work.

\begin{table}[H]
\centering
\small
\renewcommand{\arraystretch}{1.1}
\begin{tabular}{l l l}
\toprule
\textbf{Router} & \textbf{Category} & \textbf{Key Configuration} \\
\midrule
LR & Traditional ML & L2 regularization \\
SVM & Traditional ML & RBF kernel, probability=True \\
XGBoost & Traditional ML & 100 estimators, max\_depth=6 \\
MLP & Traditional ML & Hidden layers: 64, 32 \\
\midrule
DNN & Neural & 3-layer MLP (256→128→5) with BN \\
TwoTower & Neural & Query tower + paradigm embedding \\
Adaptive-RAG & Neural & LM classifier (Q+C adapted) \\
MBA-RAG & Neural & Multi-armed bandit with reward MLP \\
RouterDC & Neural & Dual contrastive learning \\
\bottomrule
\end{tabular}
\caption{Router architecture summary. All neural routers use dimension-aligned feature processing: query text is encoded via a pretrained language model and projected to 64 dimensions, corpus indicators are discretized into bins and embedded into 64 dimensions, and the resulting 128-dimensional vector serves as input.}
\label{tab:router_arch}
\end{table}

\subsection{Feature Processing}
\label{sec:app_router_feat}

For traditional ML routers, query features are encoded as a 3-dimensional one-hot vector of query type (factual, reasoning, summary), and corpus features as 6 dual-view indicators (LCC Ratio, Density, Clustering Coefficient, Intrinsic Dimension, Hubness, Average Distance). The Q+C input is their concatenation (9 dimensions).

For neural routers, query text is encoded by a pretrained language model (BERT or BGE) and projected to 64 dimensions via a linear layer. Corpus indicators are discretized into 10 bins per feature and each bin is mapped to a learnable embedding, which are concatenated and projected to 64 dimensions. This ensures balanced dimensionality between the two signal sources before concatenation into a 128-dimensional input vector.

\subsection{Evaluation Metrics}
\label{sec:app_router_metrics}

We report four complementary metrics. \textbf{Routing Accuracy (R.A)} measures the proportion of queries for which the router selects the optimal paradigm matching the ground-truth label. \textbf{Macro F1} computes the unweighted average of per-class F1 scores across all six classes, capturing balanced performance across imbalanced class distributions. \textbf{Mean Reciprocal Rank (MRR)} evaluates the ranking quality of predicted paradigm probabilities, computed as the average of $1/\text{rank}$ of the ground-truth paradigm across all queries. \textbf{Correct Rate (C.R)} measures the proportion of queries for which the paradigm selected by the router is judged as correct by LLM-as-a-Judge, where queries predicted as ``cannot answer'' fall back to the globally most frequent optimal paradigm.

\subsection{Complete Results}
\label{sec:app_router_results}

Table~\ref{tab:router_full} presents the comprehensive router evaluation across all architectures, feature configurations (query-only, corpus-only, and Q+C), and datasets. Results are reported for both BERT and BGE encoders. The main text reports BERT encoder results; BGE results serve as a robustness check confirming that the Q+C advantage holds across different pretrained representations.

\begin{table*}[t]
\centering
\renewcommand{\arraystretch}{1.05}
\caption{\textbf{Complete router evaluation across all architectures, feature configurations, and datasets.} Results are reported for both BERT and BGE encoders under query-only (Q), corpus-only (C), and joint (Q+C) feature settings. R.A: routing accuracy (\%), F1: macro F1 (\%), MRR: mean reciprocal rank, C.R: correct rate (\%). The main text reports BERT results; BGE results serve as a robustness check.}
\resizebox{\textwidth}{!}{%
\begin{tabular}{ll | cccc | cccc | cccc | cccc | cccc}
\toprule
\multirow{2}{*}{\textbf{Method}} & \multirow{2}{*}{\textbf{Feat.}} & 
\multicolumn{4}{c|}{\textbf{Avg}} & \multicolumn{4}{c|}{\textbf{MuSi.}} & \multicolumn{4}{c|}{\textbf{QuAL.}} & \multicolumn{4}{c|}{\textbf{Legal}} & \multicolumn{4}{c}{\textbf{Med.}} \\
\cmidrule(lr){3-6}
\cmidrule(lr){7-10}
\cmidrule(lr){11-14}
\cmidrule(lr){15-18}
\cmidrule(lr){19-22}
 &  & R.A & F1 & MRR & C.R & R.A & F1 & MRR & C.R & R.A & F1 & MRR & C.R & R.A & F1 & MRR & C.R & R.A & F1 & MRR & C.R \\
\midrule
\rowcolor{modeldeepseek}
\multicolumn{22}{c}{\textit{\textbf{Model: DeepSeek-V3}}} \\
\midrule
\rowcolor{gray!10} Random & -- & 20.1 & 18.9 & -- & 39.1 & 20.0 & -- & -- & 28.4 & 20.0 & -- & -- & 42.8 & 20.1 & -- & -- & 31.6 & 20.2 & -- & -- & 60.6 \\
\rowcolor{gray!10} Best-fixed & -- & 20.9 & 6.9 & -- & 39.3 & 15.5 & -- & -- & 26.4 & 30.1 & -- & -- & 48.8 & 18.7 & -- & -- & 32.2 & 26.3 & -- & -- & 61.1 \\
\midrule
\multirow{3}{*}{LR} & Q & 43.9 & 19.8 & 0.6374 & 39.3 & 48.4 & 15.2 & 0.6769 & 26.4 & 55.1 & 25.1 & 0.7152 & 48.9 & 53.1 & 22.4 & 0.6974 & 32.2 & 22.6 & 12.1 & 0.4775 & 61.1 \\
 & C & 41.9 & 18.0 & 0.6236 & 39.3 & 48.4 & 13.0 & 0.6680 & 26.4 & 39.9 & 11.4 & 0.6323 & 48.9 & 49.5 & 13.2 & 0.6762 & 32.2 & 26.3 & 8.3 & 0.5037 & 61.1 \\
 & Q+C & 46.8 & 26.3 & 0.6502 & 43.5 & 55.3 & 24.3 & 0.7106 & 35.8 & 55.1 & 25.1 & 0.7104 & 48.9 & 52.3 & 21.9 & 0.6945 & 32.8 & 22.6 & 12.1 & 0.4749 & 61.1 \\
\midrule
\multirow{3}{*}{SVM} & Q & 43.9 & 19.8 & 0.6374 & 39.3 & 48.4 & 15.2 & 0.6769 & 26.4 & 55.1 & 25.1 & 0.7152 & 48.9 & 53.1 & 22.4 & 0.6974 & 32.2 & 22.6 & 12.1 & 0.4775 & 61.1 \\
 & C & 41.9 & 18.0 & 0.6218 & 39.3 & 48.4 & 13.0 & 0.6638 & 26.4 & 39.9 & 11.4 & 0.6323 & 48.9 & 49.5 & 13.2 & 0.6762 & 32.2 & 26.3 & 8.3 & 0.5040 & 61.1 \\
 & Q+C & 47.6 & 28.0 & 0.6598 & 43.2 & 55.3 & 24.3 & 0.7135 & 35.8 & 55.1 & 25.1 & 0.7174 & 48.9 & 53.1 & 22.4 & 0.6993 & 32.2 & 25.4 & 11.2 & 0.5016 & 60.2 \\
\midrule
\multirow{3}{*}{XGBoost} & Q & 43.9 & 19.8 & 0.6374 & 39.3 & 48.4 & 15.2 & 0.6769 & 26.4 & 55.1 & 25.1 & 0.7152 & 48.9 & 53.1 & 22.4 & 0.6974 & 32.2 & 22.6 & 12.1 & 0.4775 & 61.1 \\
 & C & 41.9 & 18.0 & 0.6236 & 39.3 & 48.4 & 13.0 & 0.6680 & 26.4 & 39.9 & 11.4 & 0.6323 & 48.9 & 49.5 & 13.2 & 0.6762 & 32.2 & 26.3 & 8.3 & 0.5037 & 61.1 \\
 & Q+C & 47.6 & 28.0 & 0.6595 & 43.2 & 55.3 & 24.3 & 0.7142 & 35.8 & 55.1 & 25.1 & 0.7181 & 48.9 & 53.1 & 22.4 & 0.6999 & 32.2 & 25.4 & 11.2 & 0.4985 & 60.2 \\
\midrule
\multirow{3}{*}{MLP} & Q & 43.7 & 19.7 & 0.6305 & 39.9 & 49.8 & 17.2 & 0.6764 & 28.0 & 52.0 & 23.6 & 0.6904 & 48.5 & 53.0 & 22.4 & 0.6932 & 32.9 & 21.4 & 11.5 & 0.4693 & 60.5 \\
 & C & 41.9 & 18.0 & 0.6196 & 39.3 & 48.4 & 13.0 & 0.6638 & 26.4 & 39.9 & 11.4 & 0.6286 & 48.9 & 49.5 & 13.2 & 0.6762 & 32.2 & 26.3 & 8.3 & 0.4971 & 61.1 \\
 & Q+C & 47.2 & 27.6 & 0.6518 & 43.2 & 55.3 & 24.3 & 0.7066 & 35.8 & 55.1 & 25.1 & 0.7099 & 48.9 & 51.7 & 20.1 & 0.6902 & 32.1 & 24.8 & 10.3 & 0.4917 & 60.6 \\
\midrule
\multirow{3}{*}{\shortstack[l]{DNN\\(BERT)}} & Q & 44.2 & 31.9 & 0.6361 & 43.1 & 51.6 & 34.5 & 0.6912 & 35.2 & 45.7 & 25.5 & 0.6523 & 46.9 & 51.2 & 25.8 & 0.6767 & 33.7 & 25.3 & 22.7 & 0.5007 & 61.2 \\
 & C & 41.9 & 18.0 & 0.6238 & 39.3 & 48.4 & 13.0 & 0.6680 & 26.4 & 39.9 & 11.4 & 0.6320 & 48.9 & 49.5 & 13.2 & 0.6762 & 32.2 & 26.3 & 8.3 & 0.5048 & 61.1 \\
 & Q+C & 45.5 & 32.1 & 0.6452 & 43.4 & 53.4 & 33.6 & 0.6994 & 35.5 & 48.3 & 24.8 & 0.6729 & 48.0 & 52.8 & 26.6 & 0.6915 & 33.9 & 24.7 & 20.1 & 0.4997 & 60.9 \\
\multirow{3}{*}{\shortstack[l]{TwoTower\\(BERT)}} & Q & 44.0 & 30.4 & 0.6370 & 43.1 & 50.5 & 29.1 & 0.6833 & 33.9 & 44.8 & 23.0 & 0.6521 & 48.1 & 52.9 & 26.8 & 0.6949 & 34.0 & 25.6 & 21.5 & 0.5056 & 62.5 \\
 & C & 41.9 & 18.0 & 0.6234 & 39.3 & 48.4 & 13.0 & 0.6680 & 26.4 & 39.9 & 11.4 & 0.6323 & 48.9 & 49.5 & 13.2 & 0.6762 & 32.2 & 26.3 & 8.3 & 0.5028 & 61.1 \\
 & Q+C & 45.0 & 30.1 & 0.6424 & 42.8 & 51.4 & 28.3 & 0.6903 & 33.7 & 49.1 & 24.5 & 0.6788 & 48.2 & 53.3 & 26.1 & 0.6951 & 33.6 & 25.2 & 20.5 & 0.4992 & 61.8 \\
\multirow{3}{*}{\shortstack[l]{Adaptive-RAG\\(BERT)}} & Q & 45.0 & 30.5 & 0.6432 & 43.1 & 51.5 & 30.5 & 0.6927 & 33.8 & 47.4 & 24.0 & 0.6707 & 48.1 & 53.3 & 26.2 & 0.6951 & 34.1 & 26.1 & 21.7 & 0.5032 & 62.6 \\
 & C & 41.9 & 18.0 & 0.6237 & 39.3 & 48.4 & 13.0 & 0.6680 & 26.4 & 39.9 & 11.4 & 0.6323 & 48.9 & 49.5 & 13.2 & 0.6762 & 32.2 & 26.3 & 8.3 & 0.5040 & 61.1 \\
 & Q+C & 46.2 & 28.2 & 0.6523 & 42.8 & 53.0 & 26.7 & 0.7006 & 34.3 & 51.9 & 24.8 & 0.6975 & 49.0 & 53.1 & 25.0 & 0.6981 & 33.4 & 25.9 & 14.8 & 0.5070 & 60.6 \\
\multirow{3}{*}{\shortstack[l]{MBA-RAG\\(BERT)}} & Q & 44.1 & 29.5 & 0.6290 & 42.6 & 50.8 & 27.5 & 0.6766 & 33.5 & 45.6 & 22.9 & 0.6440 & 46.8 & 53.0 & 26.8 & 0.6886 & 34.1 & 25.4 & 19.3 & 0.4950 & 61.9 \\
 & C & 41.9 & 18.0 & 0.6194 & 39.3 & 48.4 & 13.0 & 0.6635 & 26.4 & 39.9 & 11.4 & 0.6299 & 48.9 & 49.5 & 13.2 & 0.6744 & 32.2 & 26.3 & 8.3 & 0.4976 & 61.1 \\
 & Q+C & 45.2 & 26.6 & 0.6405 & 42.6 & 51.5 & 25.1 & 0.6848 & 33.7 & 50.2 & 23.4 & 0.6834 & 48.8 & 52.5 & 22.6 & 0.6867 & 32.6 & 25.9 & 13.0 & 0.5034 & 61.3 \\
\multirow{3}{*}{\shortstack[l]{RouterDC\\(BERT)}} & Q & 43.3 & 30.8 & 0.6309 & 42.7 & 50.9 & 33.4 & 0.6845 & 34.9 & 42.9 & 24.1 & 0.6364 & 47.1 & 51.3 & 25.1 & 0.6816 & 33.1 & 24.8 & 20.0 & 0.4984 & 60.1 \\
 & C & 41.5 & 17.8 & 0.6218 & 39.0 & 48.4 & 13.0 & 0.6667 & 26.4 & 39.9 & 11.4 & 0.6322 & 48.9 & 49.5 & 13.2 & 0.6757 & 32.2 & 25.0 & 8.0 & 0.4988 & 59.6 \\
 & Q+C & 46.0 & 33.5 & 0.6484 & 43.6 & 52.6 & 35.0 & 0.6961 & 35.5 & 51.6 & 27.9 & 0.6926 & 48.6 & 53.2 & 26.8 & 0.6951 & 33.8 & 25.7 & 19.7 & 0.5046 & 61.7 \\
\midrule
\multirow{3}{*}{\shortstack[l]{DNN\\(BGE)}} & Q & 42.2 & 32.2 & 0.6238 & 42.9 & 50.1 & 34.0 & 0.6788 & 34.0 & 40.9 & 23.3 & 0.6200 & 47.1 & 49.3 & 28.7 & 0.6693 & 33.4 & 24.2 & 23.6 & 0.4968 & 62.5 \\
 & C & 41.9 & 18.0 & 0.6233 & 39.3 & 48.4 & 13.0 & 0.6680 & 26.4 & 39.9 & 11.4 & 0.6323 & 48.9 & 49.5 & 13.2 & 0.6762 & 32.2 & 26.3 & 8.3 & 0.5025 & 61.1 \\
 & Q+C & 44.0 & 33.7 & 0.6359 & 43.3 & 51.0 & 35.1 & 0.6864 & 35.0 & 44.8 & 24.9 & 0.6424 & 46.9 & 51.0 & 27.5 & 0.6820 & 33.0 & 26.1 & 24.8 & 0.5101 & 62.9 \\
\multirow{3}{*}{\shortstack[l]{TwoTower\\(BGE)}} & Q & 42.2 & 28.2 & 0.6216 & 41.7 & 49.4 & 27.4 & 0.6732 & 31.9 & 44.0 & 22.8 & 0.6425 & 48.3 & 51.0 & 25.7 & 0.6770 & 32.9 & 22.3 & 18.5 & 0.4794 & 60.8 \\
 & C & 41.9 & 18.0 & 0.6235 & 39.3 & 48.4 & 13.0 & 0.6680 & 26.4 & 39.9 & 11.4 & 0.6323 & 48.9 & 49.5 & 13.2 & 0.6762 & 32.2 & 26.3 & 8.3 & 0.5034 & 61.1 \\
 & Q+C & 43.3 & 30.3 & 0.6299 & 42.2 & 49.7 & 29.0 & 0.6763 & 32.3 & 46.0 & 23.5 & 0.6559 & 48.3 & 52.5 & 25.6 & 0.6887 & 33.2 & 24.0 & 20.1 & 0.4911 & 62.2 \\
\multirow{3}{*}{\shortstack[l]{Adaptive-RAG\\(BGE)}} & Q & 42.2 & 31.1 & 0.6222 & 42.8 & 47.8 & 31.3 & 0.6645 & 33.4 & 43.5 & 25.1 & 0.6344 & 47.2 & 51.4 & 27.0 & 0.6798 & 33.8 & 25.3 & 22.4 & 0.5005 & 62.6 \\
 & C & 41.9 & 18.0 & 0.6236 & 39.3 & 48.4 & 13.0 & 0.6680 & 26.4 & 39.9 & 11.4 & 0.6321 & 48.9 & 49.5 & 13.2 & 0.6762 & 32.2 & 26.3 & 8.3 & 0.5037 & 61.1 \\
 & Q+C & 44.0 & 30.6 & 0.6359 & 42.4 & 50.0 & 29.6 & 0.6783 & 33.2 & 47.5 & 25.0 & 0.6670 & 48.9 & 51.8 & 25.1 & 0.6877 & 32.6 & 25.6 & 21.9 & 0.5053 & 61.2 \\
\multirow{3}{*}{\shortstack[l]{MBA-RAG\\(BGE)}} & Q & 39.7 & 29.2 & 0.5984 & 41.5 & 47.8 & 30.8 & 0.6546 & 31.9 & 38.5 & 22.2 & 0.5922 & 46.8 & 44.6 & 23.3 & 0.6273 & 31.2 & 23.0 & 20.4 & 0.4836 & 62.1 \\
 & C & 41.9 & 18.0 & 0.6221 & 39.3 & 48.4 & 13.0 & 0.6646 & 26.4 & 39.9 & 11.4 & 0.6323 & 48.9 & 49.5 & 13.2 & 0.6762 & 32.2 & 26.3 & 8.3 & 0.5036 & 61.1 \\
 & Q+C & 43.3 & 29.4 & 0.6215 & 41.6 & 48.5 & 26.7 & 0.6566 & 30.7 & 47.6 & 23.6 & 0.6568 & 48.7 & 50.1 & 27.4 & 0.6678 & 33.0 & 26.8 & 21.4 & 0.5060 & 62.2 \\
\multirow{3}{*}{\shortstack[l]{RouterDC\\(BGE)}} & Q & 41.2 & 31.1 & 0.6148 & 42.6 & 48.8 & 33.8 & 0.6687 & 33.6 & 39.3 & 21.8 & 0.6092 & 46.3 & 49.0 & 26.2 & 0.6643 & 33.5 & 23.6 & 22.7 & 0.4887 & 62.3 \\
 & C & 41.9 & 18.0 & 0.6242 & 39.3 & 48.4 & 13.0 & 0.6680 & 26.4 & 39.9 & 11.4 & 0.6323 & 48.9 & 49.5 & 13.2 & 0.6756 & 32.2 & 26.3 & 8.3 & 0.5063 & 61.1 \\
 & Q+C & 43.8 & 32.5 & 0.6323 & 42.8 & 50.2 & 34.6 & 0.6766 & 34.2 & 44.8 & 23.8 & 0.6454 & 47.1 & 52.9 & 27.5 & 0.6936 & 32.7 & 25.9 & 22.7 & 0.5044 & 62.3 \\
\midrule
\rowcolor{gray!10} Oracle & -- & 100.0 & 100.0 & 1.0000 & 60.8 & 100.0 & -- & -- & 51.6 & 100.0 & -- & -- & 59.9 & 100.0 & -- & -- & 50.5 & 100.0 & -- & -- & 84.7 \\
\midrule
\rowcolor{modelllama}
\multicolumn{22}{c}{\textit{\textbf{Model: LLaMA-3.1-8B}}} \\
\midrule
\rowcolor{gray!10} Random & -- & 24.9 & 21.9 & -- & 26.6 & 24.9 & -- & -- & 16.6 & 25.2 & -- & -- & 26.6 & 24.9 & -- & -- & 26.6 & 24.9 & -- & -- & 44.1 \\
\rowcolor{gray!10} Best-fixed & -- & 14.4 & 6.3 & -- & 25.8 & 12.8 & -- & -- & 18.7 & 9.3 & -- & -- & 21.2 & 13.5 & -- & -- & 25.0 & 21.1 & -- & -- & 41.7 \\
\midrule
\multirow{3}{*}{LR} & Q & 61.1 & 30.0 & 0.7549 & 25.8 & 78.8 & 36.2 & 0.8605 & 18.7 & 59.2 & 28.3 & 0.7527 & 21.3 & 60.0 & 29.6 & 0.7485 & 25.1 & 31.8 & 20.4 & 0.5731 & 41.7 \\
 & C & 59.4 & 18.6 & 0.7497 & 25.8 & 73.2 & 21.1 & 0.8382 & 18.7 & 60.2 & 18.8 & 0.7597 & 21.3 & 58.8 & 18.5 & 0.7408 & 25.1 & 34.8 & 12.9 & 0.5916 & 41.7 \\
 & Q+C & 63.5 & 38.7 & 0.7684 & 27.7 & 78.8 & 36.2 & 0.8661 & 18.7 & 69.2 & 35.8 & 0.8036 & 30.6 & 59.9 & 29.5 & 0.7487 & 24.2 & 35.3 & 22.1 & 0.5859 & 44.0 \\
\midrule
\multirow{3}{*}{SVM} & Q & 61.1 & 30.0 & 0.7531 & 25.8 & 78.8 & 36.2 & 0.8365 & 18.7 & 59.2 & 28.3 & 0.7867 & 21.3 & 60.0 & 29.6 & 0.7481 & 25.1 & 31.8 & 20.4 & 0.5866 & 41.7 \\
 & C & 59.4 & 18.6 & 0.7495 & 25.8 & 73.2 & 21.1 & 0.8382 & 18.7 & 60.2 & 18.8 & 0.7587 & 21.3 & 58.8 & 18.5 & 0.7408 & 25.1 & 34.8 & 12.9 & 0.5916 & 41.7 \\
 & Q+C & 63.1 & 36.3 & 0.7687 & 27.2 & 78.8 & 36.2 & 0.8691 & 18.7 & 69.2 & 35.8 & 0.7907 & 30.6 & 59.4 & 28.8 & 0.7431 & 24.1 & 33.8 & 16.1 & 0.5934 & 42.4 \\
\midrule
\multirow{3}{*}{XGBoost} & Q & 61.1 & 30.0 & 0.7549 & 25.8 & 78.8 & 36.2 & 0.8605 & 18.7 & 59.2 & 28.3 & 0.7527 & 21.3 & 60.0 & 29.6 & 0.7485 & 25.1 & 31.8 & 20.4 & 0.5731 & 41.7 \\
 & C & 59.4 & 18.6 & 0.7497 & 25.8 & 73.2 & 21.1 & 0.8382 & 18.7 & 60.2 & 18.8 & 0.7597 & 21.3 & 58.8 & 18.5 & 0.7408 & 25.1 & 34.8 & 12.9 & 0.5916 & 41.7 \\
 & Q+C & 63.1 & 36.3 & 0.7700 & 27.2 & 78.8 & 36.2 & 0.8691 & 18.7 & 69.2 & 35.8 & 0.8052 & 30.6 & 59.4 & 28.8 & 0.7444 & 24.1 & 33.8 & 16.1 & 0.5890 & 42.4 \\
\midrule
\multirow{3}{*}{MLP} & Q & 60.8 & 27.7 & 0.7526 & 25.8 & 77.6 & 33.0 & 0.8528 & 18.7 & 59.8 & 26.7 & 0.7558 & 21.3 & 60.0 & 27.6 & 0.7481 & 25.1 & 32.2 & 18.8 & 0.5756 & 41.7 \\
 & C & 59.4 & 18.6 & 0.7477 & 25.8 & 73.2 & 21.1 & 0.8382 & 18.7 & 60.2 & 18.8 & 0.7494 & 21.3 & 58.8 & 18.5 & 0.7408 & 25.1 & 34.8 & 12.9 & 0.5897 & 41.7 \\
 & Q+C & 63.0 & 37.4 & 0.7665 & 27.3 & 78.8 & 36.2 & 0.8652 & 18.7 & 67.2 & 32.3 & 0.7921 & 28.5 & 58.9 & 26.5 & 0.7427 & 24.1 & 35.3 & 22.1 & 0.5913 & 44.0 \\
\midrule
\multirow{3}{*}{\shortstack[l]{DNN\\(BERT)}} & Q & 59.3 & 37.8 & 0.7461 & 26.9 & 73.4 & 40.8 & 0.8371 & 18.6 & 61.5 & 36.0 & 0.7601 & 25.7 & 58.7 & 33.9 & 0.7411 & 25.5 & 33.3 & 27.4 & 0.5788 & 43.5 \\
 & C & 59.4 & 18.6 & 0.7493 & 25.8 & 73.2 & 21.1 & 0.8382 & 18.7 & 60.2 & 18.8 & 0.7592 & 21.3 & 58.8 & 18.5 & 0.7398 & 25.1 & 34.8 & 12.9 & 0.5909 & 41.7 \\
 & Q+C & 61.5 & 37.9 & 0.7591 & 27.4 & 74.7 & 38.1 & 0.8452 & 18.8 & 65.2 & 35.9 & 0.7825 & 27.9 & 61.8 & 35.7 & 0.7600 & 25.7 & 35.3 & 27.1 & 0.5903 & 43.4 \\
\multirow{3}{*}{\shortstack[l]{TwoTower\\(BERT)}} & Q & 60.8 & 33.2 & 0.7541 & 26.3 & 74.5 & 34.2 & 0.8431 & 18.1 & 62.9 & 33.3 & 0.7676 & 24.8 & 60.9 & 31.9 & 0.7504 & 25.3 & 35.0 & 24.0 & 0.5898 & 42.6 \\
 & C & 59.4 & 18.6 & 0.7496 & 25.8 & 73.2 & 21.1 & 0.8382 & 18.7 & 60.2 & 18.8 & 0.7595 & 21.3 & 58.8 & 18.5 & 0.7403 & 25.1 & 34.8 & 12.9 & 0.5916 & 41.7 \\
 & Q+C & 61.8 & 34.2 & 0.7604 & 26.6 & 75.0 & 33.6 & 0.8458 & 18.3 & 63.8 & 34.3 & 0.7720 & 27.5 & 62.0 & 32.4 & 0.7607 & 24.7 & 36.9 & 23.9 & 0.6011 & 42.3 \\
\multirow{3}{*}{\shortstack[l]{Adaptive-RAG\\(BERT)}} & Q & 61.6 & 34.0 & 0.7587 & 26.4 & 75.1 & 33.9 & 0.8469 & 18.0 & 63.6 & 34.0 & 0.7702 & 26.1 & 60.6 & 31.5 & 0.7520 & 25.2 & 36.7 & 23.3 & 0.5985 & 42.4 \\
 & C & 59.4 & 18.6 & 0.7497 & 25.8 & 73.2 & 21.1 & 0.8382 & 18.7 & 60.2 & 18.8 & 0.7597 & 21.3 & 58.8 & 18.5 & 0.7408 & 25.1 & 34.8 & 12.9 & 0.5916 & 41.7 \\
 & Q+C & 62.7 & 35.3 & 0.7673 & 27.2 & 75.4 & 32.3 & 0.8510 & 18.7 & 67.8 & 34.8 & 0.8007 & 28.8 & 61.5 & 34.1 & 0.7565 & 25.0 & 37.6 & 23.2 & 0.6047 & 42.7 \\
\multirow{3}{*}{\shortstack[l]{MBA-RAG\\(BERT)}} & Q & 60.7 & 34.0 & 0.7501 & 26.2 & 74.4 & 34.7 & 0.8387 & 18.0 & 63.3 & 32.8 & 0.7669 & 24.8 & 60.0 & 30.5 & 0.7450 & 24.6 & 35.0 & 24.1 & 0.5849 & 42.7 \\
 & C & 59.4 & 18.6 & 0.7462 & 25.8 & 73.2 & 21.1 & 0.8334 & 18.7 & 60.2 & 18.8 & 0.7593 & 21.3 & 58.8 & 18.5 & 0.7394 & 25.1 & 34.8 & 12.9 & 0.5870 & 41.7 \\
 & Q+C & 62.9 & 34.9 & 0.7647 & 27.1 & 75.8 & 31.4 & 0.8476 & 18.7 & 66.9 & 33.8 & 0.7899 & 27.7 & 62.4 & 35.1 & 0.7578 & 25.5 & 37.7 & 23.8 & 0.6058 & 42.8 \\
\multirow{3}{*}{\shortstack[l]{RouterDC\\(BERT)}} & Q & 60.0 & 35.5 & 0.7488 & 26.5 & 73.2 & 37.2 & 0.8341 & 18.7 & 63.5 & 36.1 & 0.7673 & 24.9 & 61.3 & 34.4 & 0.7556 & 25.2 & 33.3 & 25.2 & 0.5804 & 42.3 \\
 & C & 59.4 & 18.6 & 0.7474 & 25.8 & 73.2 & 21.1 & 0.8382 & 18.7 & 60.2 & 18.8 & 0.7595 & 21.3 & 58.8 & 18.5 & 0.7372 & 25.1 & 34.8 & 12.9 & 0.5851 & 41.7 \\
 & Q+C & 61.3 & 37.3 & 0.7593 & 27.2 & 73.8 & 37.5 & 0.8404 & 18.9 & 65.6 & 38.7 & 0.7865 & 28.0 & 61.4 & 34.8 & 0.7568 & 25.3 & 36.5 & 26.4 & 0.5997 & 42.7 \\
\midrule
\multirow{3}{*}{\shortstack[l]{DNN\\(BGE)}} & Q & 58.0 & 37.6 & 0.7378 & 27.6 & 71.4 & 38.4 & 0.8238 & 18.9 & 58.7 & 33.6 & 0.7433 & 26.2 & 57.9 & 34.7 & 0.7386 & 25.9 & 33.8 & 30.8 & 0.5810 & 45.1 \\
 & C & 59.4 & 18.6 & 0.7495 & 25.8 & 73.2 & 21.1 & 0.8382 & 18.7 & 60.2 & 18.8 & 0.7592 & 21.3 & 58.8 & 18.5 & 0.7403 & 25.1 & 34.8 & 12.9 & 0.5916 & 41.7 \\
 & Q+C & 58.3 & 37.6 & 0.7424 & 27.4 & 74.1 & 37.5 & 0.8305 & 18.7 & 61.4 & 35.4 & 0.7561 & 27.6 & 60.4 & 35.3 & 0.7398 & 25.1 & 35.5 & 28.5 & 0.5793 & 44.3 \\
\multirow{3}{*}{\shortstack[l]{TwoTower\\(BGE)}} & Q & 59.4 & 32.4 & 0.7459 & 26.4 & 72.7 & 30.2 & 0.8313 & 18.2 & 61.2 & 31.4 & 0.7565 & 24.2 & 60.0 & 30.9 & 0.7461 & 25.3 & 34.4 & 26.0 & 0.5875 & 43.3 \\
 & C & 59.4 & 18.6 & 0.7496 & 25.8 & 73.2 & 21.1 & 0.8382 & 18.7 & 60.2 & 18.8 & 0.7592 & 21.3 & 58.8 & 18.5 & 0.7408 & 25.1 & 34.8 & 12.9 & 0.5916 & 41.7 \\
 & Q+C & 60.0 & 33.1 & 0.7497 & 27.0 & 74.2 & 30.7 & 0.8339 & 18.6 & 62.8 & 35.1 & 0.7729 & 26.2 & 61.3 & 32.5 & 0.7554 & 25.1 & 35.6 & 23.4 & 0.5812 & 43.9 \\
\multirow{3}{*}{\shortstack[l]{Adaptive-RAG\\(BGE)}} & Q & 60.2 & 33.8 & 0.7500 & 26.4 & 73.3 & 32.6 & 0.8330 & 18.3 & 61.8 & 31.9 & 0.7591 & 24.9 & 60.0 & 30.9 & 0.7492 & 24.9 & 36.1 & 28.3 & 0.5973 & 42.9 \\
 & C & 59.4 & 18.6 & 0.7493 & 25.8 & 73.2 & 21.1 & 0.8382 & 18.7 & 60.2 & 18.8 & 0.7590 & 21.3 & 58.8 & 18.5 & 0.7393 & 25.1 & 34.8 & 12.9 & 0.5916 & 41.7 \\
 & Q+C & 60.3 & 33.9 & 0.7518 & 26.8 & 75.6 & 31.3 & 0.8368 & 18.7 & 62.5 & 31.4 & 0.7665 & 26.1 & 61.1 & 34.2 & 0.7569 & 25.1 & 35.7 & 25.6 & 0.5879 & 43.1 \\
\multirow{3}{*}{\shortstack[l]{MBA-RAG\\(BGE)}} & Q & 56.4 & 34.3 & 0.7190 & 26.8 & 71.2 & 35.0 & 0.8133 & 18.0 & 58.7 & 32.9 & 0.7371 & 26.1 & 56.2 & 32.3 & 0.7195 & 25.6 & 28.7 & 25.1 & 0.5401 & 43.9 \\
 & C & 59.4 & 18.6 & 0.7471 & 25.8 & 73.2 & 21.1 & 0.8328 & 18.7 & 60.2 & 18.8 & 0.7587 & 21.3 & 58.8 & 18.5 & 0.7412 & 25.1 & 34.8 & 12.9 & 0.5907 & 41.7 \\
 & Q+C & 60.2 & 33.5 & 0.7390 & 27.1 & 75.8 & 29.7 & 0.8265 & 18.5 & 62.5 & 31.8 & 0.7629 & 27.2 & 60.5 & 32.3 & 0.7327 & 25.7 & 35.5 & 24.3 & 0.5733 & 43.4 \\
\multirow{3}{*}{\shortstack[l]{RouterDC\\(BGE)}} & Q & 57.7 & 36.5 & 0.7359 & 27.3 & 72.0 & 37.0 & 0.8279 & 18.4 & 57.8 & 32.4 & 0.7358 & 25.2 & 58.9 & 33.4 & 0.7424 & 25.7 & 31.6 & 29.0 & 0.5689 & 45.7 \\
 & C & 59.4 & 18.6 & 0.7488 & 25.8 & 73.2 & 21.1 & 0.8382 & 18.7 & 60.2 & 18.8 & 0.7596 & 21.3 & 58.8 & 18.5 & 0.7389 & 25.1 & 34.8 & 12.9 & 0.5892 & 41.7 \\
 & Q+C & 60.0 & 35.5 & 0.7502 & 27.1 & 74.0 & 38.2 & 0.8397 & 18.9 & 61.5 & 32.9 & 0.7524 & 26.6 & 59.5 & 35.0 & 0.7442 & 25.6 & 35.3 & 25.8 & 0.5938 & 42.8 \\
\midrule
\rowcolor{gray!10} Oracle & -- & 100.0 & 100.0 & 1.0000 & 40.6 & 100.0 & -- & -- & 26.8 & 100.0 & -- & -- & 39.7 & 100.0 & -- & -- & 41.2 & 100.0 & -- & -- & 65.1 \\
\bottomrule
\end{tabular}%
}
\label{tab:router_full}
\end{table*}

\section{Limitations}
\label{sec:limitations}

Our benchmark currently covers four corpus domains, which limits the diversity of corpus-level features available to the router and may constrain generalizability to other domains. The gap between the best router and Oracle indicates that the current feature space has not fully captured all routing-relevant signals, and more expressive corpus representations or fine-tuned encoders may further improve routing performance. Additionally, while our query augmentation ensures logical soundness, synthetic queries may not fully capture the noise distribution characteristic of real-world user interactions. Finally, our router evaluation assumes a fixed set of five RAG paradigms; extending the routing framework to accommodate new or user-defined paradigms remains an open direction.

\end{document}